\documentclass[11pt,a4paper]{article}
\usepackage{jheppub}

\usepackage{graphicx}
\usepackage{amsmath}

\newcommand{\onloop}[1]{\lfloor #1\rfloor}
\newcommand{\oploop}[1]{\lceil #1\rceil}
\newcommand{\oneloop}[1]{\langle #1\rangle}
\newcommand{\sunset}[1]{\langle\langle #1\rangle\rangle}
\newcommand{\sunsetx}[1]{[[ #1 ]]}
\newcommand{\be}{\begin{equation}}
\newcommand{\ee}{\end{equation}}
\newcommand{\ba}{\begin{align}}
\newcommand{\ea}{\end{align}}

\preprint{\begin{minipage}{4cm}\begin{flushright}
LU TP 13-40\\NT@UW-13-26
\\November 2013
\end{flushright}
\end{minipage}
}

\title{Two-loop Sunset Integrals at Finite Volume}

\author[a]{Johan Bijnens,}
\author[a]{Emil Bostr\"om,}
\author[b,c]{and Timo A. L\"ahde}
\affiliation[a]{Department of Astronomy and Theoretical Physics,
Lund University,\\
S\"olvegatan 14A, SE - 223 62 Lund, Sweden}
\affiliation[b]{
Institute for Advanced Simulation, Institut~f\"{u}r Kernphysik, and
J\"{u}lich Center for Hadron Physics, Forschungszentrum~J\"{u}lich,
D-52425~J\"{u}lich, Germany}
\affiliation[c]{Department of Physics, University of Washington, Seattle,
Washington 98195-1560, USA}

\emailAdd{bijnens@thep.lu.se}
\emailAdd{emile.bostrom@gmail.com}
\emailAdd{t.laehde@fz-juelich.de}

\abstract{
We show how to compute the two-loop sunset integrals at finite volume, 
for non-degenerate masses and non-zero momentum. We present results for all
integrals that appear in the Chiral Perturbation Therory ($\chi$PT) calculation
of the pseudoscalar meson masses and decay constants at NNLO, including
the case of Partially Quenched $\chi$PT. We also provide numerical
implementations of the finite-volume sunset integrals, and review the results
for one-loop integrals at finite volume.}


\begin{document}

\maketitle

\section{Introduction}

An analytical, {\it ab initio} description of Quantum Chromodynamics~(QCD)
in the hadronic low-energy regime remains elusive. One of the most promising
alternatives involves numerical evaluation of the functional
integral of QCD on a discretized space-time lattice. Known as
Lattice QCD, this approach has long been restricted for computational reasons
to large and unphysical values of the light quark masses. Recently,
due to improvements
in computing power and algorithmics, calculations with significantly smaller
quark masses have become possible. A side effect of the lowered quark 
masses is an increase in the size of finite-volume corrections, and a 
detailed treatment of such effects is thus called for.

Fortunately, in many cases the finite-volume corrections can be evaluated
analytically using Chiral Perturbation Theory ($\chi$PT)~\cite{Weinberg,GL1}, 
which is the low-energy effective theory of QCD. The application of $\chi$PT
at finite volume was first performed by Gasser and Leutwyler in Ref.~\cite{GL2},
and a review of recent work in this area can be found in Ref.~\cite{Colangelo}.
As
many Lattice QCD simulations are performed with unequal valence and sea-quark
masses~\cite{Sharpe}, the properties of the light pseudoscalar mesons have also been 
calculated to next-to-next-to-leading order (NNLO) in Partially Quenched $\chi$PT (PQ$\chi$PT) in 
Refs.~\cite{Lahde1,Lahde2,Lahde3}. Therefore, it is also of interest to extend 
the finite-volume description of the relevant loop integrals to account for the 
appearance of double poles in the PQ$\chi$PT propagators. It should be noted 
that $\chi$PT is applicable at finite volume as soon as the
typical momenta of a given process are sufficiently small.
This imposes the restriction $F_\pi L > 1$, where $F_\pi$ is the pion 
decay constant and the volume $V \equiv L^3$. This study deals with
the so-called $p$-regime, in which $V$ is sufficiently large for
zero-momentum fluctuations of the meson fields to be treated perturbatively,
which introduces the additional requirement $m_\pi^2 F_\pi^2 V \gg 1$, where $m_\pi^{}$ is the pion mass.
A multitude of finite-volume calculations exist at one-loop or next-to-leading order (NLO), and it 
should also be noted that some work at NNLO has recently appeared.
This includes Ref.~\cite{BG}, where the finite-volume 
corrections to the quark condensate were calculated, and Ref.~\cite{CH} which
considered $m_\pi^{}$ for the case of degenerate quark masses.

Our main objective is to show how the integrals needed in $\chi$PT 
calculations of pseudoscalar meson properties at NNLO and finite volume can be performed. 
As a starting point, the known results at one-loop order are reviewed, and we also show how 
these can be extended to higher order in $d-4$. The methods for the one-loop integrals are 
then applied to the two-loop ``sunset'' integrals for arbitrary masses and momenta. We focus
here on the integrals necessary for the calculation of form factors to NLO, and
for the calculation of masses, decay constants and two-point functions to two loops (NNLO).

This paper is structured as follows: Section~\ref{preliminaries} discusses a few preliminaries.
In Section~\ref{Oneloop}, the derivation of the one-loop
integrals at finite volume is revisited, with emphasis on the 
treatment of PQ$\chi$PT calculations at NLO. In Section~\ref{Twoloop}, the
two-loop sunset integrals are considered, and explicit expressions are given
for the finite and divergent parts, for arbitrary values of the quark
masses and with the propagator structure of PQ$\chi$PT fully 
accounted for. Section~\ref{Numerics} contains a numerical overview
of the integrals presented in this study, along with a concluding
discussion in Section~\ref{Conclusions}. The appendices summarize the ingredients involving 
modified Bessel functions and theta functions, along with basic integrals in $d$ dimensions and 
comments on the notational conventions in earlier work. Some preliminary results related to
this study have been presented in Refs.~\cite{Thesis,latticetalk}.


\section{Preliminaries}
\label{preliminaries}

\subsection{Finite-volume sums}

At finite volume in a cubic box, integrals over momenta should be replaced by
sums over the allowed momenta. In one dimension of length $L$, with periodic boundary 
conditions,\footnote{We do not consider twisted boundary conditions as 
discussed in Ref.~\cite{twisted}. These can be treated by adding a shift to the
allowed momenta, relative to the summations used here.} this entails a summation over the allowed momenta
$p_n \equiv 2\pi n/L$, with $n \in \mathbf{Z}$ integer. The integrals 
over momenta should thus be replaced according to
\begin{equation}
\int\frac{dp}{2\pi}\,F(p) \to \frac{1}{L}\,\sum_{n \in \mathbf{Z}} 
F(p_n) \:\equiv\:
\int_V \frac{dp}{2\pi}\,F(p),
\label{replacement1}
\end{equation}
where the latter notation will be used to indicate a finite-volume summation
in the remainder of this paper. Infinities will be treated by dimensional 
regularization, using the convention $d \equiv 4-2\varepsilon$. The infinite-volume 
integrals have been treated extensively in the literature, see {\it e.g.} Ref.~\cite{ABT1} including
appendices and references therein. 

In practice, 
it is often desirable to study deviations from the infinite-volume limit, and we shall
therefore use a framework in which the infinite-volume
contribution can be easily identified. This can be achieved by application of 
the Poisson summation formula to Eq.~(\ref{replacement1}), yielding
\begin{equation}
\frac{1}{L}\sum_{n\in\mathbf{Z}} F(p_n) =
\sum_{l_p}\int\frac{dp}{2\pi}\,e^{il_pp}_{}\,F(p),
\label{replacement2}
\end{equation}
where the summation over $l_p$ spans a set of vectors of
length $nL$ such that $n\in\mathbf{Z}$. The term with $n = 0$ then represents 
the infinite-volume result, while the sum of all the other terms is the 
finite-volume correction.

In the case of loop integrals over momenta in higher 
dimensions, Eq.~(\ref{replacement2}) should be applied to all dimensions 
which have a finite extent. The four-vector $l_{p\mu}$ then has components 
$(0,n_1L, n_2L, n_3L)$ when three of the dimensions have a finite extent $L$. 
The loop integrals in this paper are performed throughout in Euclidean space, 
with metric $g_{\mu\nu} = \delta_{\mu\nu}$ and signature $(+,+,+,+)$.
Throughout this paper, one of the dimensions (the ``time'' dimension) is assumed
to be much larger in extent than the other three dimensions, which is the
usual situation encountered in Lattice QCD.

\subsection{Passarino-Veltman reduction}

At infinite volume, a general 
method was developed by Passarino and Veltman~\cite{PV} to obtain a minimal set
of integrals by reduction of the tensor integrals $H_{\mu\nu}$ to a set of scalar integrals.
This method relies on separation of the integrals into components that are
scalars under Lorentz transformations and prefactors that contain 
$\delta_{\mu\nu}$ and various momenta. Although Lorentz-invariance is 
explicitly broken by the introduction of a finite size, it is still possible,
in the frame where $p\cdot l_p = 0$, to rewrite the integrals in scalar 
components, provided that a four-vector
\begin{equation}
t_\mu \equiv (1,0,0,0)
\end{equation}
is introduced. The situation $p\cdot l_p = 0$ is referred to as the 
``center-of-mass'' (cms) frame, which is a situation often realized in
Lattice QCD. Because of the remaining symmetries, $t_\mu$ is 
the only additional object required to rewrite the integrals in scalar components,
but we also introduce the tensor
\begin{equation}
 t_{\mu\nu} \equiv \delta_{\mu\nu}-t_\mu t_\nu = \mathrm{diag}(0,1,1,1)
\end{equation}
as a convenient additional abbreviation.


\section{One-loop integrals at finite volume}
\label{Oneloop}

In general, the one-loop integrals in the NNLO expressions for the 
pseudoscalar meson masses and decay constants contain a maximum of two propagators with 
distinct masses. The simplest case with one propagator is denoted~$A$,
whereas the case with two distinct propagators is denoted~$B$. In PQ$\chi$PT, some 
three-propagator integrals denoted~$C$ also appear. These are
due to the mixing of different lowest-order states in PQ$\chi$PT, and they can always be re-expressed in terms
of the~$B$ integrals.

All of the integrals mentioned above have been extensively treated in
the literature, see {\it e.g.} Refs.~\cite{GL2,HL,Beane,Becirevic}.
However, it is instructive to review certain aspects of their derivation and numerical
evaluation here, since they form building blocks in the calculation of the 
two-loop sunset integrals at finite volume. 

\subsection{One-propagator integrals}
\label{onepropagator}

The basic one-loop, one-propagator integrals are
\begin{equation}
\onloop{X} = \int_V \frac{d^dr}{(2\pi)^d} \:
\frac{X}{{(r^2+m^2)}^{n}},
\label{A_int1}
\end{equation}
where $X = 1, r_\mu$ and $r_\mu r_\nu$. By application of the 
Poisson summation formula for the finite dimensions, Eq.~(\ref{A_int1}) 
may be written as
\begin{equation}
\onloop{X} = \sum_{l_r} \int \frac{d^dr}{(2\pi)^d} \:
\frac{X \: e^{il_r\cdot r}}{{(r^2+m^2)}^{n}},
\label{A_int2}
\end{equation}
where the term with $l_r = 0$ represents the infinite-volume contribution.
In order to isolate the finite-volume part, Eq.~(\ref{A_int2}) is decomposed according to
\begin{equation}
\onloop{X} \equiv \onloop{X}^\infty_{} + \onloop{X}^V_{}, 
\label{splitA}
\end{equation}
where the first term represents the infinite-volume result and
will not be considered further. The second term represents the finite-volume correction, and is
free from divergences.

First, we consider the case of $X=1$.
We rewrite Eq.~(\ref{A_int1}) using Eq.~(\ref{prop}) as
\be
\label{onelooplambda}
\onloop{1}^V = \frac{1}{\Gamma(n)}\sum_{l_r}^\prime \int \frac{d^dr}{(2\pi)^d} 
\int_0^\infty d\lambda \,\lambda^{n-1}\,
e^{il_r\cdot r} e^{-\lambda(r^2+m^2)},
\ee
where the primed sum
indicates that the term with $l_r = 0$ is excluded.
We next substitute $r\equiv\bar r +il_r/(2\lambda)$, and obtain
\be
\onloop{1}^V = \frac{1}{\Gamma(n)}\sum_{l_r}^\prime
\int_0^\infty d\lambda \, \lambda^{n-1}\,
e^{-\lambda m^2-\frac{l_r^2}{4\lambda}} 
\int \frac{d^d\bar r}{(2\pi)^d}  \,e^{-\lambda \bar r^2},
\ee
where the $\bar r$ integral can be performed using Eq.~(\ref{simpint}) and by rescaling
$\bar r \equiv \tilde r/\sqrt{\lambda}$, which gives
\be
\label{resultA1}
\onloop{1}^V = \frac{1}{(4\pi)^{d/2}\Gamma(n)}\sum_{l_r}^\prime
\int_0^\infty d\lambda \,\lambda^{n-\frac{d}{2}-1}\:
e^{-\lambda m^2-\frac{l_r^2}{4\lambda}}.
\ee

The (triple) sum and integral can be evaluated in different ways.
The technique used in Refs.~\cite{GL2,HL} is to employ Eq.~(\ref{lambdaint}), which yields
\be
\onloop{1}^V = \frac{1}{(4\pi)^{d/2}\Gamma(n)} \sum_{l_r}^\prime
\mathcal{K}_{n-\frac{d}{2}}\left(\frac{l_r^2}{4}, m^2\right),
\ee
where
the modified Bessel functions $\mathcal{K}_\nu$ are defined in App.~\ref{AppBessel}.
The triple sum can be simplified by observing that $l_r^2 = k L^2$, with
$k$ integer. We further define the factor $x(k)$, which indicates the number of times
each value of $k \equiv n_1^2+n_2^2+n_3^2$ occurs in the triple sum. We then find
\be
\label{xktrick}
\sum_{l_r}^\prime f(l_r^2) = \sum_{k>0} x(k) f(k),
\ee
which reduces the triple sum to a single sum. The final result is
\be
\label{resultAsum}
\onloop{1}^V = \frac{1}{(4\pi)^{d/2}\Gamma(n)} \sum_{k > 0} x(k) \,
\mathcal{K}_{n-\frac{d}{2}}\left(\frac{k L^2}{4},m^2\right),
\ee
where the arguments of $\mathcal{K}_\nu$ can be modified by rescaling 
$\lambda$ before Eq.~(\ref{lambdaint}) is applied. Also, the sum over modified Bessel functions 
is found to converge fairly slowly.

The second method considered here involves performing the summation, and leaving the integral 
to be evaluated numerically, see Ref.~\cite{Becirevic}. We observe that
\be
\label{sumtheta}
\sum_{l_r}^\prime e^{-\frac{l_r^2}{4\lambda}} = \left[\sum_{l_1}
e^{-\frac{L^2}{4\lambda}l_1^2}\right]^3-1,
\ee
using the relation $l_r^2 = (l_1^2+l_2^2+l_3^2)L^2$.
The cubic power accounts for the summations over $l_1,l_2$ and $l_3$.
The remaining sum in Eq.~(\ref{sumtheta}) can be performed in terms of the theta function
$\theta_{30}$, 
which is defined in App.~\ref{AppTheta}. This gives
\be
\label{resultAtheta0}
\onloop{1}^V = \frac{1}{(4\pi)^{d/2}\Gamma(n)}
\int_0^\infty d\lambda \, \lambda^{n-\frac{d}{2}-1}\,
\left[\theta_{30}\left(e^{-L^2/(4\lambda)}\right)^3-1\right]
e^{-\lambda m^2},
\ee
where, as a final step, we rescale $\lambda$ to obtain
\be
\label{resultAtheta}
\onloop{1}^V = \frac{1}{(4\pi)^{d/2}\Gamma(n)} \left(\frac{L^2}{4}\right)^{n-\frac{d}{2}}
\int_0^\infty d\lambda \, \lambda^{n-\frac{d}{2}-1}\,
\left[\theta_{30}(e^{-1/\lambda})^3-1\right] e^{-\lambda\frac{ m^2 L^2}{4}},
\ee
which is also valid for $mL\sim 1$.

Integrals with factors of $r_\mu$ in the numerator are also required. Up to NNLO,
these are $\onloop{r_\mu}$ and $\onloop{r_\mu r_\nu}$. Proceeding as above, we obtain
\begin{align}
\onloop{ (r_\mu ; r_\mu r_\nu)}^V & = 
 \frac{1}{\Gamma(n)}\sum_{l_r}^\prime
\int_0^\infty d\lambda \,\lambda^{n-1}\,
e^{-\lambda m^2-\frac{l_r^2}{4\lambda}} \nonumber \\ 
&\quad \times
\int \frac{d^d\bar r}{(2\pi)^d} 
\left[\bar r_\mu + \frac{il_{r\mu}}{2\lambda} ; 
\bigg(\bar r_\mu + \frac{il_{r\mu}}{2\lambda}\bigg)\bigg(\bar r_\nu + \frac{il_{r\nu}}{2\lambda}\bigg)
\right]  e^{-\lambda \bar r^2},
\end{align}
where we note that integrals odd in $\bar r$ vanish, and that
\begin{equation}
\label{simplifyrelationsr}
\int d^d \bar r \, r_\mu r_\nu \,f(\bar r^2) = \frac{\delta_{\mu\nu}}{d}
\int d^d \bar r \, \bar r^2 f(\bar r^2).
\end{equation}
The summations over the components of $l_r$ include both positive and negative contributions,
and are symmetric under interchange of spatial directions. The sums which are odd in the components of
$l_r$ then vanish, and
\begin{equation}
\label{simplifyrelationsl}
\sum_{l_r} l_{r\mu} l_{r\nu} f(l_r^2) = \frac{1}{3}  t_{\mu\nu}
\sum_{l_r} l_r^2 f(l_r^2).
\end{equation}
Thus, the final results for the $\onloop{r_\mu}$ and $\onloop{r_\mu r_\nu}$ integrals are
\begin{align}
\onloop{r_\mu}^V & = 0,
\nonumber\\
\onloop{r_\mu r_\nu}^V & = \frac{1}{\Gamma(n)}\sum_{l_r}^\prime
\int_0^\infty d\lambda \lambda^{n-1}\:
e^{-\lambda m^2-\frac{l_r^2}{4\lambda}}
\int \frac{d^d\bar r}{(2\pi)^d} 
\left(\frac{\delta_{\mu\nu}}{d} \bar r^2 -\frac{ t_{\mu\nu}}{12\lambda^2}
l_r^2 \right)  e^{-\lambda \bar r^2},
\nonumber\\
& = \frac{1}{(4\pi)^{d/2}\Gamma(n)}\sum_{l_r}^\prime
\int_0^\infty d\lambda \,\lambda^{n-\frac{d}{2}-1}\,
\left(\frac{\delta_{\mu\nu}}{2\lambda} -\frac{ t_{\mu\nu}}{12\lambda^2}
l_r^2 \right)
e^{-\lambda m^2-\frac{l_r^2}{4\lambda}},
\end{align}
where the remaining integration can again be performed in terms of the modified Bessel functions, giving
\be
\label{resultArrsum}
\onloop{r_\mu r_\nu}^V = \frac{1}{(4\pi)^{d/2}\Gamma(n)} \sum_{k > 0} x(k)
\left[
\frac{\delta_{\mu\nu}}{2}
\mathcal{K}_{n-\frac{d}{2}-1}\left(\frac{k L^2}{4},m^2\right)
-\frac{ t_{\mu\nu}}{12}kL^2
\mathcal{K}_{n-\frac{d}{2}-2}\left(\frac{k L^2}{4},m^2\right)\right].
\ee
For the second method which involves the theta functions, we rewrite the sum using the identity
\be
\label{sumtheta2}
\sum_{n\in\mathbb{Z}^3} n^2 q^{(n^2)}
= q\frac{\partial}{\partial q}
\left[\sum_{n\in\mathbb{Z}^3} q^{(n^2)}\right]
= q\frac{\partial}{\partial q}\bigg[\theta_{30}(q)^3\bigg]
= 3 \theta_{32}(q)\theta_{30}(q)^2,
\ee
which is also valid for the primed sums, as the term with $n = 0$ does not contribute.
After rescaling $\lambda$, this gives
\begin{align}
\label{resultArrtheta}
\onloop{r_\mu r_\nu}^V & = 
 \frac{1}{(4\pi)^{d/2}\Gamma(n)} \left(\frac{L^2}{4}\right)^{n-\frac{d}{2}-1}
\int_0^\infty d\lambda \, \lambda^{n-\frac{d}{2}-2}\,
e^{-\lambda\frac{ m^2 L^2}{4}} \nonumber \\
&\quad \times
\left\{\frac{\delta_{\mu\nu}}{2}\left[\theta_{30}\left(e^{-1/\lambda}\right)^3-1\right]
-\frac{t_{\mu\nu}}{\lambda} \left[\theta_{32}\left(e^{-1/\lambda}\right)\,\theta_{30}\left(e^{-1/\lambda}\right)^2 \right]
\right\},
\end{align}
and following the same steps as before, we also find
\be
\onloop{r_\mu r_\nu r_\alpha}^V = 0.
\ee
%


\subsection{Two-propagator integrals}
\label{twopropagator}

The basic one-loop, two-propagator integrals are
\begin{align}
\oneloop{X} =& \int_V \frac{d^dr}{(2\pi)^d} \:
\frac{X}{{(r^2+m_1^2)}^{n_1}{((r-p)^2+m_2^2)}^{n_2}},
\label{B_int1}
\end{align}
where $X = 1,r_\mu,r_\mu r_\nu$ and $r_\mu r_\nu r_\alpha$. By application of the 
Poisson summation formula for the finite dimensions, Eq.~(\ref{B_int1}) may be written as
\begin{align}
\oneloop{X} =& \sum_{l_r} \int \frac{d^dr}{(2\pi)^d} \:
\frac{X \: e^{il_r\cdot r}}{{(r^2+m_1^2)}^{n_1}{((r-p)^2+m_2^2)}^{n_2}},
\label{B_int2}
\end{align}
where the term with $l_r = 0$ represents the infinite-volume contribution.
We again decompose Eq.~(\ref{B_int2}) into the infinite-volume part and the finite-volume correction using
\begin{equation}
\oneloop{X} \equiv \oneloop{X}^\infty_{} + \oneloop{X}^V_{}, 
\label{splitB}
\end{equation}
where the latter term is obtained from Eq.~(\ref{B_int2}) by replacing the unprimed sum with the primed one,
indicating that the term with $l_r = 0$ is excluded.

The methods of Sect.~\ref{onepropagator} also apply here. We begin by introducing Gaussian parameterizations
for both propagators in Eq.~(\ref{B_int2}) in terms of the integration variables $\lambda_1$ and $\lambda_2$.
In a second step, we switch to a new set of variables ($\lambda,x$)
with $\lambda_1 \equiv x\lambda$ and $\lambda_2 \equiv (1-x)\lambda$. Alternatively,
we may first combine the two propagators using the Feynman parameterization 
\be
\frac{1}{a^m b^n} = 
\frac{\Gamma(m+n)}{\Gamma(m)\Gamma(n)}
\int_0^1 dx \,\frac{x^{m-1} y^{n-1}}{(ax+yb)^{m+n}},
\ee
where $y=1-x$, and then treat the denominator according to Eq.~(\ref{prop}). In both cases, the result is
\begin{align}
\oneloop{X}^V
& = \frac{1}{\Gamma(n_1)\Gamma(n_2)} \sum_{l_r}^\prime 
\int_0^1 dx \int \frac{d^dr}{(2\pi)^d} 
\nonumber \\
& \quad \times \int_0^\infty d\lambda\,\lambda^{n_1+n_2-1}\,
x^{n_1-1} y^{n_2-1} X\, e^{il_r\cdot r} e^{-\lambda[x(r^2+m_1^2)+y((r-p)^2+m_2^2)]},
\end{align}
which is equivalent to Eq.~(\ref{onelooplambda}). We now shift the integration variable to
$r \equiv \bar r +il/(2\lambda)+yp$
and obtain for the simplest case 
\begin{align}
\oneloop{1}^V
& = \frac{1}{\Gamma(n_1)\Gamma(n_2)} \sum_{l_r}^\prime
\int_0^1 dx \int_0^\infty d\lambda \, \lambda^{n_1+n_2-1}
x^{n_1-1} y^{n_2-1} \, e^{i y l_r\cdot p} e^{-\lambda\tilde m^2-\frac{l_r^2}{4\lambda}} \, 
 \int \frac{d^dr}{(2\pi)^d} \: e^{-\lambda \bar r^2} \nonumber\\
& = \frac{1}{(4\pi)^{d/2}\Gamma(n_1)\Gamma(n_2)} \sum_{l_r}^\prime
\int_0^1 dx \int_0^\infty 
d\lambda \, \lambda^{n_1+n_2-\frac{d}{2}-1} \, x^{n_1-1} y^{n_2-1} \, e^{i y l_r\cdot p} e^{-\lambda\tilde m^2-\frac{l_r^2}{4\lambda}},
\end{align}
where
\be
\label{defmtilde}
\tilde m^2 = x m_1^2+ym_2^2+xy p^2,
\ee
which differs from Eq.~(\ref{resultA1}) by the integration over $x$
and the factor $e^{iy l_r\cdot p}$. Due to the summation over components of $l_{r\mu}$ with alternating signs,
this factor always produces real-valued results. For the remaining integrals, we obtain
\begin{align}
\oneloop{X}^V & = 
\frac{1}{(4\pi)^{d/2}\Gamma(n_1)\Gamma(n_2)} \sum_{l_r}^\prime
\int_0^1 dx \int_0^\infty d\lambda \, \lambda^{n_1+n_2-\frac{d}{2}-1}
 x^{n_1-1} y^{n_2-1} \oploop{X} \, e^{i y l_r\cdot p} e^{-\lambda\tilde m^2-\frac{l_r^2}{4\lambda}},
\end{align}
with
\begin{align}
\oploop{r_\mu} & = yp_\mu+\frac{il_{r\mu}}{2\lambda}, \nonumber \\
\oploop{r_\mu r_\nu} & = \frac{\delta_{\mu\nu}}{2} + y^2 p_\mu p_\nu + \frac{iy}{2\lambda}
\{l_r,p\}_{\mu\nu}-\frac{l_{r\mu}l_{r\nu}}{4\lambda^2}, \nonumber\\
\oploop{r_\mu r_\nu r_\alpha} & = \frac{1}{2}\left[
 \delta_{\mu\nu}\left(yp_\alpha+\frac{il_{r\alpha}}{2\lambda}\right)
+\delta_{\mu\alpha}\left(yp_\nu+\frac{il_{r\nu}}{2\lambda}\right)
+\delta_{\nu\alpha}\left(yp_\mu+\frac{il_{r\mu}}{2\lambda}\right)\right] \nonumber \\ 
& \quad + \left(yp_\mu+\frac{il_{r\mu}}{2\lambda}\right)
\left(yp_\nu+\frac{il_{r\nu}}{2\lambda}\right)
\left(yp_\alpha+\frac{il_{r\alpha}}{2\lambda}\right),
\end{align}
where $\{a,b\}_{\mu\nu} \equiv a_\mu b_\nu+b_\mu a_\nu$.

\subsubsection{Center-of-mass frame}

In the cms frame, $p=(p,0,0,0)$ such that $p\cdot l_r = 0$ for all $l_r$.
The integrals in the cms frame can be computed similarly to the one-propagator integrals, giving
\begin{align}
\oneloop{1}^V_{n_1 n_2} & =
\frac{\Gamma(n_1+n_2)}{\Gamma(n_1)\Gamma(n_2)}
\int_0^1 dx \, x^{n_1-1} y^{n_2-1} \, \onloop{1}^V_{n_1+n_2},
\nonumber\\
\oneloop{r_\mu}^V_{n_1 n_2} & =
\frac{\Gamma(n_1+n_2)}{\Gamma(n_1)\Gamma(n_2)}
\int_0^1 dx \, x^{n_1-1} y^{n_2} \, p_\mu \onloop{1}^V_{n_1+n_2},
\nonumber\\
\oneloop{r_\mu r_\nu}^V_{n_1 n_2} & =
\frac{\Gamma(n_1+n_2)}{\Gamma(n_1)\Gamma(n_2)}
\int_0^1 dx \, x^{n_1-1} y^{n_2-1} \bigg(
 \onloop{r_\mu r_\nu}^V_{n_1+n_2} + y^2 p_\mu p_\nu \onloop{1}^V_{n_1+n_2} \bigg),
\nonumber\\
\oneloop{r_\mu r_\nu r_\alpha}^V_{n_1 n_2} & =
\frac{\Gamma(n_1+n_2)}{\Gamma(n_1)\Gamma(n_2)}
\int_0^1 dx \, x^{n_1-1} y^{n_2-1} \bigg(
 yp_\alpha \onloop{r_\mu r_\nu}^V_{n_1+n_2} 
+yp_\mu \onloop{r_\nu r_\alpha}^V_{n_1+n_2} 
\nonumber \\
& \quad +yp_\nu \onloop{r_\alpha r_\mu}^V_{n_1+n_2} 
+y^3 p_\mu p_\nu p_\alpha\onloop{1}^V_{n_1+n_2} \bigg),
\end{align}
where the subscripts of the $\onloop{X}^V$ indicate the value of $n$ in the one-propagator integrals given
in Sect.~\ref{onepropagator}. Also, the one-propagator integrals in the above expressions are functions
of $\tilde m^2$ rather than $m^2$. We may then compute the integral over $\lambda$ in $\onloop{X}^V$
and obtain a sum over modified Bessel functions. We are finally left with a single summation
and an integral over $x$, to be performed numerically.

The method introduced in Sect.~\ref{onepropagator} where the summations are performed in terms of
theta functions is also applicable here, and yields a double integral over $\lambda$ and $x$. In that case,
the integral over $x$ can be performed analytically.
By setting
\begin{align}
\tilde m^2 & = -p^2\left(x-\frac{m_1^2-m_2^2+p^2}{2 p^2}\right)^2
+ m_2^2 + \frac{\left(m_1^2-m_2^2+p^2\right)^2}{4p^2}, \nonumber \\
z & = x-\frac{m_1^2-m_2^2+p^2}{2p^2},
\end{align}
the resulting integral with no additional powers of $z$ is related to Dawson's integral or the error function (erf), 
depending on the sign of $p^2$. The other cases are related to the (complex-valued) incomplete Gamma function
by the substitution $z^2=u$. However, a straightforward numerical evaluation of the double integral converges sufficiently
fast for practical purposes.

\subsubsection{Moving frame}

In a general ``moving frame'', $p$ can have non-zero components in the dimensions of finite length.
In this case, the sums with odd powers of components of $l_r$ no longer vanish.
In general, the finite-volume corrections can depend on all components of $p$, and no
simple way of writing the result in terms of scalar functions of $p^2$ exists, as only
a discrete subgroup of the three-dimensional rotation group remains as a symmetry in a finite cubic volume.

Nevertheless, the relevant expressions can be evaluated numerically, albeit with some additional complications.
For the formulation in terms of modified Bessel functions, the summation is no longer exclusively dependent on 
$l_r^2$, and thus the reduction of the triple sums using Eq.~(\ref{xktrick}) is no longer possible. 
For the formulation in terms of theta functions, the summation over $l_r$ can still be performed separately for 
each dimension, provided that the factors of $\theta_{30}^3$ are replaced by the product 
$\theta_3(u_1,q) \, \theta_3(u_2,q)\theta_3(u_3,q)$, where $u_i \equiv y p_i L/(2\pi)$ and $q \equiv e^{-1/\lambda}$.
When factors of $r_\mu$ appear in the integrands, derivatives {\it w.r.t.} $u$ and $q$, as well as
uncontracted factors of $l_{r\mu}$, also need to be accounted for.

\subsection{Summary of one-loop results}
\label{oneloopsummary}

Next, we discuss the relations between the various one-loop integrals and summarize the explicit 
expressions in a concise form. With the definition of Eq.~(\ref{A_int1}) in mind, we introduce the more conventional
notation
\begin{align}
\onloop{1}^V & = A^V,
\nonumber\\
\onloop{r_\mu}^V & = 0,
\nonumber\\
\onloop{r_\mu r_\nu}^V & = \delta_{\mu\nu} A_{22}^V+ t_{\mu\nu} A_{23}^V,
\nonumber\\
\onloop{r_\mu r_\nu r_\alpha}^V & = 0,
\label{Adef}
\end{align}
where only the finite-volume correction has been retained.
As discussed above, no simple rewriting in scalar components is possible for the momentum-dependent integrals, 
except in the cms frame with $p=(p,0,0,0)$. In that frame, we define
\begin{align}
\label{Bdef}
\left.\oneloop{1}^V\right|_{\mathrm{cms}} & = B^V,
\nonumber\\
\left.\oneloop{r_\mu}^V\right|_{\mathrm{cms}} & = p_\mu B^V_1,
\nonumber\\
\left.\oneloop{r_\mu r_\nu}^V\right|_{\mathrm{cms}} & =
 p_\mu p_\nu B_{21}^V+\delta_{\mu\nu}B_{22}^V +B_{23}^V  t_{\mu\nu},
\nonumber\\
\left.\oneloop{r_\mu r_\nu r_\alpha}^V\right|_{\mathrm{cms}} & =
 p_\mu p_\nu p_\alpha B_{31}^V 
+\left(\delta_{\mu\nu}p_\alpha+\delta_{\mu\alpha}p_\nu
+\delta_{\nu\alpha}p_\mu\right)B_{32}^V \nonumber \\
& \quad +\left( t_{\mu\nu}p_\alpha+ t_{\mu\alpha}p_\nu
+ t_{\nu\alpha}p_\mu\right)B_{33}^V,
\end{align}
which correspond to the usual definitions at infinite volume, except for the
terms involving $t_{\mu\nu}$, which appear only in the finite-volume contribution.

The Passarino-Veltman construction~\cite{PV} produces relations between
the various integrals upon multiplication with $p_\mu$ or $\delta_{\mu\nu}$. Using
\begin{equation}
2 p\cdot r = (r^2+m_1^2)-[(r-p)^2+m_2^2]-m_1^2+m_2^2,
\end{equation}
a number of relations can be obtained. These are
%
\begin{align}
& d A_{22}^V(n)+3 A_{23}^V(n)+m^2 A^V(n) = A^V(n-1),
\nonumber\\
& p^2 B_1^V(n_1,n_2)+\frac{1}{2}(m_1^2-m_2^2-p^2) B^V(n_1,n_2)
= \frac{1}{2} B^V(n_1-1,n_2)-\frac{1}{2} B^V(n_1,n_2-1),
\nonumber\\
& p^2 B_{21}^V(n_1,n_2)+ d B_{22}^V(n_1,n_2)+ 3 B_{23}^V(n_1,n_2)
+ m_1^2 B^V(n_1,n_2) = B^V(n_1-1,n_2),
\nonumber\\
& p^2 B_{21}^V(n_1,n_2)+ B_{22}^V(n_1,n_2)
+ \frac{1}{2}(m_1^2-m_2^2-p^2) B_1^V(n_1,n_2) \nonumber \\
& \quad = \frac{1}{2} B_1^V(n_1-1,n_2)-\frac{1}{2}B_1^V(n_1,n_2-1),
\label{PVrelations1}
\end{align}
and
\begin{align}
& p^2 B_{31}^V(n_1,n_2)+ (d+2) B_{32}^V(n_1,n_2)+ 3 B_{33}^V(n_1,n_2)
+ m_1^2 B_1^V(n_1,n_2) = B_1^V(n_1-1,n_2),
\nonumber\\
& p^2 B_{31}^V(n_1,n_2)+ 2B_{32}^V(n_1,n_2)
+ \frac{1}{2}(m_1^2-m_2^2-p^2) B_{21}^V(n_1,n_2) \nonumber \\
& \quad = \frac{1}{2}B_{21}^V(n_1-1,n_2)-\frac{1}{2}B_{21}^V(n_1,n_2-1),
\nonumber\\
& p^2 B_{32}^V(n_1,n_2)+
 \frac{1}{2}(m_1^2-m_2^2-p^2) B_{22}^V(n_1,n_2) =\frac{1}{2} B_{22}^V(n_1-1,n_2)-\frac{1}{2}B_{22}^V(n_1,n_2-1),
\nonumber\\
& p^2 B_{33}^V(n_1,n_2)
+ \frac{1}{2}(m_1^2-m_2^2-p^2) B_{23}^V(n_1,n_2) =\frac{1}{2} B_{23}^V(n_1-1,n_2)-\frac{1}{2}B_{23}^V(n_1,n_2-1),
\label{PVrelations2}
\end{align}
%
where we note that the relations in Eq.~(\ref{PVrelations2}) are linearly dependent. 
Up to the order considered here, this leaves $A^V, A_{23}^V, B^V$ and $B_{23}^V$ as independent functions.
We have checked the validity of the above relations numerically for $n_1,n_2 = 1,2$. 

At NNLO in $\chi$PT, all one-loop integrals should be expanded around $d=4$
up to and including terms of $\mathcal{O}(\varepsilon)$. This is necessary,
since products of two one-loop integrals appear throughout the NNLO expressions,
including the factorizable parts of the two-loop sunset integrals. 
We thus define 
\begin{align}
A^{V} & \equiv \bar A^{V} + \varepsilon\bar A^{V\varepsilon} \,+\, \mathcal{O}(\varepsilon^2), \nonumber \\
B^{V} & \equiv \bar B^{V} + \varepsilon\bar B^{V\varepsilon} \,+\, \mathcal{O}(\varepsilon^2),
\end{align}
with similar expansions for all functions $A^V_i$ and $B^V_i$ in Eqs.~(\ref{Adef}) and (\ref{Bdef}).
The one-propagator integrals can then be written as
\begin{equation}
\bar A^V = \frac{1}{16\pi^2\Gamma(n)}\sum_{k>0} x(k) \hat A^V
=  \frac{1}{16\pi^2\Gamma(n)}\left(\frac{L^2}{4}\right)^{n-2} \int_0^\infty d\lambda \, \lambda^{n-3}e^{-\lambda\frac{m^2 L^2}{4}}
\tilde A^V,
\end{equation}
using Eqs.~(\ref{resultAsum}), (\ref{resultAtheta}), (\ref{resultArrsum})
and~(\ref{resultArrtheta}). The integrands can be expressed either in terms of modified Bessel functions or theta functions, and are
in each case given by
\begin{align}
\label{Afinalresult}
\hat A^V & = \mathcal{K}_{n-2}\left(\frac{k L^2}{4},m^2\right),
&\tilde A^V & = \theta_{30}\left(e^{-1/\lambda}\right)^3-1,
\nonumber\\
\hat A_{22}^V & = \frac{1}{2} \, \mathcal{K}_{n-3}\left(\frac{k L^2}{4},m^2\right),
&\tilde A_{22}^V & = \frac{2}{\lambda L^2}
\left[\theta_{30}\left(e^{-1/\lambda}\right)^3-1\right],
\nonumber\\
\hat A_{23}^V & = -\frac{1}{12}k L^2 \, \mathcal{K}_{n-4}\left(\frac{k L^2}{4},m^2\right),
&\tilde A_{23}^V & = -\frac{4}{\lambda^2 L^2} \,
\theta_{32}\left(e^{-1/\lambda}\right)\theta_{30}\left(e^{-1/\lambda}\right)^2.
\end{align}

The expansion in $\varepsilon = (4-d)/2$ can be performed using
\begin{align}
\label{epsexpansion}
(4\pi)^\varepsilon& = 1+\varepsilon \log(4\pi)+\mathcal{O}(\varepsilon^2), \nonumber \\
\mathcal{K}_{m+\varepsilon}& = \mathcal{K}_{m}
+\varepsilon \tilde{\mathcal{K}}_{m}+\mathcal{O}(\varepsilon^2), \nonumber \\
(4\pi\lambda L^2)^\varepsilon& = 1+\varepsilon\log(4\pi\lambda L^2)
 +\mathcal{O}(\varepsilon^2),
\end{align}
where the functions $\tilde{\mathcal{K}}_m$ are related to the modified Bessel functions and are
defined in App.~\ref{AppBessel}. For all quantities in Eq.~(\ref{Afinalresult}), the above results lead to
\begin{align}
\label{Aepsilon}
\hat A^{V\varepsilon} & = \log(4\pi) \, \hat A^V
 + A^V(\mathcal{K}_m \to \tilde{\mathcal{K}}_m), \nonumber \\
\tilde A^{V\varepsilon} & = [\log(4\pi)+\log(\lambda)+2\log(L)] \,\tilde A^V,
\end{align}
where $\mathcal{K}_m\to \tilde{\mathcal{K}}_m$ indicates that the functions $\mathcal{K}_m$
should be replaced by the corresponding expressions for $\tilde{\mathcal{K}}_m$.

For the one-loop two-propagator integrals, we find similar results, given by
\begin{align}
\bar B^V & = \frac{1}{16\pi^2\Gamma(n_1)\Gamma(n_2)}
\sum_{k>0}x(k) 
\int_0^1dx \:x^{n_1-1}y^{n_2-1}
\hat B^V \nonumber \\
 & = \frac{1}{16\pi^2\Gamma(n_1)\Gamma(n_2)}\int_0^1dx \:x^{n_1-1}y^{n_2-1}
\left(\frac{L^2}{4}\right)^{n_1+n_2-2} \int_0^\infty d\lambda \, \lambda^{n_1+n_2-3}e^{-\lambda\frac{\tilde m^2 L^2}{4}}
\tilde B^V,
\end{align}
with $\tilde m^2 = x m_1^2+(1-x)m_2^2+xyp^2$ and $y = 1-x$, where $x(k)$ is defined in Eq.~(\ref{xktrick}). The explicit expressions 
for the integrands are
\begin{align}
\label{Bfinalresult}
\hat B^V & = \mathcal{K}_{n_1+n_2-2}\left( \frac{k L^2}{4},\tilde m^2\right),
&\tilde B^V & = \theta_{30}\left(e^{-1/\lambda}\right)^3-1,
\nonumber\\
\hat B_1^V & = y\,\mathcal{K}_{n_1+n_2-2}\left( \frac{k L^2}{4},\tilde m^2\right),
&\tilde B_1^V & = y\left[\theta_{30}\left(e^{-1/\lambda}\right)^3-1\right],
\nonumber\\
\hat B_{21}^V & = y^2\mathcal{K}_{n_1+n_2-2}\left( \frac{k L^2}{4},\tilde m^2\right),
&\tilde B_{21}^V & = y^2\left[\theta_{30}\left(e^{-1/\lambda}\right)^3-1\right],
\nonumber\\
\hat B_{22}^V & = \frac{1}{2} \mathcal{K}_{n_1+n_2-3}\left( \frac{k L^2}{4},\tilde m^2\right),
&\tilde B_{22}^V & = \frac{2}{\lambda L^2}
\left[\theta_{30}\left(e^{-1/\lambda}\right)^3-1\right],
\nonumber\\
\hat B_{23}^V & = -\frac{1}{12}k L^2 \mathcal{K}_{n_1+n_2-4}\left( \frac{k L^2}{4},\tilde m^2\right),
&\tilde B_{23}^V & = -\frac{4}{\lambda^2 L^2} \,
\theta_{32}\left(e^{-1/\lambda}\right)\theta_{30}\left(e^{-1/\lambda}\right)^2.
\nonumber\\
\hat B_{31}^V & = y^3\mathcal{K}_{n_1+n_2-2}\left(\frac{k L^2}{4},\tilde m^2\right),
&\tilde B_{31}^V & = y^3\left[\theta_{30}\left(e^{-1/\lambda}\right)^3-1\right],
\nonumber\\
\hat B_{32}^V & = \frac{y}{2} \mathcal{K}_{n_1+n_2-3}\left(\frac{k L^2}{4},\tilde m^2\right),
&\tilde B_{32}^V & = y\frac{2}{\lambda L^2}
\left[\theta_{30}\left(e^{-1/\lambda}\right)^3-1\right],
\nonumber\\
\hat B_{33}^V &  -\frac{y}{12}k L^2 \mathcal{K}_{n_1+n_2-4}\left(\frac{k L^2}{4},\tilde m^2\right),
&\tilde B_{33}^V & = -\frac{4y}{\lambda^2 L^2} \,
\theta_{32}\left(e^{-1/\lambda}\right)\theta_{30}\left(e^{-1/\lambda}\right)^2,
\end{align}
where each case has again been given in terms of modified Bessel functions or theta functions.
The functions $\bar B^{V\varepsilon}$ can be obtained from the above expressions using the equivalent of
Eq.~(\ref{epsexpansion}), along with corresponding changes in Eq.~(\ref{Aepsilon}). However, the functions 
$\bar A^{V\varepsilon}$ and $\bar B^{V\varepsilon}$ are expected to cancel completely in a full calculation within 
the $\overline{MS}$ scheme. This cancellation has already been demonstrated at NNLO for the scalar condensate 
in Ref.~\cite{BG}, and for $m_\pi^{}$ in two-flavour ChPT in Ref.~\cite{CH}.


\section{Two-loop sunset integrals at finite volume}
\label{Twoloop}

First, we recall that some NNLO work at finite volume 
already exists. In Ref.~\cite{BG}, the finite-volume corrections were 
calculated for the quark condensate, and in Ref.~\cite{CH} for $m_\pi^{}$. 
The former only involved products of one-loop integrals, while the latter 
only required consideration of the sunset integrals with degenerate masses.
In this section, we provide completely general expressions for the sunset integrals,
for arbitrary, non-degenerate masses. At finite volume, we define the basic sunset integral as
\begin{align}
\label{defsunset}
\sunset{X} & \equiv \int_V 
\frac{d^dr}{(2\pi)^d} \frac{d^ds}{(2\pi)^d} \: \frac{X}
{{(r^2+m_1^2)}^{n_1^{}}{(s^2+m_2^2)}^{n_2^{}}{((r+s-p)^2+m_3^2)}^{n_3^{}}}, 
\end{align}
where the required operators $X$ are
$1, r_\mu^{}, s_\mu^{}, r_\mu^{} r_\nu^{}, r_\mu^{} s_\nu^{}$ and 
$s_\mu^{} s_\nu^{}$. In Eq.~(\ref{defsunset}), the $n_i^{}$ are always
non-zero and positive. If one of the $n_i^{}$ is zero or negative, the integral 
becomes separable into a product of two one-loop integrals, which we have already
dealt with in Section~\ref{Oneloop}.

Application of the
Poisson summation formula for all momenta in a finite dimension yields
\begin{align}
\label{defsunset2}
\sunset{X} & =  
\sum_{l_r^{},l_s^{}} \int 
\frac{d^dr}{(2\pi)^d}
\frac{d^ds}{(2\pi)^d} \:
\frac{X \: e^{il_r^{} \cdot r}e^{il_s^{} \cdot s}}
{{(r^2+m_1^2)}^{n_1^{}}{(s^2+m_2^2)}^{n_2^{}}{((r+s-p)^2+m_3^2)}^{n_3^{}}},
\end{align}
where $\sunset{X}(1,2,3)$ will be used as a short-hand notation
indicating which of the arguments ($n_i^{}$,$m_i^2$) are associated with the first, second
and third propagators in Eq.~(\ref{defsunset2}), respectively. 
The vectors $l_r,l_s$ are of the form $(0,k_1L,k_2L,k_3L)$ with $k_i\in \mathbf{Z}$. 
Eq.~(\ref{defsunset2}) can then be decomposed according to
\begin{align}
\label{split1}
\sunset{X} & \equiv \sunset{X}_{}^\infty + \sunset{X}_{}^V,
\end{align}
where $\sunset{X}_{}^\infty$ denotes the infinite-volume result with $l_r=l_s=0$.
The sunset integrals at infinite volume have been evaluated
in several different ways (see {\it e.g.} Refs.~\cite{ABT2,GS,Korner,bokasun})
and will not be considered further here. The second term in Eq.~(\ref{split1}) represents the 
finite-volume correction. The present approach to the finite-volume correction
is along the lines of Refs.~\cite{ABT2,GS}, combined with an extension of the methods
for the one-loop integrals in Section~\ref{Oneloop}.

We further decompose $\sunset{X}_{}^V$ into terms
where one of the possible loop momenta is not quantized
and a contribution where both are quantized, according to
\begin{align}
\sunset{X}^V & \equiv \sunset{X}_r^{}
+ \sunset{X}_s^{}
+ \sunset{X}_t^{}
+ \sunset{X}_{rs}^{},
\label{ssplit}
\end{align}
with
\begin{align}
\sunset{X}_r & = \sum_{l_r}^\prime \int 
\frac{d^dr}{(2\pi)^d}
\frac{d^ds}{(2\pi)^d} \:
\frac{X \: e^{il_r \cdot r}}
{{(r^2+m_1^2)}^{n_1^{}}{(s^2+m_2^2)}^{n_2^{}}{((r+s-p)^2+m_3^2)}^{n_3^{}}},
\nonumber\\
\sunset{X}_s & = \sum_{l_r}^\prime \int 
\frac{d^dr}{(2\pi)^d}
\frac{d^ds}{(2\pi)^d} \:
\frac{X \: e^{il_s \cdot s}}
{{(r^2+m_1^2)}^{n_1^{}}{(s^2+m_2^2)}^{n_2^{}}{((r+s-p)^2+m_3^2)}^{n_3^{}}},
\nonumber\\
\sunset{X}_t & = \sum_{l_t}^\prime \int 
\frac{d^dr}{(2\pi)^d}
\frac{d^ds}{(2\pi)^d} \:
\frac{X \: e^{il_t^{} \cdot (p-r-s)}}
{{(r^2+m_1^2)}^{n_1^{}}{(s^2+m_2^2)}^{n_2^{}}{((r+s-p)^2+m_3^2)}^{n_3^{}}},
\nonumber\\
\sunset{X}_{rs} & = \sum_{l_r,l_s}^{\prime\prime} \int 
\frac{d^dr}{(2\pi)^d}
\frac{d^ds}{(2\pi)^d} \:
\frac{X \: e^{il_r^{} \cdot r}e^{il_s^{} \cdot s}}
{{(r^2+m_1^2)}^{n_1^{}}{(s^2+m_2^2)}^{n_2^{}}{((r+s-p)^2+m_3^2)}^{n_3^{}}},
\label{defsunset3}
\end{align}
where a ``singly primed'' sum indicates that the term with $l = 0$ has been
excluded. For the ``doubly primed'' sums, all contributions with $l_r^{} = 0$, 
$l_s^{} = 0$ or $l_r^{} = l_s^{}$ have been removed, {\it i.e.} the retained terms
satisfy $l_r\ne 0,l_s\ne 0$ and $l_r\ne l_s$.
The sum of all the terms in Eq.~(\ref{defsunset3}) reproduces the full sum in Eq.~(\ref{defsunset2}).
Here, it should be taken into account that $p$ is also quantized
in the finite dimensions, such that the spatial momentum components satisfy
\begin{equation}
\label{piquantized}
p_i^{} \equiv \frac{2\pi j_i^{}}{L},
\quad \quad e^{il_r\cdot p} = e^{il_s\cdot p} = e^{il_t\cdot p} = 1.
\end{equation}

We note that $\sunset{X}_{rs}$ is 
always finite, whereas $\sunset{X}_r$, $\sunset{X}_s$ and 
$\sunset{X}_t$ may contain a non-local divergence, depending on the 
operator $X$ and the values of the $n_i$. If these integrals should
be finite, they can be included in $\sunset{X}_{rs}^{}$ by summation over all values
of $l_r^{}$ and $l_s^{}$ (except of course $l_r^{} = l_s^{} = 0$).

\subsection{Simplest sunset integral}

We first restrict ourselves to the simplest case of $\sunset{1}$ with $n_1=n_2=n_3=1$, which allows
us to outline our procedure in a straightforward way. We will then proceed to give the
expressions for the general case using the formalism established here.

From Eqs.~(\ref{defsunset}), (\ref{defsunset2}) and (\ref{defsunset3}), and keeping in mind Eq.~(\ref{piquantized}), 
we find that the sunset integrals exhibit a high degree of symmetry with
respect to interchanges of $r$, $s$ and $t=p-r-s$, together with $l_r$, $l_s$ and $l_t$.
Substituting $(r,s) \to (s,r)$ and $(r,t) \to (t,r)$, including the respective $l_i^{}$, leads to
the relations
\begin{align}
\label{relationssunset1}
\sunset{1}(1,2,3) & = \sunset{1}(2,1,3)=\sunset{1}(3,2,1),
\nonumber\\
\sunset{1}^\infty(1,2,3) & = \sunset{1}^\infty(2,1,3)=\sunset{1}^\infty(3,2,1),
\nonumber\\
\sunset{1}^V(1,2,3) & = \sunset{1}^V(2,1,3)=\sunset{1}^V(3,2,1),
\nonumber\\
\sunset{1}_{rs}(1,2,3) & = \sunset{1}_{rs}(2,1,3)=\sunset{1}_{rs}(3,2,1),
\nonumber\\
\sunset{1}_r(1,2,3) & = \sunset{1}_r(1,3,2),
\nonumber\\
\sunset{1}_r(1,2,3) & = \sunset{1}_s(2,1,3)=\sunset{1}_t(3,2,1),
\end{align}
where we recall that the notation $(1,2,3)$ refers to the propagators, as 
exhibited in Eq.~(\ref{defsunset2}). From the last relation in Eq.~(\ref{relationssunset1}), 
we find that the evaluation of $\sunset{1}_r$ and $\sunset{1}_{rs}$ suffices to obtain the full result.

\subsubsection{Simplest sunset integral with one quantized loop momentum}
\label{simplestr}

First, we calculate $\sunset{1}_r$. We begin by combining two of the propagators
with a Feynman parameter $x$, giving
\begin{align}
\sunset{1}_r & = \sum_{l_r}^\prime \int 
\frac{d^dr}{(2\pi)^d}
\frac{d^ds}{(2\pi)^d} \: 
\frac{ e^{il_r^{} \cdot r}}{(r^2+m_1^2)(s^2+m_2^2)((r+s-p)^2+m_3^2)} \nonumber \\
& = \sum_{l_r}^\prime\int
\frac{d^dr}{(2\pi)^d} \:
\frac{e^{il_r^{} \cdot r}}{{(r^2+m_1^2)}}\:
\int_0^1 dx
\int \!\!\frac{d^d\tilde s}{(2\pi)^d} \:
\frac{1}{\left(\tilde s^2 + \overline m^2\right)^2},
\end{align}
where we have shifted the integration variable according to
$s_\mu \equiv \tilde s_\mu - x (r-p)_\mu$, and defined
\begin{equation}
\overline m^2 \equiv (1-x) m_2^2 + x m_3^2 + 
x(1-x)(r-p)^2.
\end{equation} 
The integration over $\tilde s$ may then be performed in terms
of standard $d$-dimensional integrals in Euclidean space, given in App.~\ref{feynmanintegrals}. This gives
\begin{equation}
\sunset{1}_r = \sum_{l_r}^\prime\int 
\frac{d^dr}{(2\pi)^d}
\frac{e^{il_r^{} \cdot r}}{(r^2+m_1^2)}
\int_0^1 dx
\frac{\Gamma\left(2-\frac{d}{2}\right)}{(4\pi)^{\frac{d}{2}}}\:
(\overline m^2)^{\frac{d}{2}-2},
\end{equation}
where the expansion
to $\mathcal{O}(\varepsilon)$ may be performed using
\begin{align}
\label{expandepsilon}
\frac{\Gamma\left(2-\frac{d}{2}\right)}{(4\pi)^{\frac{d}{2}}}\:
(\overline m^2)^{\frac{d}{2}-2}_{} & = \frac{1}{16\pi^2}\left[\lambda_0^{} 
- 1 - \log(\overline m^2)\right] \:+\: \mathcal{O}(\varepsilon), 
\end{align}
where $\lambda_0^{} \equiv 1/\varepsilon + \log(4\pi) + 1 - \gamma$. The term proportional
to $\lambda_0$ involves the one-loop integral
$A^V$, which has been treated in Sect.~\ref{Oneloop}. This also contains
the nonlocal divergence, and contributes
\begin{equation}
\label{defsunsetA}
\sunset{1}_{r,A} = \frac{\lambda_0}{16\pi^2}
\onloop{1}^V(1,m_1^2)
\end{equation}
to $\sunset{1}_r$. For clarity, we have added the arguments $n_1=1$ and $m_1^2$
to the notation for the one-loop integral. The remaining terms in Eq.~(\ref{expandepsilon}) contribute
%
\begin{equation}
\label{defsunsetF}
\sunset{1}_{r,F} =
-\frac{1}{16\pi^2}\sum_{l_r}^\prime\int 
\frac{d^dr}{(2\pi)^d}
\frac{e^{il_r^{} \cdot r}}{(r^2+m_1^2)}
\int_0^1 dx\left[1+\log(\overline m^2)\right],
\end{equation}
where we can set $d = 4$ directly.
In order to deal with the dependence of $\overline m^2$ or $r$,
we perform a partial integration in $x$ to obtain
\begin{align}
\sunset{1}_{r,F} & = 
-\frac{1}{16\pi^2}\sum_{l_r}^\prime\int 
\frac{d^4r}{(2\pi)^4}
\frac{e^{il_r^{} \cdot r}}{(r^2+m_1^2)}
\Bigg[1+\log(m_3^2)
\nonumber\\
 & \quad -\int_0^1 dx \, x \, \frac{m_3^2-m_2^2+(1-2x)(r-p)^2}
{\overline m^2}\Bigg].
\end{align}
Here, the first two terms once more contain a one-loop integral, and
we refer to this part as $\sunset{1}_{r,G}$, with the remainder labeled
$\sunset{1}_{r,H}$. Further, we introduce the Gaussian parameters
$\lambda_1$ and $\lambda_4$ according to Eq.~(\ref{prop}) for the denominators 
$(r^2+m_1^2)$ and $\overline m^2$, respectively. This gives
\begin{align}
\label{defsunsetGH}
\sunset{1}_{r,F} & \equiv \sunset{1}_{r,G}+\sunset{1}_{r,H},
\nonumber\\
\sunset{1}_{r,G} & = -\frac{1+\log(m_3^2)}{16\pi^2} \onloop{1}^V(1,m_1^2),
\nonumber\\
\sunset{1}_{r,H} & = \frac{1}{16\pi^2}\sum_{l_r}^\prime \int \frac{d^4r}{(2\pi)^4}
\int_0^\infty\! d\lambda_1 d\lambda_4 \int_0^1 dx 
\nonumber\\
& \quad \times
x \left[m_3^2-m_2^2+(1-2x)(r-p)^2\right]
e^{il_r^{} \cdot r
-\lambda_1(r^2+m_1^2)-\lambda_4\overline m^2},
\end{align}
where we may complete the square in the exponential factor by substituting
\begin{align}
r & \equiv \frac{1}{\sqrt{\lambda_5}} \, \tilde r +\frac{i l_r}{2\lambda_5}
+\frac{x(1-x)\lambda_4}{\lambda_5} \, p, 
\nonumber \\
\lambda_5 & \equiv \lambda_1+x(1-x)\lambda_4.
\label{complete}
\end{align}
The $\tilde r$ integral can then be performed using Eq.~(\ref{simpint}), which gives
\begin{align}
\sunset{1}_{r,H} & = \frac{1}{(16\pi^2)^2}\sum_{l_r}^\prime\int_0^\infty\! d\lambda_1 d\lambda_4 
\int_0^1 dx \, \frac{x}{\lambda_5^2}
\nonumber\\
& \quad \times
\left[m_3^2-m_2^2+\frac{1-2x}{\lambda_5^2}
\left(2\lambda_5+\lambda_1^2 p^2-\frac{l_r^2}{4}-i\lambda_1 l_r\cdot p\right)\right]
\nonumber\\
& \quad \times
e^{-\left(\lambda_1 m_1^2+\lambda_4(1-x) m_2^2+ \lambda_4 x m_3^2
+\frac{\lambda_1\lambda_4 x(1-x)}{\lambda_5}p^2
+\frac{l_r^2}{4\lambda_5}-i\frac{\lambda_4x(1-x)}{\lambda_5}l_r\cdot p\right)}.
\end{align}
Here, a more symmetric form can be obtained by substituting $\lambda_2 \equiv (1-x)\lambda_4$
and $\lambda_3 \equiv x \lambda_4$ as integration variables, giving
\begin{align}
\sunset{1}_{r,H} & = \frac{1}{(16\pi^2)^2}\sum_{l_r}^\prime\int_0^\infty
d\lambda_1 d\lambda_2d\lambda_3 \, \frac{\lambda_3}{\tilde \lambda^2} 
\left[m_3^2-m_2^2+\frac{\lambda_2-\lambda_3}{\tilde\lambda}
\left(2+\frac{\lambda_3+\lambda_2}{\tilde\lambda}
\tilde p^2
\right)\right] e^{-M^2},
\label{defM}
\end{align} 
with
\begin{align}
M^2 & \equiv \lambda_1 m_1^2+\lambda_2 m_2^2+\lambda_3 m_3^2
+\frac{\lambda_1 \lambda_2 \lambda_3}{\tilde\lambda} p^2
+\frac{\lambda_2+\lambda_3}{\tilde\lambda}\frac{l_r^2}{4}
-i\frac{\lambda_2\lambda_3}{\tilde\lambda}l_r\cdot p,
\nonumber \\
\tilde\lambda & \equiv \lambda_1\lambda_2+\lambda_2\lambda_3+\lambda_3\lambda_1,
\nonumber \\
\tilde p & \equiv \frac{i l_r}{2}-\lambda_1 p,
\end{align} 
which can be evaluated numerically with the methods discussed in Sect.~\ref{simplestnumerics}.

\subsubsection{Simplest sunset integral with two quantized loop momenta}
\label{simplestrs}

Second, we calculate $\sunset{1}_{rs}$. We introduce Gaussian parameterizations for
all three propagators using Eq.~(\ref{prop}) and set $d=4$, giving
\begin{align}
\sunset{1}_{rs} & = \sum_{l_r,l_s}^{\prime\prime}\int_0^\infty d\lambda_1 d\lambda_2 d\lambda_3 
\int \frac{d^4r}{(2\pi)^4} \frac{d^4s}{(2\pi)^4} 
\nonumber \\
& \quad \times
e^{-\left(\lambda_1 m_1^2+\lambda_2 m_2^2 +\lambda_3 m_3^2-il_r \cdot r- l_s \cdot s
+\lambda_1 r^2 +\lambda_2 s^2+\lambda_3 (r+s-p)^2\right)},
\end{align}
after which we perform the redefinition
\begin{equation}
\label{shiftr}
 r \equiv \frac{1}{\sqrt{\lambda_1+\lambda_3}} \tilde r
       -\frac{\lambda_3}{\lambda_1+\lambda_3}(s-p)
       +\frac{i}{2(\lambda_1+\lambda_3)}l_r,
\end{equation}
and shift $s$ by
\begin{equation}
\label{shifts}
s \equiv \frac{\sqrt{\lambda_1+\lambda_3}}{\sqrt{\tilde\lambda}}\tilde s
  +\frac{\lambda_1\lambda_3}{\tilde\lambda} p
  +\frac{i(\lambda_1+\lambda_3)}{2\tilde\lambda}l_s
  -\frac{i\lambda_3}{2\tilde\lambda}l_r,
\end{equation}
where we have again made use of 
$\tilde \lambda \equiv \lambda_1\lambda_2+\lambda_2\lambda_3+\lambda_3\lambda_1$.
We note that an analogous transformation results by first redefining $s$ and then shifting $r$.
The result is
\begin{align}
\sunset{1}_{rs} & = \sum_{l_r,l_s}^{\prime\prime}\int_0^\infty d\lambda_1 d\lambda_2
d\lambda_3 \int \frac{d^4\tilde r}{(2\pi)^4}\frac{d^4\tilde s}{(2\pi)^4} \:
\tilde\lambda^{-2}e^{-\tilde r^2 -\tilde s^2-\tilde M^2}
\nonumber\\
& = \frac{1}{(16\pi^2)^2}
\sum_{l_r,l_s}^{\prime\prime}\int_0^\infty d\lambda_1 d\lambda_2 d\lambda_3 \:
\tilde\lambda^{-2}e^{-\tilde M^2},
\label{defMtilde}
\end{align}
with
\begin{align}
\tilde M^2 & \equiv \lambda_1 m_1^2+\lambda_2 m_2^2+\lambda_3 m_3^2
+\frac{\lambda_1 \lambda_2 \lambda_3}{\tilde\lambda} p^2
+\frac{\lambda_2}{\tilde\lambda}\frac{l_r^2}{4}
+\frac{\lambda_1}{\tilde\lambda}\frac{l_s^2}{4}
+\frac{\lambda_3}{\tilde\lambda}\frac{(l_r-l_s)^2}{4} \nonumber\\
& \quad
-i\frac{\lambda_2\lambda_3}{\tilde\lambda}l_r\cdot p
-i\frac{\lambda_1\lambda_3}{\tilde\lambda}l_s\cdot p.
\end{align}
We note that the arguments of the exponential functions in Eqs.~(\ref{defM}) and~(\ref{defMtilde}) coincide
when $l_s=0$.

\subsubsection{Numerical evaluation}
\label{simplestnumerics}

Next, we discuss the numerical evaluation of Eq.~(\ref{defMtilde}).
For this purpose, it is convenient to switch to the variables
$x,y,z$ and $\lambda$, 
\begin{equation}
\label{xyzlambda}
\lambda_1 \equiv x\lambda, \quad 
\lambda_2 \equiv y\lambda, \quad 
\lambda_3 \equiv (1-x-y)\lambda = z\lambda,  \quad
\tilde\lambda = \lambda^2(xy+yz+zx) \equiv \lambda^2\sigma,
\end{equation}
where $\sigma \equiv xy+yz+zx$ and $x+y+z = 1$. We also introduce the quantities
\begin{align}
\label{rsdefs}
l_n & \equiv l_r-l_s,
\nonumber\\
S_{rs} & \equiv -\frac{yz}{\sigma}l_r\cdot p-\frac{xz}{\sigma}l_s\cdot p,
\nonumber\\
Y_{rs} & \equiv \frac{y}{4\sigma}l_r^2+\frac{x}{4\sigma}l_s^2
+\frac{z}{4\sigma}l_n^2,
\nonumber\\
Z_{rs} & \equiv x m_1^2+y m_2^2 + z m_3^2+\frac{xyz}{\sigma}p^2,
\end{align}
which brings Eq.~(\ref{defMtilde}) into the form
\begin{equation}
\label{xyzintegral}
\sunset{1}_{rs} = \frac{1}{(16\pi^2)^2}
\sum_{l_r,l_s}^{\prime\prime}\int_0^\infty d\lambda\int_0^1 dx \int_0^{1-x} dy \,
\sigma^{-2}\lambda^{-2} \,
e^{-\lambda Z_{rs}-\frac{Y_{rs}}{\lambda}}e^{iS_{rs}}.
\end{equation}

As for the one-loop integrals, we may either perform the summations in terms
of theta functions, or the $\lambda$ integration in terms of modified Bessel functions.
In terms of the latter, the result is
\begin{equation}
\sunset{1}_{rs} =
\frac{1}{(16\pi^2)^2}
\sum_{l_r,l_s}^{\prime\prime}\int_0^1 dx \int_0^{1-x} dy \,
\sigma^{-2} \,
\mathcal{K}_{-1}\left(Y_{rs},Z_{rs}\right)e^{iS_{rs}},
\end{equation}
where we note that 
in the cms frame where $S_{rs}=0$, we may write
\begin{align}
\sum_{l_r^{},l_s^{}}^{\prime\prime} f(l_r^2,l_s^2,l_n^2) & = \!\!\!\!
\sum_{k_r^{},k_s^{},k_n^{} = 1}^{\infty} 
x(k_r^{},k_s^{},k_n^{}) \:\times\: f(k_r^{} L^2, k_s^{} L^2, k_n^{} L^2),
\label{xktrick2}
\end{align}
similarly to Eq.~(\ref{xktrick}). Here, the factor $x(k_r^{},k_s^{},k_n^{})$ denotes the number of times
a given triplet of squares appears when the components of $l_r^{}$ and $l_s^{}$ are varied over all 
positive and negative integer values. In terms of theta functions, we find in the cms frame
\begin{align}
\label{resultHrssimple}
\sunset{1}_{rs} & = \frac{1}{(16\pi^2)^2}
\int_0^1 dx \int_0^{1-x} dy\int_0^\infty d\lambda \, \frac{e^{-\lambda Z_{rs}}}{(\sigma\lambda)^2}
\Bigg[ \theta^{(2)}_0\left(\frac{y L^2}{4\sigma\lambda},\frac{x L^2}{4\sigma\lambda},\frac{z L^2}{4\sigma\lambda}\right)^3
\nonumber \\
& \quad
-\theta_{30}\left(e^{-\frac{(x+z)L^2}{4\sigma\lambda}}\right)^3
-\theta_{30}\left(e^{-\frac{(y+z)L^2}{4\sigma\lambda}}\right)^3
-\theta_{30}\left(e^{-\frac{(x+y)L^2}{4\sigma\lambda}}\right)^3 + 2
\Bigg],
\end{align}
where the contributions with $l_r^2, l_s^2$ or $l_n^2$ equal to zero have been subtracted. 
The Jacobi and Riemann theta functions are defined in App.~\ref{AppTheta}, see also
Eq.~(\ref{sumtheta4}) and the accompanying discussion.

The expression for $\sunset{1}_{r,H}$ in Eq.~(\ref{defM}) is clearly similar
and can be treated along the same lines. In terms of modified Bessel functions, 
the terms with a single sum over $l_r^{}$ may be treated similarly to the one-loop integrals
using Eq.~(\ref{xktrick}). Alternatively,  the summation can be performed in terms
of theta functions. The relevant expressions will be given when we summarize the full results
for the sunset integrals.

\subsection{Permutation properties}
\label{permutation}

The finite-volume sunset integrals satisfy a number of relations which simplify
the calculations, and provide useful checks on the numerics. These are the more general 
versions of Eq.~(\ref{relationssunset1}).
When applied to the full sunset integrals $\sunset{X}$, the variable interchanges $(s,r)$, $(r,t)$
and $(s,t)$, with $t=p-r-s$, yield the relations
\begin{align}
\label{relationssunset3}
& \sunset{1}(1,2,3) = \sunset{1}(2,1,3) = \sunset{1}(3,2,1), 
\nonumber \\
& \sunset{r_\mu^{}}(1,2,3) = \sunset{r_\mu^{}}(1,3,2),
\nonumber \\
& \sunset{s_\mu^{}}(1,2,3) = \sunset{r_\mu^{}}(2,1,3),
\nonumber \\
& \sunset{r_\mu^{}r_\nu^{}}(1,2,3) = \sunset{r_\mu^{}r_\nu^{}}(1,3,2),
\nonumber \\
& \sunset{s_\mu^{}s_\nu^{}}(1,2,3) = \sunset{r_\mu^{}r_\nu^{}}(2,1,3),
\nonumber \\
& \sunset{r_\mu^{}s_\nu^{}}(1,2,3) = \sunset{r_\mu^{}s_\nu^{}}(2,1,3),
\end{align}
where the notation $(1,2,3)$ is explained in the context of Eq.~(\ref{defsunset2}), and refers to the masses 
$m_i^2$ and powers $n_i^{}$ of the propagators in Eq.~(\ref{defsunset}).

Further, we may derive the relations
\begin{align}
\label{relationssunset4}
p_\mu^{} \sunset{1}(1,2,3)
& = \sunset{r_\mu^{}}(1,2,3) + \sunset{r_\mu^{}}(2,1,3) + \sunset{r_\mu^{}}(3,1,2),
\nonumber\\
\sunset{r_\mu^{} s_\nu^{} + s_\mu^{} r_\nu^{}}(1,2,3) & = 
\sunset{r_\mu^{} r_\nu^{}}(3,1,2) - \sunset{r_\mu^{} r_\nu^{}}(1,2,3) -
\sunset{r_\mu^{} r_\nu^{}}(2,1,3)
\nonumber \\ 
& \quad -p_\mu^{} \sunset{r_\nu^{}}(3,1,2) - p_\nu^{} \sunset{r_\mu^{}}(3,1,2) +
p_\mu^{} p_\nu^{} \sunset{1}(1,2,3), 
\end{align}
where the latter one follows from the identity
\begin{align}
r_\mu^{} s_\nu^{} + s_\mu^{} r_\nu^{}
& = (r+s-p)_\mu^{} (r+s-p)_\nu^{} - r_\mu^{} r_\nu^{} - s_\mu^{} s_\nu^{} - p_\mu^{}(-r-s+p)_\nu^{}
\nonumber\\ 
& \quad - (-r-s+p)_\mu^{} \, p_\nu^{} + p_\mu^{} p_\nu^{},
\label{rsrel}
\end{align}
from which it also follows that all parts of $\sunset{r_\mu^{} s_\nu^{}}$ that
are symmetric in $\mu$ and $\nu$ can be rewritten in terms of other integrals. In particular, at infinite volume
$\sunset{r_\mu^{} s_\nu^{}}$ can be expressed in terms of $\sunset{r_\mu^{} r_\nu^{}}$ using various 
permutations of the $m_i^2$ and $n_i^{}$. This also holds for the case of $m_1^{}=m_2^{}$ and $n_1^{}=n_2^{}$.
The relations (\ref{relationssunset3}) and (\ref{relationssunset4})
are also separately valid for $\sunset{X}^\infty$, $\sunset{X}^V$ and 
$\sunset{X}_{rs}$, but not for the other components of Eq.~(\ref{ssplit}).

From the above considerations, we can deduce what integrals should be calculated in order
to obtain a complete description.
As $\sunset{X}_r$, $\sunset{X}_s$ and $\sunset{X}_t$ are closely 
related, we can obtain the required cases of $\sunset{X}_s^{}$ using
\begin{align}
\sunset{1}_s^{}(1,2,3) & = 
\sunset{1}_r^{}(2,1,3;l_r^{} \to l_s^{}),
\nonumber \\
\sunset{r_\mu^{}}_s^{}(1,2,3) & = 
\sunset{s_\mu^{}}_r^{}(2,1,3;l_r^{} \to l_s^{}),
\nonumber \\
\sunset{r_\mu^{} r_\nu^{}}_s^{}(1,2,3) & = 
\sunset{s_\mu^{} s_\nu^{}}_r^{}(2,1,3;l_r^{} \to l_s^{}),
\nonumber \\
\sunset{r_\mu^{} s_\nu^{}}_s^{}(1,2,3) & = 
\sunset{s_\mu^{} r_\nu^{}}_r^{}(2,1,3;l_r^{} \to l_s^{}),
\label{srel}
\end{align}
and for the $\sunset{X}_t^{}$ we find\footnote{Here, we used the fact that the spatial components of
$p$ satisfy periodic boundary conditions, and hence $e^{il_t^{}\cdot p} = 1$.}
\begin{align}
\sunset{1}_t^{}(1,2,3) & =
\sunset{1}_r^{}(3,2,1;l_r^{} \to -l_t^{}),
\nonumber \\
\sunset{r_\mu^{}}_t^{}(1,2,3) & =
\sunset{-r_\mu^{}-s_\mu^{}+p_\mu^{}}_r^{}(3,2,1;l_r^{} \to -l_t^{}),
\nonumber \\
\sunset{r_\mu^{} r_\nu^{}}_t^{}(1,2,3) & =
\sunset{(r+s-p)_\mu^{} (r+s-p)_\nu^{}}_r^{}(3,2,1;l_r^{} \to -l_t^{}),
\nonumber \\
\sunset{r_\mu^{} s_\nu^{}}_t^{}(1,2,3) & =
\sunset{-r_\mu^{} s_\nu^{} - s_\mu^{} s_\nu^{} + p_\mu^{} s_\nu^{}}_r^{}
(3,2,1;l_r^{} \to -l_t^{}),
\label{trel}
\end{align}
from which we conclude that a complete description entails the calculation of
$\sunset{X}_{rs}^{}$ for $X = 1, r_\mu^{}, 
r_\mu^{} r_\nu^{}$ and $r_\mu^{} s_\nu^{}$, and of $\sunset{X}_r^{}$ for
$X = 1, r_\mu^{}, s_\mu^{}, r_\mu^{} r_\nu^{}, r_\mu^{} s_\nu^{}$ and 
$s_\mu^{} s_\nu^{}$. We also note that the $\sunset{X}_r^{}$ are symmetric 
under the interchange $(m_2^{},n_2^{}) \leftrightarrow (m_3^{},n_3^{})$ for 
$X = 1, r_\mu^{}$ and $r_\mu^{} r_\nu^{}$.

For conciseness, we now introduce a set of functions to be used
in the remainder of the text. In an arbitrary frame, we define
\begin{align}
\sunset{1}_{}^V &\equiv H^V_{},
\nonumber \\
\sunset{r_\mu^{}}_{}^V &\equiv H_1^V \,p_\mu^{} + H_{3\mu}^V, 
\nonumber \\
\sunset{s_\mu^{}}_{}^V &\equiv H_2^V \,p_\mu^{} + H_{4\mu}^V, 
\nonumber \\
\sunset{r_\mu^{}r_\nu^{}}_{}^V &\equiv H_{21}^V \,p_\mu^{}p_\nu^{}
+ H_{22}^V \,\delta_{\mu\nu}^{} + H_{27\mu\nu}^V, 
\nonumber \\
\sunset{r_\mu^{}s_\nu^{}}_{}^V &\equiv H_{23}^V \,p_\mu^{}p_\nu^{}
+ H_{24}^V \,\delta_{\mu\nu}^{} + H_{28\mu\nu}^V,
\nonumber\\
\sunset{s_\mu^{}s_\nu^{}}_{}^V &\equiv H_{25}^V \,p_\mu^{}p_\nu^{}
+ H_{26}^V \,\delta_{\mu\nu}^{} + H_{29\mu\nu}^V,
\end{align}
where the $H_{3\mu}^V$, $H_{4\mu}^V$, $H_{27\mu\nu}^V$, $H_{26\mu\nu}^V$ and $H_{28\mu\nu}^V$
contain instances of the vectors $l_r$ or $l_s$ with uncontracted Lorentz indices.
In the cms frame, such contributions with one Lorentz index vanish, and the bilinear ones
become proportional to $t_{\mu\nu}^{}$. In the cms frame, we therefore have a simplified set of functions
\begin{align}
\label{defH}
\sunset{1}^V & \equiv H^V, 
\nonumber \\
\sunset{r_\mu^{}}^V & \equiv H_1^V \,p_\mu^{}, 
\nonumber \\
\sunset{s_\mu^{}}^V & \equiv H_2^V \,p_\mu^{}, 
\nonumber \\
\sunset{r_\mu^{}r_\nu^{}}_{}^V & \equiv H_{21}^V \,p_\mu^{}p_\nu^{}
+ H_{22}^V \,\delta_{\mu\nu}^{}
+ H_{27}^V \, t_{\mu\nu}^{}, 
\nonumber \\
\sunset{r_\mu^{}s_\nu^{}}_{}^V & \equiv H_{23}^V \,p_\mu^{}p_\nu^{}
+ H_{24}^V \,\delta_{\mu\nu}^{}
+ H_{28}^V \, t_{\mu\nu}^{},
\nonumber \\
\sunset{s_\mu^{}s_\nu^{}}_{}^V & \equiv H_{24}^V \,p_\mu^{}p_\nu^{}
+ H_{25}^V \,\delta_{\mu\nu}^{}
+ H_{29}^V \, t_{\mu\nu}^{}.
\end{align}
Because of this structure, $\sunset{r_\mu s_\nu}$ is symmetric in $\mu$, $\nu$
and can be obtained using Eq.~(\ref{relationssunset4}). Still, we include $\sunset{r_\mu s_\nu}$
as a useful check on our numerics, and because it appears in the expressions for the sunset integrals
with one quantized loop momentum. Our numbering scheme for the sunset
integrals has been chosen to be consistent with Ref.~\cite{ABT2}.
We also refer to the components of the functions $H_{i}$ by appending the indices
$(r,G)$, $(r,H)$ {\it etc.}, which were introduced in the detailed treatment of the simplest sunset integral.

\subsection{Sunset integrals with one quantized loop momentum}

Here, we follow along the lines of Sect.~\ref{simplestr} and account for all needed
cases of $\sunset{X}_r$ with $X=1,r_\mu,s_\mu,$ $r_\mu r_\nu,$ $r_\mu s_\nu$ and $s_\mu s_\nu$.
Again, the first step is to combine the last two propagators with a Feynman parameter
$x$ and shift the integration variable by $s_\mu \equiv \tilde s - x(r-p)_\mu$. The integral over $\tilde s$
can then be performed using Eq.~(\ref{feynmanintegrals}).
Using the notation $f(r_\alpha)$ for additional factors of $r_\mu,r_\nu$, this gives
\begin{align}
{\langle\langle f(r_\alpha^{}) \rangle\rangle}_r^{} & = \int 
\frac{d^dr}{(2\pi)^d}
\frac{e^{il_r^{} \cdot r}\,f(r_\alpha^{})}{{(r^2+m_1^2)}^{n_1^{}}}
\int_0^1 dx\,
\frac{\Gamma\left(2-\frac{d}{2}\right)}{(4\pi)^{\frac{d}{2}}}\:
(\overline m^2)^{\frac{d}{2}-2}_{},
\label{req} \\
{\langle\langle f(r_\alpha^{})s_\mu^{} \rangle\rangle}_r^{} & = \int 
\frac{d^dr}{(2\pi)^d}
\frac{e^{il_r^{} \cdot r}\:f(r_\alpha^{})}{{(r^2+m_1^2)}^{n_1^{}}}
 \int_0^1 dx\,
\frac{\Gamma\left(2-\frac{d}{2}\right)}{(4\pi)^{\frac{d}{2}}}\:
(\overline m^2)^{\frac{d}{2}-2}_{} (-x)(r-p)_\mu^{},
\label{seq} \\
{\langle\langle s_\mu^{}s_\nu^{} \rangle\rangle}_r^{} & = \int 
\frac{d^dr}{(2\pi)^d}
\frac{e^{il_r^{} \cdot r}}{{(r^2+m_1^2)}^{n_1^{}}}
\int_0^1 dx
\left[\frac{\Gamma\left(2-\frac{d}{2}\right)}{(4\pi)^{\frac{d}{2}}}\:
(\overline m^2)^{\frac{d}{2}-2}_{}\: 
x^2(r-p)_\mu^{}(r-p)_\nu^{} \right.
 \nonumber \\ & \quad
 +\left.\frac{\Gamma\left(1-\frac{d}{2}\right)}{(4\pi)^{\frac{d}{2}}}\:
(\overline m^2)^{\frac{d}{2}-1}_{}\,\frac{\delta_{\mu\nu}^{}}{2}
\right],
\label{sseq}
\end{align}
where the remaining
integral over $r$ is always finite because of the factor $e^{i l_r\cdot r}$.
It is then sufficient to expand the $\tilde s$ integral in $\varepsilon$, while
keeping only the singular and $\mathcal{O}(1)$ terms as in Eq.~(\ref{expandepsilon}).
We rewrite the singular terms using $\lambda_0^{} \equiv 1/\varepsilon + \ln(4\pi) + 1 - \gamma$,
and define the components of the sunset integrals proportional to $\lambda_0^{}$ with the 
subscript~$A$ as in Eq.~(\ref{defsunsetA}). 
In terms of the one-loop integrals defined in Sect.~\ref{Oneloop}, we find for the non-zero cases
with $n_2,n_3=1,2$ the expressions
\begin{align}
\sunset{1}^{n_1 1 1}_{r,A} & = \frac{\lambda_0}{16\pi^2}\onloop{1}^V(n_1,m_1^2),
\nonumber\\
\sunset{s_\mu}^{n_1 1 1}_{r,A} & = \frac{\lambda_0}{16\pi^2}\frac{p_\mu}{2}\onloop{1}^V(n_1,m_1^2),
\nonumber\\
\sunset{r_\mu r_\nu}^{n_1 1 1}_{r,A} & = \frac{\lambda_0}{16\pi^2}\onloop{r_\mu r_\nu}^V(n_1,m_1^2),
\nonumber\\
\sunset{r_\mu s_\nu}^{n_1 1 1}_{r,A} & = \frac{\lambda_0}{16\pi^2}\frac{-1}{2}\onloop{r_\mu r_\nu}^V(n_1,m_1^2),
\nonumber\\
\sunset{s_\mu s_\nu}^{n_1 1 1}_{r,A} & = \frac{\lambda_0}{16\pi^2}
\Bigg\{\left[
\delta_{\mu\nu}\left(-\frac{m_2^2}{4}-\frac{m_3^2}{4}-\frac{p^2}{12}\right)
+\frac{p_\mu p_\nu}{3}\right]
\onloop{1}^V(n_1,m_1^2)
\nonumber\\
& \quad +\left(\frac{1}{3}\delta_{\mu\alpha}\delta_{\nu\beta}
         -\frac{1}{12}\delta_{\mu\nu}\delta_{\alpha\beta}\right)
        \onloop{r_\alpha r_\beta}^V(n_1,m_1^2)\Bigg\},
\nonumber\\
\sunset{s_\mu s_\nu}^{n_1 2 1}_{r,A} & =  
\sunset{s_\mu s_\nu}^{n_1 1 2}_{r,A} =\frac{\lambda_0}{16\pi^2}\frac{\delta_{\mu\nu}}{4}
      \onloop{1}^V(n_1,m_1^2),
\end{align}
where the superscripts denote the $n_i^{}$ in the sunset integrals.
Also, the one-loop integrals now show explicitly the $m_i^2$ and $n_i^{}$
of the denominator they involve. As the above integrals contain a non-local divergence, they should 
always cancel in physical results.

We now proceed to treat the terms containing $\log(\overline m^2)$. As before,
we first perform a partial integration in $x$, giving
\begin{align}
\int_0^1 dx\, x^n\log(\overline m^2) & =
\frac{1}{n+1}\log(m_3^2) \nonumber \\
& \quad -\frac{1}{n+1}\int_0^1dx\, x^{n+1}
\bigg[m_3^2-m_2^2+(1-2x)(r-p)^2\bigg]
\, \frac{1}{\overline m^2},
\end{align}
after which we denote the terms with negative powers of $\overline m^2$ as $\sunset{X}_{r,H}$, and the 
others as $\sunset{X}_{r,G}$, as defined in Eq.~(\ref{defsunsetGH}) for the case of the simplest sunset integral. 
We note that the $\sunset{X}_{r,G}$ can again
be expressed in terms of one-loop integrals. For $n_2,n_3=1,2$, the non-zero cases are
\begin{align}
\label{Gintegral}
\sunset{1}^{n_1 1 1}_{r,G} & = 
\frac{1}{16\pi^2}\left(-1-\log(m_3^2)\right)\onloop{1}^V\!(n_1^{},m_1^2),
\nonumber\\
\sunset{s_\mu}^{n_1 1 1}_{r,G} & = 
\frac{1}{16\pi^2}\frac{p_\mu}{2}\left(-1-\log(m_3^2)\right)\onloop{1}^V\!(n_1^{},m_1^2),
\nonumber\\
\sunset{r_\mu r_\nu}^{n_1 1 1}_{r,G} & = \frac{1}{16\pi^2}\left(-1-\log(m_3^2)\right)
\onloop{r_\mu r_\nu}^V(n_1^{},m_1^2),
\nonumber\\
\sunset{r_\mu s_\nu}^{n_1 1 1}_{r,G} & = \frac{1}{16\pi^2}\frac{1}{2}
 \left(1+\log(m_3^2)\right)\onloop{r_\mu r_\nu}^V\!(n_1^{},m_1^2),
\nonumber\\
\sunset{s_\mu s_\nu}^{n_1 1 1}_{r,G} & = \frac{1}{16\pi^2}
\left\{\left[
\delta_{\mu\nu}\log(m_3^2)
 \left(\frac{m_2^2}{4}+\frac{m_3^2}{4}+\frac{p^2}{12}\right)
-\frac{p_\mu p_\nu}{3}\left(1+\log(m_3^2)\right)\right] \!
\onloop{1}^V\!(n_1^{},m_1^2) \right.
\nonumber\\
& \left. \quad
+\left[\frac{-1}{3}\delta_{\mu\alpha}\delta_{\nu\beta}\left(1+\log(m_3^2)\right)
         +\frac{1}{12}\delta_{\mu\nu}\delta_{\alpha\beta}\log(m_3^2)\right] \!
        \onloop{r_\alpha r_\beta}^V\!(n_1,m_1^2)\right\},
\nonumber\\
\sunset{s_\mu s_\nu}^{n_1 2 1}_{r,G} & =  
\sunset{s_\mu s_\nu}^{n_1 1 2}_{r,G} =\frac{1}{16\pi^2}\frac{-\delta_{\mu\nu}}{4}
   \left(1+\log(m_3^2)\right)   \onloop{1}^V\!(n_1^{},m_1^2).
\end{align}
We note that the decomposition of the parts of the sunset integrals which do not depend on $\lambda_0^{}$
into $\sunset{X}_{r,G}$ and $\sunset{X}_{r,H}$ is clearly not unique, as it depends on
the choice of Feynman parameterization.
For example, had we chosen $y=1-x$ instead of $x$ as the Feynman parameter, we would have
obtained terms containing $\log(m_2^2)$ in the $\sunset{X}_{r,G}$. Also, the decomposition does not
commute with derivatives {\it w.r.t.} masses, note {\it e.g.} that
$\sunset{1}_{r,G}^{n_1 1 2} = 0 \ne -(\partial/\partial m_3^2)\sunset{1}_{r,G}^{n_1 1 1}$.

The remaining part $\sunset{X}_{r,H}$ is algebraically the most complicated, but again follows exactly the procedure
for the simplest sunset integral. First, we introduce Gaussian parameterizations for the negative powers of
$\overline m^2$ and $(r^2+m_1^2)$ using Eq.~(\ref{prop}) with parameters $\lambda_4$ and
$\lambda_1$, respectively. While the expressions corresponding to $\sunset{1}_{r,H}$ in 
Eq.~(\ref{defsunsetGH}) are relatively lengthy, they all share the same basic structure. In particular,
they all contain the same exponential factor, for which we may complete the square using the substitutions
of Eq.~(\ref{complete}). The resulting integrals can then be performed by means of Eq.~(\ref{simpint}).
Finally, we define $\lambda_2 \equiv (1-x)\lambda_4$ and $\lambda_3 \equiv x\lambda_4$ and perform
the substitutions of Eq.~(\ref{xyzlambda}) to obtain an integral in terms of $x,y,z$ and $\lambda$.

Before we give explicit expressions for $\sunset{X}_{r,H}$, we briefly discuss the
methods used to obtain them. Due to the complexity of the required analytical manipulations,
we have found it convenient to use {\tt FORM}~\cite{Form} according to the procedure outlined
above. Alternatively, as described in Ref.~\cite{Thesis}, a number of tricks can be used to considerably
simplify the task. For example, powers of $r_\mu$ can be introduced into the numerators of the sunset
integrals by taking derivatives {\it w.r.t.} $l_r$, giving
\begin{align}
\sunset{r_\mu}_r & = -i\frac{\partial}{\partial l_{r\mu}} \sunset{1}_r.
\end{align}
It is also noteworthy that integrals such as $\sunset{s_\mu}_r$ are very similar to the case of $\sunset{1}_r$,
differing only in an additional factor of $x(r-p)_\mu$. This leads to relations such as
\begin{align}
\sunset{s_\mu} & = \sunset{x r_\mu}-p_\mu\sunset{x 1},
\end{align}
where the factor of $x$ is understood to be included in the respective integrals. Due to the length and complexity of the
resulting expressions for $\sunset{X}_{r,H}$, we make use of the auxiliary quantities
\begin{align}
\delta & \equiv \frac{y-z}{\sigma}, &
A & \equiv m_3^2-m_2^2+\delta\rho x^2 p^2,
\nonumber\\
\rho & \equiv \frac{y+z}{\sigma}, & 
B & \equiv ix\delta\rho \, l_r\cdot p,
\nonumber\\
\sigma & \equiv xy+yz+zx, & 
C & \equiv \frac{\delta\rho}{4}l_r^2,
\nonumber\\
\tau & \equiv \frac{yz}{\sigma}, &
D & \equiv A-\frac{B}{\lambda}-\frac{C}{\lambda^2},
\end{align}
and
\begin{equation}
Y \equiv \frac{\rho}{4} \, l_r^2, \qquad
Z \equiv x m_1^2+y m_2^2+ z m_3^2+\frac{xyz}{\sigma}p^2,
\end{equation}
and we also introduce the notation
\begin{align}
\sunset{X}^{n_1^{} n_2^{} n_3^{}}_{r,H} & = \frac{1}{\Gamma(n_1)(16\pi^2)^2}
\sum^\prime_{l_r} \int_0^1dx\int_0^{1-x}dy
\int_0^\infty d\lambda
\nonumber \\
& \quad \times 
\frac{(x\lambda)^{n_1^{}-1}}{\lambda\sigma^2} \,
\sunsetx{X}_{r,H}^{n_1^{} n_2^{} n_3^{}} \,
e^{-\lambda Y-\frac{Z}{\lambda}+\frac{iyz}{\sigma}l_r\cdot p}.
\end{align}
With these abbreviations, we obtain
\begin{align}
\label{resultHr1}
\sunsetx{1}_{r,H}^{n_1^{} 1 1} & =
z\left(D+\frac{2\delta}{\lambda}\right),
\nonumber\\
\sunsetx{1}^{n_1^{} 2 1}_{r,H} & = y,
\nonumber\\
\sunsetx{1}^{n_1^{} 1 2}_{r,H} & = z,
\nonumber\\
\sunsetx{1}^{n_1^{} 2 2}_{r,H} & = yz\lambda,
\end{align}
for the simplest sunset integral, and
\begin{align}
\label{resultHr2}
\sunsetx{r_\mu^{}}_{r,H}^{n_1^{} 1 1} & =
\sunsetx{1}_{r,H}^{n_1^{} 1 1} \left(\frac{i\rho l_{r\mu}}{2\lambda}+\tau p_\mu\right)
  +\frac{z\rho\delta}{\lambda} \left(\frac{i l_{r\mu}}{2\lambda}-x p_\mu\right),
\nonumber\\
\sunsetx{r_\mu^{}}^{n_1^{} 2 1}_{r,H} & = y\left(\frac{i\rho l_{r\mu}}{2\lambda}+\tau p_\mu\right),
\nonumber\\
\sunsetx{r_\mu^{}}^{n_1^{} 1 2}_{r,H} & = z\left(\frac{i\rho l_{r\mu}}{2\lambda}+\tau p_\mu\right),
\nonumber\\
\sunsetx{r_\mu^{}}^{n_1^{} 2 2}_{r,H} & = yz\lambda\left(\frac{i\rho l_{r\mu}}{2\lambda}+\tau p_\mu\right),
\end{align}
\begin{align}
\label{resultHr3}
\sunsetx{s_\mu^{}}^{n_1^{} 1 1}_{r,H} & = \frac{-z^2}{2\sigma}
\left(D+\frac{3\delta}{\lambda}\right)
 \left(\frac{i l_{r\mu}}{2\lambda}-x p_\mu\right),
\nonumber\\
\sunsetx{s_\mu^{}}^{n_1^{} 2 1}_{r,H} & = \frac{-zy}{\sigma}
 \left(\frac{i l_{r\mu}}{2\lambda}-x p_\mu\right),
\nonumber\\
\sunsetx{s_\mu^{}}^{n_1^{} 1 2}_{r,H} & = \frac{-z^2}{\sigma}
 \left(\frac{i l_{r\mu}}{2\lambda}-x p_\mu\right),
\nonumber\\
\sunsetx{s_\mu^{}}^{n_1^{} 2 2}_{r,H} & = \frac{-yz^2\lambda}{\sigma}
 \left(\frac{i l_{r\mu}}{2\lambda}-x p_\mu\right),
\end{align}
for $X = r_\mu, s_\mu$. 
With $\{a,b\}_{\mu\nu} = a_\mu b_\nu + a_\nu b_\mu$, we find for the bilinear operators
\begin{align}
\label{resultHr4}
\sunsetx{r_\mu^{} r_\nu^{}}^{n_1^{} 1 1}_{r,H} & = z
\Bigg\{
\frac{\rho}{2\lambda}\left(D+\frac{3\delta}{\lambda}\right)\delta_{\mu\nu}
+\tau\left[\tau D+\frac{2\delta}{\lambda}\left(\tau-\rho x\right)\right]
         p_\mu p_\nu
\nonumber\\ & \quad
+\frac{i\rho}{2\lambda}
     \left[\tau D+\frac{\delta}{\lambda}\left(3\tau-\rho x\right)\right]
     \{p,l_r\}_{\mu\nu}
-\frac{\rho^2}{4\lambda^2}\left(D+\frac{4\delta}{\lambda}\right)l_{r\mu}l_{r\nu}
\Bigg\},
\nonumber\\
\sunsetx{r_\mu^{} r_\nu^{}}^{n_1^{} 2 1}_{r,H} & =
y\left[\frac{\rho}{2\lambda}\delta_{\mu\nu}
+\tau^2 p_\mu p_\nu +\frac{i\rho\tau}{2\lambda}\{p,l_r\}_{\mu\nu}
-\frac{\rho^2}{4\lambda^2}l_{r\mu}l_{r\nu}\right],
\nonumber\\
\sunsetx{r_\mu^{} r_\nu^{}}^{n_1^{} 1 2}_{r,H} & =
z\left[\frac{\rho}{2\lambda}\delta_{\mu\nu}
+\tau^2 p_\mu p_\nu +\frac{i\rho\tau}{2\lambda}\{p,l_r\}_{\mu\nu}
-\frac{\rho^2}{4\lambda^2}l_{r\mu}l_{r\nu}\right],
\nonumber\\
\sunsetx{r_\mu^{} r_\nu^{}}^{n_1^{} 2 1}_{r,H} & =
yz\lambda\left[\frac{\rho}{2\lambda}\delta_{\mu\nu}
+\tau^2 p_\mu p_\nu +\frac{i\rho\tau}{2\lambda}\{p,l_r\}_{\mu\nu}
-\frac{\rho^2}{4\lambda^2}l_{r\mu}l_{r\nu}\right],
\end{align}
\begin{align}
\label{resultHr5}
\sunsetx{r_\mu^{} s_\nu^{}}^{n_1^{} 1 1}_{r,H} & = \frac{z^2}{2\sigma}\Bigg\{
\frac{-1}{2\lambda}\left(D+\frac{3\delta}{\lambda}\right)\delta_{\mu\nu}
+\left[\tau D+\frac{\delta}{\lambda}\left(3\tau-\rho x\right)\right]
        \left(x p_\mu p_\nu-\frac{i}{2\lambda}p_\mu l_{r\nu}\right)
\nonumber\\ &
+\frac{\rho}{4\lambda^2}\left(D+\frac{4\delta}{\lambda}\right)
\left(l_{r\mu}l_{r\nu}+2ix\lambda l_{r\mu}p_{\nu}\right)
\Bigg\},
\nonumber\\
\sunsetx{r_\mu^{} s_\nu^{}}^{n_1^{} 2 1}_{r,H} & =
\frac{yz}{\sigma}\left[\frac{-1}{2\lambda}\delta_{\mu\nu}
+\tau x p_\mu p_\nu -\frac{i\tau}{2\lambda} p_\mu l_{r\nu}
+\frac{i\rho x}{2}l_{r\mu}p_\nu
+\frac{\rho}{4\lambda^2}l_{r\mu}l_{r\nu}\right],
\nonumber\\
\sunsetx{r_\mu^{} s_\nu^{}}^{n_1^{} 1 2}_{r,H} & =
\frac{z^2}{\sigma}\left[\frac{-1}{2\lambda}\delta_{\mu\nu}
+\tau x p_\mu p_\nu -\frac{i\tau}{2\lambda} p_\mu l_{r\nu}
+\frac{i\rho x}{2}l_{r\mu}p_\nu
+\frac{\rho}{4\lambda^2}l_{r\mu}l_{r\nu}\right],
\nonumber\\
\sunsetx{r_\mu^{} s_\nu^{}}^{n_1^{} 2 2}_{r,H} & =
\frac{yz^2\lambda}{\sigma}\left[\frac{-1}{2\lambda}\delta_{\mu\nu}
+\tau x p_\mu p_\nu -\frac{i\tau}{2\lambda} p_\mu l_{r\nu}
+\frac{i\rho x}{2}l_{r\mu}p_\nu
+\frac{\rho}{4\lambda^2}l_{r\mu}l_{r\nu}\right],
\end{align}
and
\begin{align}
\label{resultHr6}
\sunsetx{s_\mu^{} s_\nu^{}}^{n_1^{} 1 1}_{r,H} & = \frac{z^3}{3\sigma^2}
\left(D+\frac{4\delta}{\lambda}\right)
\left(x^2 p_\mu p_\nu-\frac{i x}{2\lambda}\{p,l_r\}_{\mu\nu}-\frac{1}{4\lambda^2}l_{r\mu}l_{r\nu}\right)
\nonumber\\ & \quad
+\delta_{\mu\nu}\Bigg\{
-\frac{m_2^2z}{2}\left(D+\frac{2\delta}{\lambda}\right)
-\frac{Az^2}{4\rho\sigma}\left(D+\frac{2\delta}{\lambda}\right)
-\frac{\tau z}{2\rho\lambda\sigma}\left(D+\frac{3\delta}{\lambda}\right)
\nonumber\\ & \quad
+\frac{z^2}{\sigma^2}\left(D+\frac{4\delta}{\lambda}\right)
\left[
 \frac{z+3y}{12}\left(\frac{i x p\cdot l_r}{\lambda}
                 +\frac{l_r^2}{4\lambda^2}\right)
-\frac{z x^2 p^2}{3}
\right]
-\frac{x^2z^2\delta^2p^2}{2\lambda\sigma^2}
\Bigg\},
\nonumber\\ 
\sunsetx{s_\mu^{} s_\nu^{}}^{n_1^{} 2 1}_{r,H} & = \frac{y z^2}{\sigma^2}
\left(x^2 p_\mu p_\nu-\frac{i x}{2\lambda}\{p,l_r\}_{\mu\nu}
     -\frac{1}{4\lambda^2}l_{r\mu}l_{r\nu}\right)
\nonumber\\ & \quad
+\delta_{\mu\nu}\left[\frac{1}{4}\left(\frac{\tau}{\rho}+z\right)D
                      +\frac{z\delta}{\lambda}+\frac{z^3}{\rho\sigma^2\lambda}
              \right],
\nonumber\\ 
\sunsetx{s_\mu^{} s_\nu^{}}^{n_1^{} 1 2}_{r,H} & = \frac{z^3}{\sigma^2}
\left(x^2 p_\mu p_\nu-\frac{i x}{2\lambda}\{p,l_r\}_{\mu\nu}
     -\frac{1}{4\lambda^2}l_{r\mu}l_{r\nu}\right)
\nonumber\\ & \quad
+\delta_{\mu\nu}\left[\frac{1}{4}\left(-\frac{\tau}{\rho}+z\right)D
                      +\frac{z^2}{\sigma\lambda}
                 -\frac{z^3}{2\rho\sigma^2\lambda}
              \right],
\nonumber\\ 
\sunsetx{s_\mu^{} s_\nu^{}}^{n_1^{} 2 2}_{r,H} & = \frac{yz^3\lambda}{\sigma^2}
\left(x^2 p_\mu p_\nu-\frac{i x}{2\lambda}\{p,l_r\}_{\mu\nu}
     -\frac{1}{4\lambda^2}l_{r\mu}l_{r\nu}\right)
+\delta_{\mu\nu}\left[\frac{\tau}{2\rho}(1-\tau)+\frac{z\tau}{2}
              \right].
\end{align}

Given these expressions for $\sunset{X}_{r,H}$, we may proceed as for the
one-loop integrals and choose between performing the summations in terms of theta functions,
or evaluating the $\lambda$ integral in terms of modified Bessel functions.
The results quoted in Eqs.~(\ref{resultHr1})-(\ref{resultHr6}) make no assumptions
on the momentum $p$. Below, we restrict ourselves to the
cms frame where $p\cdot l_r=0$ or $p=(p,0,0,0)$.
This case is the most commonly encountered, and the expressions for a moving frame can be obtained
along similar lines.

\subsubsection{Center-of-mass frame: Bessel functions}
\label{Hrbessel}

Here, we have performed the integration over $\lambda$ in terms of the
functions $\mathcal{K}_\nu(Y,Z)$ defined in App.~(\ref{lambdaint}). We note that the summation only
depends on $l_r^2$, such that Eq.~(\ref{xktrick}) is applicable. We have suppressed the arguments $(Y,Z)$ in 
order to keep the expressions short and concise.
The expressions always contain the abbreviated part
\newcommand{\intrH}{\int\hskip-2.5ex B}
\begin{align}
\intrH & = \frac{1}{\Gamma(n_1)(16\pi^2)^2} \sum_{l_r}^\prime
\int_0^1dx\int_0^{1-x}dy\,\frac{x^{n_1-1}}{\sigma^2},
\end{align}
and numerical results for selected examples are given in Sect.~\ref{Numerics}.
For the simplest sunset integrals, we find
\begin{align}
H^{r,H;n_1 1 1} =& \intrH z
\left(A\mathcal{K}_{n_1-1}+2\delta \mathcal{K}_{n_1-2}
 -C\mathcal{K}_{n_1-3}\right),
\nonumber\\
H^{r,H;n_1 2 1} =& \intrH y \, \mathcal{K}_{n_1-1},
\nonumber\\
H^{r,H;n_1 1 2} =& \intrH z \, \mathcal{K}_{n_1-1},
\nonumber\\
H^{r,H;n_1 2 2} =& \intrH zy \, \mathcal{K}_{n_1},
\end{align}
and for $X = r_\mu, s_\mu$ we find
\begin{align}
H^{r,H;n_1 1 1}_1 =& \intrH z
\left(\tau A\mathcal{K}_{n_1-1}+(2\tau-\rho x)\delta \mathcal{K}_{n_1-2}
 -\tau C\mathcal{K}_{n_1-3}\right),
\nonumber\\
H^{r,H;n_1 2 1}_1 =& \intrH y\tau \, \mathcal{K}_{n_1-1},
\nonumber\\
H^{r,H;n_1 1 2}_1 =& \intrH z\tau \, \mathcal{K}_{n_1-1},
\nonumber\\
H^{r,H;n_1 2 2}_1 =& \intrH zy\tau \, \mathcal{K}_{n_1},
\end{align}
\begin{align}
H^{r,H;n_1 1 1}_2 =& \intrH \frac{x z^2}{2\sigma}
\left( A\mathcal{K}_{n_1-1}+3\delta \mathcal{K}_{n_1-2}
 -C\mathcal{K}_{n_1-3}\right),
\nonumber\\
H^{r,H;n_1 2 1}_2 =& \intrH \frac{xyz}{\sigma} \, \mathcal{K}_{n_1-1},
\nonumber\\
H^{r,H;n_1 1 2}_2 =& \intrH \frac{xz^2}{\sigma} \, \mathcal{K}_{n_1-1},
\nonumber\\
H^{r,H;n_1 2 2}_2 =& \intrH \frac{xyz^2}{\sigma} \, \mathcal{K}_{n_1},
\end{align}
respectively. For $X = r_\mu r_\nu$, we have
\begin{align}
H^{r,H;n_1 1 1}_{21} =& \intrH z\tau
\left(\tau A\mathcal{K}_{n_1-1}+2(\tau-\rho x)\delta \mathcal{K}_{n_1-2}
 -\tau C\mathcal{K}_{n_1-3}\right),
\nonumber\\
H^{r,H;n_1 2 1}_{21} =& \intrH y\tau^2 \, \mathcal{K}_{n_1-1},
\nonumber\\
H^{r,H;n_1 1 2}_{21} =& \intrH z\tau^2 \, \mathcal{K}_{n_1-1},
\nonumber\\
H^{r,H;n_1 2 2}_{21} =& \intrH yz\tau^2 \, \mathcal{K}_{n_1},
\end{align}
\begin{align}
H^{r,H;n_1 1 1}_{22} =& \intrH \frac{z\rho}{2}
\left( A\mathcal{K}_{n_1-2}+3\delta \mathcal{K}_{n_1-3}
 -C\mathcal{K}_{n_1-4}\right),
\nonumber\\
H^{r,H;n_1 2 1}_{22} =& \intrH \frac{y\rho}{2} \, \mathcal{K}_{n_1-2},
\nonumber\\
H^{r,H;n_1 1 2}_{22} =& \intrH \frac{z\rho}{2} \, \mathcal{K}_{n_1-2},
\nonumber\\
H^{r,H;n_1 2 2}_{22} =& \intrH \frac{yz\rho}{2} \, \mathcal{K}_{n_1-1},
\end{align}
\begin{align}
H^{r,H;n_1 1 1}_{27} =& \intrH \frac{-z\rho^2 l_r^2}{12}
\left( A\mathcal{K}_{n_1-3}+4\delta \mathcal{K}_{n_1-4}
 -C \mathcal{K}_{n_1-5}\right),
\nonumber\\
H^{r,H;n_1 2 1}_{27} =& \intrH  \frac{-y\rho^2 l_r^2}{12} \, \mathcal{K}_{n_1-3},
\nonumber\\
H^{r,H;n_1 1 2}_{27} =& \intrH  \frac{-z\rho^2 l_r^2}{12} \, \mathcal{K}_{n_1-3},
\nonumber\\
H^{r,H;n_1 2 2}_{27} =& \intrH \frac{-yz\rho^2 l_r^2}{12} \, \mathcal{K}_{n_1-2},
\end{align}
for $X = r_\mu s_\nu$, we find
\begin{align}
H^{r,H;n_1 1 1}_{23} =& \intrH \frac{x z^2}{2\sigma}
\left(\tau A\mathcal{K}_{n_1-1}+(3\tau-\rho x)\delta \mathcal{K}_{n_1-2}
 -\tau C\mathcal{K}_{n_1-3}\right),
\nonumber\\
H^{r,H;n_1 2 1}_{23} =& \intrH \frac{xyz\tau}{\sigma} \, \mathcal{K}_{n_1-1},
\nonumber\\
H^{r,H;n_1 1 2}_{23} =& \intrH \frac{xz^2\tau}{\sigma} \, \mathcal{K}_{n_1-1},
\nonumber\\
H^{r,H;n_1 2 2}_{23} =& \intrH \frac{xyz^2\tau}{\sigma} \, \mathcal{K}_{n_1},
\end{align}
\begin{align}
H^{r,H;n_1 1 1}_{24} =& \intrH \frac{-z^2}{4\sigma}
\left( A\mathcal{K}_{n_1-2}+3\delta \mathcal{K}_{n_1-3}
 - C\mathcal{K}_{n_1-4}\right),
\nonumber\\
H^{r,H;n_1 2 1}_{24} =& \intrH \frac{-yz}{2\sigma} \, \mathcal{K}_{n_1-2},
\nonumber\\
H^{r,H;n_1 1 2}_{24} =& \intrH \frac{-z^2}{2\sigma} \, \mathcal{K}_{n_1-2},
\nonumber\\
H^{r,H;n_1 2 2}_{24} =& \intrH \frac{-yz^2}{2\sigma} \, \mathcal{K}_{n_1-1},
\end{align}
\begin{align}
H^{r,H;n_1 1 1}_{28} =& \intrH \frac{z^2\rho l_r^2}{24\sigma}
\left( A\mathcal{K}_{n_1-3}+4\delta \mathcal{K}_{n_1-4}
 -C \mathcal{K}_{n_1-5}\right),
\nonumber\\
H^{r,H;n_1 2 1}_{28} =& \intrH  \frac{yz\rho l_r^2}{12\sigma} \, \mathcal{K}_{n_1-3},
\nonumber\\
H^{r,H;n_1 1 2}_{28} =& \intrH  \frac{z^2\rho l_r^2}{12\sigma} \, \mathcal{K}_{n_1-3},
\nonumber\\
H^{r,H;n_1 2 2}_{28} =& \intrH \frac{yz^2\rho l_r^2}{12\sigma} \, \mathcal{K}_{n_1-2},
\end{align}
and for $X = s_\mu s_\nu$, we have
\begin{align}
H^{r,H;n_1 1 1}_{25} & = \intrH \, \frac{x^2 z^3}{3\sigma^2}
\left( A\mathcal{K}_{n_1-1}+4\delta \mathcal{K}_{n_1-2}
 -C \mathcal{K}_{n_1-3}\right),
\nonumber\\
H^{r,H;n_1 2 1}_{25} & = \intrH  \, \frac{x^2yz^2}{\sigma^2} \, \mathcal{K}_{n_1-1},
\nonumber\\
H^{r,H;n_1 1 2}_{25} & = \intrH  \, \frac{x^2z^3}{\sigma^2} \, \mathcal{K}_{n_1-1},
\nonumber\\
H^{r,H;n_1 2 2}_{25} & = \intrH \, \frac{x^2yz^3}{\sigma^2} \, \mathcal{K}_{n_1},
\end{align}
\begin{align}
H^{r,H;n_1 1 1}_{26} & = \intrH \, 
\Bigg\{\frac{-z}{12} A\left(6 m_2^2+3 \frac{z}{\rho\sigma}A
                +\frac{4x^2z^2p^2}{\sigma^2}\right) \mathcal{K}_{n_1-1}
\nonumber\\ & \quad
+\left[-z\delta m_2^2
    +\frac{z^2}{6\rho\sigma^2}(m_3^2-m_2^2)\left(5z+3y\right)
    -\frac{z^2A}{6\rho\sigma^2}(2z+9y)\right]\mathcal{K}_{n_1-2}
\nonumber\\ & \quad
+\left[\frac{z^2 l_r^2}{24\sigma^2}\left((z+3y)A-2z(m_3^2-m_2^2)\right)
     -\frac{3y z^2\delta}{2\rho\sigma^2}+\frac{z}{2} m_2^2 C
 \right] \mathcal{K}_{n_1-3}
\nonumber\\ & \quad
 +\frac{z^2 l_r^2\delta}{24\sigma^2}(2z+9y) \mathcal{K}_{n_1-4}
-\frac{z^2C l_r^2}{48\sigma^2}(z+3y) \mathcal{K}_{n_1-5}
\Bigg\},
\nonumber\\
H^{r,H;n_1 2 1}_{26} & = \intrH \, \left[ \frac{1}{4}\left(\frac{\tau}{\rho}+z\right)
\left(A\mathcal{K}_{n_1-1}-\frac{2z}{\sigma}\mathcal{K}_{n_1-2}
-C\mathcal{K}_{n_1-3}\right)+\tau\mathcal{K}_{n_2-2}\right]
,
\nonumber\\
H^{r,H;n_1 1 2}_{26} & = \intrH \, \frac{z^2}{4\rho\sigma}
\left(A\mathcal{K}_{n_1-1}+\frac{2y}{\sigma}\mathcal{K}_{n_1-2}
-C\mathcal{K}_{n_1-3}\right),
\nonumber\\
H^{r,H;n_1 2 2}_{26} & = \intrH \, \frac{\tau}{2\rho}
     \left(1+\frac{z^2}{\sigma}\right)\mathcal{K}_{n_1-1},
\end{align}
\begin{align}
H^{r,H;n_1 1 1}_{29} & = \intrH \, \frac{-z^3 l_r^2}{36\sigma^2}
\left( A\mathcal{K}_{n_1-3}+4\delta \mathcal{K}_{n_1-4}
 -C \mathcal{K}_{n_1-5}\right),
\nonumber\\
H^{r,H;n_1 2 1}_{29} & = \intrH \,  \frac{-y z^2 l_r^2}{12\sigma^2} \, \mathcal{K}_{n_1-3},
\nonumber\\
H^{r,H;n_1 1 2}_{29} & = \intrH \,  \frac{- z^3 l_r^2}{12\sigma^2} \, \mathcal{K}_{n_1-3},
\nonumber\\
H^{r,H;n_1 2 2}_{29} & = \intrH \, \frac{-y z^3 l_r^2}{12\sigma^2} \, \mathcal{K}_{n_1-2}.
\end{align}

\subsubsection{Center-of-mass frame: Theta functions}
\label{Hrtheta}

Next, instead of computing the integrals over $x,y$ and $\lambda$, we have performed
the summation in terms of the theta functions, previously encountered for the one-loop and simplest sunset integrals.
In the cms frame, we make use of Eqs.~(\ref{sumtheta}),~(\ref{sumtheta2}), and
\begin{align}
\label{sumtheta3}
\sum_{n\in\mathbb{Z}^3} (n^2)^2 q^{(n^2)}
& = \left(q\frac{\partial}{\partial q}\right)^2
\left(\sum_{n\in\mathbb{Z}^3} q^{(n^2)}\right)
= \left(q\frac{\partial}{\partial q}\right)^2\left(\theta_{30}(q)^3\right)
\nonumber\\
& = 3 \theta_{34}(q)\theta_{30}(q)^2+6\theta_{32}(q)^2\theta_{30}(q),
\end{align}
where we note that Eq.~(\ref{sumtheta3}) can immediately be used for the primed sums by setting
$l_r = nL$, as the term with $n = 0$ does not contribute. We rescale $\lambda$ such that the 
argument of all theta functions is $e^{-1/\lambda}$, which we suppress for brevity.
Further, we introduce the abbreviation
\newcommand{\intrHtheta}{\int\hskip-2ex T}
\newcommand{\lam}{\hat\lambda}
\begin{align}
\intrHtheta & = \frac{1}{\Gamma(n_1)(16\pi^2)^2}\int_0^1dx\int_0^{1-x}dy
\int_0^\infty d\lambda \,\frac{(x\lam)^{n_1-1}}{\lambda\sigma^2} \, e^{-\lam Z},
\end{align}
where $\lam \equiv \lambda \rho L^2/4$.
For the simplest sunset integral, we have
\begin{align}
H^{r,H;n_1 1 1} & = \intrHtheta z\left[
       \left(A+\frac{2\delta}{\lam}\right) \left(\theta_{30}^3-1\right)
       -\frac{3\delta\rho}{4\lam^2}L^2\theta_{32}^{}\theta_{30}^2\right],
\nonumber\\
H^{r,H;n_1 2 1} & = \intrHtheta \, y \left(\theta_{30}^3-1\right),
\nonumber\\
H^{r,H;n_1 1 2} & = \intrHtheta \, z \left(\theta_{30}^3-1\right),
\nonumber\\
H^{r,H;n_1 2 2} & = \intrHtheta \, yz\lam \left(\theta_{30}^3-1\right),
\end{align}
and for the $H_{1}^{r,H}$ and $H_{2}^{r,H}$, we find
\begin{align}
H^{r,H;n_1 1 1}_1 & = \intrHtheta \, z\left[
       \left(\tau A+\frac{2\tau\delta}{\lam}-\frac{x\delta\rho}{\lam}\right)
                \left(\theta_{30}^3-1\right)
       -\frac{3\delta\rho\tau}{4\lam^2}L^2\theta_{32}^{}\theta_{30}^2\right],
\nonumber\\
H^{r,H;n_1 2 1}_1 & = \intrHtheta \, y\tau \left(\theta_{30}^3-1\right),
\nonumber\\
H^{r,H;n_1 1 2}_1 & = \intrHtheta \, z\tau \left(\theta_{30}^3-1\right),
\nonumber\\
H^{r,H;n_1 2 2}_1 & = \intrHtheta \, yz\tau\lam \left(\theta_{30}^3-1\right),
\end{align}
\begin{align}
H^{r,H;n_1 1 1}_2 & = \intrHtheta \, \frac{xz^2}{2\sigma}\left[
       \left( A+\frac{3\delta}{\lam}\right)
                \left(\theta_{30}^3-1\right)
       -\frac{3\delta\rho}{4\lam^2}L^2\theta_{32}^{}\theta_{30}^2\right],
\nonumber\\
H^{r,H;n_1 2 1}_2 & = \intrHtheta \, \frac{xyz}{\sigma} \left(\theta_{30}^3-1\right),
\nonumber\\
H^{r,H;n_1 1 2}_2 & = \intrHtheta \, \frac{xz^2}{\sigma} \left(\theta_{30}^3-1\right),
\nonumber\\
H^{r,H;n_1 2 2}_2 & = \intrHtheta \, \frac{xyz^2\lam}{\sigma} \left(\theta_{30}^3-1\right),
\end{align}
respectively. For the $H_{21}^{r,H}$, $H_{22}^{r,H}$, and $H_{27}^{r,H}$, we find
\begin{align}
H^{r,H;n_1 1 1}_{21} & = \intrHtheta \, z\tau\left[
       \left(\tau A+\frac{2\tau\delta}{\lam}-\frac{2x\delta\rho}{\lam}\right)
                \left(\theta_{30}^3-1\right)
       -\frac{3\tau\delta\rho}{4\lam^2}L^2\theta_{32}^{}\theta_{30}^2\right],
\nonumber\\
H^{r,H;n_1 2 1}_{21} & = \intrHtheta \, y\tau^2 \left(\theta_{30}^3-1\right),
\nonumber\\
H^{r,H;n_1 1 2}_{21} & = \intrHtheta \, z\tau^2 \left(\theta_{30}^3-1\right),
\nonumber\\
H^{r,H;n_1 2 2}_{21} & = \intrHtheta \, yz\tau^2\lam \left(\theta_{30}^3-1\right),
\end{align}
\begin{align}
H^{r,H;n_1 1 1}_{22} = & \intrHtheta \, \frac{z\rho}{2\lam}\left[
       \left(A+\frac{3\delta}{\lam}\right)
                \left(\theta_{30}^3-1\right)
       -\frac{3\delta\rho}{4\lam^2}L^2\theta_{32}^{}\theta_{30}^2\right],
\nonumber\\
H^{r,H;n_1 2 1}_{22} = & \intrHtheta \, \frac{y\rho}{2\lam}\left(\theta_{30}^3-1\right),
\nonumber\\
H^{r,H;n_1 1 2}_{22} = & \intrHtheta \, \frac{z\rho}{2\lam}\left(\theta_{30}^3-1\right),
\nonumber\\
H^{r,H;n_1 2 2}_{22} = & \intrHtheta \, \frac{yz\rho}{2} \left(\theta_{30}^3-1\right),
\end{align}
\begin{align}
H^{r,H;n_1 1 1}_{27} = & \intrHtheta \, \frac{-z\rho^2}{4\lam^2}\left[
       \left(A+\frac{4\delta}{\lam}\right)
             L^2\theta_{32}^{}\theta_{30}^2
       -\frac{\delta\rho}{4\lam^2}
          L^4\left(\theta_{34}^{}\theta_{30}^2+2\theta_{32}^2\theta_{30}^{}\right)
          \right],
\nonumber\\
H^{r,H;n_1 2 1}_{27} = & \intrHtheta \, \frac{-y\rho^2}{4\lam^2} L^2\theta_{32}^{}\theta_{30}^2,
\nonumber\\
H^{r,H;n_1 1 2}_{27} = & \intrHtheta \, \frac{-z\rho^2}{4\lam^2} L^2\theta_{32}^{}\theta_{30}^2,
\nonumber\\
H^{r,H;n_1 2 2}_{27} = & \intrHtheta \, \frac{-yz\rho^2}{4\lam} L^2\theta_{32}^{}\theta_{30}^2,
\end{align}
respectively, and for the $H_{23}^{r,H}$, $H_{24}^{r,H}$, and $H_{28}^{r,H}$, we have
\begin{align}
H^{r,H;n_1 1 1}_{23} = & \intrHtheta \, \frac{xz^2}{2\sigma}\left[
       \left(\tau A+\frac{3\tau\delta}{\lam}-\frac{x\delta\rho}{\lam}\right)
                \left(\theta_{30}^3-1\right)
       -\frac{3\tau\delta\rho}{4\lam^2}L^2\theta_{32}^{}\theta_{30}^2\right],
\nonumber\\
H^{r,H;n_1 2 1}_{23} = & \intrHtheta \, \frac{xyz\tau}{\sigma} \left(\theta_{30}^3-1\right),
\nonumber\\
H^{r,H;n_1 1 2}_{23} = & \intrHtheta \, \frac{xz^2\tau}{\sigma} \left(\theta_{30}^3-1\right),
\nonumber\\
H^{r,H;n_1 2 2}_{23} = & \intrHtheta \, \frac{xyz^2\tau\lam}{\sigma} \left(\theta_{30}^3-1\right),
\end{align}
\begin{align}
H^{r,H;n_1 1 1}_{24} = & \intrHtheta \, \frac{-z^2}{4\sigma\lam}\left[
       \left(A+\frac{3\delta}{\lam}\right)
                \left(\theta_{30}^3-1\right)
       -\frac{3\delta\rho}{4\lam^2}L^2\theta_{32}^{}\theta_{30}^2\right],
\nonumber\\
H^{r,H;n_1 2 1}_{24} = & \intrHtheta \, \frac{-yz}{2\sigma\lam}\left(\theta_{30}^3-1\right),
\nonumber\\
H^{r,H;n_1 1 2}_{24} = & \intrHtheta \, \frac{-z^2}{2\sigma\lam}\left(\theta_{30}^3-1\right),
\nonumber\\
H^{r,H;n_1 2 2}_{24} = & \intrHtheta \, \frac{-yz^2}{2\sigma} \left(\theta_{30}^3-1\right),
\end{align}
\begin{align}
H^{r,H;n_1 1 1}_{28} = & \intrHtheta \, \frac{z^2\rho}{8\sigma\lam^2}\left[
       \left(A+\frac{4\delta}{\lam}\right)
             L^2\theta_{32}^{}\theta_{30}^2
       -\frac{\delta\rho}{4\lam^2}
          L^4\left(\theta_{34}^{}\theta_{30}^2+2\theta_{32}^2\theta_{30}\right)
          \right],
\nonumber\\
H^{r,H;n_1 2 1}_{28} = & \intrHtheta \, \frac{yz\rho}{4\sigma\lam^2} L^2\theta_{32}^{}\theta_{30}^2,
\nonumber\\
H^{r,H;n_1 1 2}_{28} = & \intrHtheta \, \frac{z^2\rho}{4\sigma\lam^2} L^2\theta_{32}^{}\theta_{30}^2,
\nonumber\\
H^{r,H;n_1 2 2}_{28} = & \intrHtheta \, \frac{yz^2\rho}{4\sigma\lam} L^2\theta_{32}^{}\theta_{30}^2,
\end{align}
respectively. Finally, for the $H_{25}^{r,H}$, $H_{26}^{r,H}$, and $H_{29}^{r,H}$, we find
\begin{align}
H^{r,H;n_1 1 1}_{25} = & \intrHtheta \, \frac{x^2z^3}{3\sigma^2}\left[
       \left(A+\frac{4\delta}{\lam}\right)
                \left(\theta_{30}^3-1\right)
       -\frac{3\delta\rho}{4\lam^2}L^2\theta_{32}^{}\theta_{30}^2\right],
\nonumber\\
H^{r,H;n_1 2 1}_{25} = & \intrHtheta \, \frac{x^2yz^2}{\sigma^2} \left(\theta_{30}^3-1\right),
\nonumber\\
H^{r,H;n_1 1 2}_{25} = & \intrHtheta \, \frac{x^2z^3}{\sigma^2} \left(\theta_{30}^3-1\right),
\nonumber\\
H^{r,H;n_1 2 2}_{25} = & \intrHtheta \, \frac{x^2yz^3\lam}{\sigma^2} \left(\theta_{30}^3-1\right),
\end{align}
\begin{align}
H^{r,H;n_1 1 1}_{26} & = \intrHtheta \, \Bigg\{
\Bigg[-\frac{z m_2^2}{2}\left(A+\frac{2\delta}{\lam}\right)
     +\frac{z^2}{6\rho\sigma^2\lam}(5z+3y)\left(m_3^2-m_2^2\right) 
     -\frac{3yz^2}{2\rho\sigma^2\lam}\left(A+\frac{\delta}{\lam}\right)
\nonumber\\ & \quad
     -\frac{z^2 A}{12\rho\sigma}\left(3A+\frac{4z}{\lam\sigma}
                  +\frac{4x^2z\rho p^2}{\sigma}\right)
 \Bigg]   \left(\theta_{30}^3-1\right)
\nonumber\\ & \quad
    +  \left[\frac{z^2\delta}{8\sigma^2\lam^3}(2z+9y)
            +\frac{z^2}{8\sigma^2\lam^2}\left((z+3y)A-2z(m_3^2-m_2^2)\right)
            +\frac{3z\delta\rho}{8\lam^2}m_2^2
      \right] L^2\theta_{32}^{}\theta_{30}^2
\nonumber\\ & \quad
 -\frac{z^2\delta\rho}{64\sigma^2\lam^4}(z+3y)L^4
       \left(\theta_{34}^{}\theta_{30}^2+2\theta_{32}^2\theta_{30}^{}\right)
\Bigg\},
\nonumber\\
H^{r,H;n_1 2 1}_{26} & = \intrHtheta \, \left\{
   \left[\frac{1}{4}\left(z+\frac{\tau}{\rho}\right)A
         +\frac{z\delta}{\lam}+\frac{z^3}{2\rho\sigma^2\lam}\right]
\left(\theta_{30}^3-1\right)
-\frac{3\delta\rho}{16\lam^2}\left(z+\frac{\tau}{\rho}\right)
   L^2\theta_{32}^{}\theta_{30}^2
\right\},
\nonumber\\
H^{r,H;n_1 1 2}_{26} & = \intrHtheta \, \left\{
   \left[\frac{1}{4}\left(z-\frac{\tau}{\rho}\right)A
         +\frac{z^2}{2\sigma\lam}-\frac{z^3}{2\rho\sigma^2\lam}\right]
\left(\theta_{30}^3-1\right)
-\frac{3\delta\rho}{16\lam^2}\left(z-\frac{\tau}{\rho}\right)
   L^2\theta_{32}^{}\theta_{30}^2
\right\},
\nonumber\\
H^{r,H;n_1 2 2}_{26} & = \intrHtheta \, \frac{\tau}{2\rho}
                      \left(1+\frac{z^2}{\sigma}\right)
                      \left(\theta_{30}^3-1\right),
\end{align}
and
\begin{align}
H^{r,H;n_1 1 1}_{29} & = \intrHtheta \, 
      \frac{-z^3}{12\sigma^2\lam^2}\left[
       \left(A+\frac{4\delta}{\lam}\right)
             L^2\theta_{32}^{}\theta_{30}^2
       -\frac{\delta\rho}{4\lam^2}
          L^4\left(\theta_{34}^{}\theta_{30}^2+2\theta_{32}^2\theta_{30}^{}\right)
          \right],
\nonumber\\
H^{r,H;n_1 2 1}_{29} & = \intrHtheta  \, \frac{-yz^2}{4\sigma^2\lam^2} L^2\theta_{32}^{}\theta_{30}^2,
\nonumber\\
H^{r,H;n_1 1 2}_{29} & = \intrHtheta  \, \frac{-z^3}{4\sigma^2\lam^2} L^2\theta_{32}^{}\theta_{30}^2,
\nonumber\\
H^{r,H;n_1 2 2}_{29} & = \intrHtheta \, \frac{-yz^3}{4\sigma^2\lam} L^2\theta_{32}^{}\theta_{30}^2.
\end{align}

\subsection{Sunset integrals with two quantized loop momenta}

Here, we follow the treatment of Sect.~\ref{simplestrs}, and generalize to
all integrals $\sunset{X}_{rs}$ with $X=1,r_\mu,s_\mu,r_\mu r_\nu,r_\mu s_\nu$ and $s_\mu s_\nu$.
All of these are not needed for completeness, but the redundant ones 
enable a check on our results by means of the relations given in Sect.~\ref{permutation}.
We again introduce Gaussian parameterizations for the propagators using Eq.~(\ref{prop}),
and then shift the momenta using Eqs.~(\ref{shiftr}) and~(\ref{shifts}). This leads to
\begin{align}
\sunset{X}_{rs} & =
\frac{1}{\Gamma(n_1)\Gamma(n_2)\Gamma(n_3)(4\pi)^d}\sum^{\prime\prime}_{l_r,l_s}
\int_0^\infty d\lambda_1 d\lambda_2 d\lambda_3 
\, \frac{\lambda_1^{n_1-1}\lambda_2^{n_2-1}\lambda_3^{n_3-1}}{\tilde\lambda^{d/2}}
\, \sunsetx{X}_{rs} \,
e^{-\tilde M^2},
\end{align}
where $\tilde M^2$ is defined in Eq.~(\ref{defMtilde}), and $\tilde\lambda
\equiv \lambda_1\lambda_2+ \lambda_2\lambda_3 + \lambda_3\lambda_1$.
For the $\sunsetx{X}_{rs}$, we find
\begin{align}
\sunsetx{1}_{rs} & = 1 ,
\nonumber\\
\sunsetx{r_\mu}_{rs} & = \frac{1}{\tilde\lambda}
  \left(\lambda_2\lambda_3 p_\mu+\frac{i}{2}\lambda_2 l_{r\mu}
              +\frac{i}{2}\lambda_3 l_{n\mu}\right),
\nonumber\\
\sunsetx{s_\mu}_{rs} & = \frac{1}{\tilde\lambda}
  \left(\lambda_1\lambda_3 p_\mu+\frac{i}{2}\lambda_1 l_{s\mu}
              -\frac{i}{2}\lambda_3 l_{n\mu}\right),
\nonumber\\
\sunsetx{r_\mu r_\nu}_{rs} & =
\frac{\lambda_2^2\lambda_3^2}{\tilde\lambda^2} p_\mu p_\nu
+\frac{\lambda_2+\lambda_3}{2\tilde\lambda}\delta_{\mu\nu}
+\frac{i\lambda_2^2\lambda_3}{2\tilde\lambda^2}\{p,l_r\}_{\mu\nu}
+\frac{i\lambda_2\lambda_3^2}{2\tilde\lambda^2}\{p,l_n\}_{\mu\nu}
\nonumber\\
& \quad -\frac{1}{4\tilde\lambda^2}\left(\lambda_2^2 l_{r\mu}l_{r\nu}
      +\lambda_2\lambda_3\{l_r,l_n\}_{\mu\nu}+\lambda_3^2 l_{n\mu}l_{n\nu}\right),
\nonumber\\
\sunsetx{r_\mu s_\nu}_{rs} & = 
\frac{\lambda_1\lambda_2\lambda_3^2}{\tilde\lambda^2} p_\mu p_\nu
-\frac{\lambda_3}{2\tilde\lambda}\delta_{\mu\nu}
-\frac{i\lambda_2\lambda_3^2}{2\tilde\lambda^2}p_\mu l_{n\nu}
+\frac{i\lambda_1\lambda_3^2}{2\tilde\lambda^2}p_\nu l_{n\mu}
+\frac{i\lambda_1\lambda_2\lambda_3}{2\tilde\lambda^2}\left(p_\mu l_{s\nu}
         +p_{\nu}l_{r\mu}\right)
\nonumber\\
& \quad +\frac{1}{4\tilde\lambda^2}\left(\lambda_3^2 l_{n\mu}l_{n\nu}
      +\lambda_2\lambda_3 l_{r\mu} l_{n\mu}
      -\lambda_1\lambda_3 l_{n\mu}l_{s\nu}
      -\lambda_1\lambda_2 l_{r\mu}l_{s\nu}\right),
\nonumber\\
\sunsetx{s_\mu s_\nu}_{rs} & =  
\frac{\lambda_1^2\lambda_3^2}{\tilde\lambda^2} p_\mu p_\nu
+\frac{\lambda_1+\lambda_3}{2\tilde\lambda}\delta_{\mu\nu}
+\frac{i\lambda_1^2\lambda_3}{2\tilde\lambda^2}\{p,l_s\}_{\mu\nu}
-\frac{i\lambda_1\lambda_3^2}{2\tilde\lambda^2}\{p,l_n\}_{\mu\nu}
\nonumber\\
& \quad -\frac{1}{4\tilde\lambda^2}\left(\lambda_1^2 l_{s\mu}l_{s\nu}
      -\lambda_1\lambda_3\{l_s,l_n\}_{\mu\nu}+\lambda_3^2 l_{n\mu}l_{n\nu}\right),
\end{align}
where $l_n \equiv l_r-l_s$. We may now switch integration variables to 
to $x,y,z \equiv 1-x-y$ and $\lambda$ as in Eq.~(\ref{xyzlambda}), which
gives us an integral similar to Eq.~(\ref{xyzintegral}).

In what follows, we restrict ourselves to the cms frame with
$p\cdot l_r = p\cdot l_s = 0$, which simplifies the expressions greatly.
The results for a moving frame can again be obtained
using the same methods. In the cms frame, the exponential factors depend only 
on the components of $l_r$ and $l_s$ via $l_r^2, l_s^2$ and $l_n^2$. This allows us to write
\begin{align}
\sum^{\prime\prime}_{l_r,l_s} l_{r\mu} f(l_r^2,l_s^2,l_n^2) & =
\sum^{\prime\prime}_{l_r,l_s} l_{r\mu} f(l_r^2,l_s^2,l_n^2) = 0,
\nonumber\\
\sum^{\prime\prime}_{l_r,l_s} l_{r\mu}l_{r\nu} f(l_r^2,l_s^2,l_n^2) & =
\frac{t_{\mu\nu}}{3}\sum^{\prime\prime}_{l_r,l_s} l_r^2 f(l_r^2,l_s^2,l_n^2),
\nonumber\\
\sum^{\prime\prime}_{l_r,l_s} l_{s\mu}l_{s\nu} f(l_r^2,l_s^2,l_n^2) & =
\frac{t_{\mu\nu}}{3}\sum^{\prime\prime}_{l_r,l_s} l_s^2 f(l_r^2,l_s^2,l_n^2),
\nonumber\\
\sum^{\prime\prime}_{l_r,l_s} l_{r\mu}l_{s\nu} f(l_r^2,l_s^2,l_n^2) & =
\frac{t_{\mu\nu}}{3}\sum^{\prime\prime}_{l_r,l_s} l_r\cdot l_s
 f(l_r^2,l_s^2,l_n^2),
\nonumber\\
l_r\cdot l_s & = \frac{1}{2}\left(l_r^2+l_s^2-l_n^2\right).
\end{align}

\subsubsection{Center-of-mass frame: Bessel functions}
\label{Hrsbessel}

As for the sunset integrals with one quantized loop momentum, the 
integral over $\lambda$ can again be performed in terms of the modified Bessel functions
$\mathcal{K}_\nu(Y_{rs},Z_{rs})$, where $Y_{rs}$ and $Z_{rs}$ are defined in Eq.~(\ref{rsdefs}).
These arguments will be suppressed for brevity. While the sextuple summation 
over the components of $l_r$ and $l_s$ can be reduced to a triple sum using Eq.~(\ref{xktrick2}),
we find that the remaining summations converge fairly slowly for moderate values of $m_i^{} L$.
In the following expressions, we set $d=4$ since no divergences appear.
Using the notation
\newcommand{\intHrs}{\int\hskip-2.5ex D}
\begin{align}
\intHrs  \: & \equiv \:
\frac{1}{\Gamma(n_1)\Gamma(n_2)\Gamma(n_3)(16\pi^2)^2}
\sum^{\prime\prime}_{l_r,l_s}\int_0^1 dx \int_0^{1-x} dy
 \, \frac{x^{n_1-1} y^{n_2-1} z^{n_3-1}}{\sigma^2},
\end{align}
and $m \equiv n_1+n_2+n_3-4$, we obtain
\begin{align}
H^{rs; n_1^{} n_2^{} n_3^{}} & = \intHrs \, \mathcal{K}_{m}^{},
\nonumber\\
H^{rs; n_1^{} n_2^{} n_3^{}}_1 & = \intHrs \, \frac{yz}{\sigma} \, \mathcal{K}_{m}^{},
\nonumber\\
H^{rs; n_1^{} n_2^{} n_3^{}}_2 & = \intHrs \, \frac{xz}{\sigma} \, \mathcal{K}_{m}^{},
\end{align}
for the simplest sunset integral and the scalar components of the integrals with one Lorentz index. For the
components of the sunset integrals with two Lorentz indices, we find
\begin{align}
H^{rs; n_1^{} n_2^{} n_3^{}}_{21} & = \intHrs \, \frac{y^2z^2}{\sigma^2} \, \mathcal{K}_{m}^{},
\nonumber\\
H^{rs; n_1^{} n_2^{} n_3^{}}_{22} & = \intHrs \, \frac{y+z}{2\sigma} \, \mathcal{K}_{m-1}^{},
\nonumber\\
H^{rs; n_1^{} n_2^{} n_3^{}}_{27} & = \intHrs \, \frac{1}{12\sigma^2}
\Big[\!-\!y(y+z) \, l_r^2 + yz \, l_s^2 -z(y+z) \, l_n^2\Big] \, \mathcal{K}_{m-2}^{},
\end{align}
\begin{align}
H^{rs; n_1^{} n_2^{} n_3^{}}_{23} & = \intHrs \, \frac{xyz^2}{\sigma^2} \, \mathcal{K}_{m}^{},
\nonumber\\
H^{rs; n_1^{} n_2^{} n_3^{}}_{24} & = \intHrs \, \frac{-z}{2\sigma} \, \mathcal{K}_{m-1}^{},
\nonumber\\
H^{rs; n_1^{} n_2^{} n_3^{}}_{28} & = \intHrs \, \frac{1}{24\sigma^2}
\Big[(2yz-\sigma) \, l_r^2 + (2xz-\sigma) \, l_s^2 + (2z^2+\sigma) \, l_n^2\Big] \, \mathcal{K}_{m-2}^{},
\end{align}
and
\begin{align}
H^{rs; n_1^{} n_2^{} n_3^{}}_{25} & = \intHrs \, \frac{x^2z^2}{\sigma^2} \, \mathcal{K}_{m}^{},
\nonumber\\
H^{rs; n_1^{} n_2^{} n_3^{}}_{26} & = \intHrs \, \frac{x+z}{2\sigma} \, \mathcal{K}_{m-1}^{},
\nonumber\\
H^{rs; n_1^{} n_2^{} n_3^{}}_{29} & = \intHrs \, \frac{1}{12\sigma^2}
\Big[xz \, l_r^2-x(x+z) \, l_s^2-z(x+z) \, l_n^2\Big] \, \mathcal{K}_{m-2}^{}.
\end{align}

\subsubsection{Center-of-mass frame: Theta functions}
\label{Hrstheta}

In the cms frame, the double summation can be performed in terms
of the theta functions, as encountered in the treatment of the simplest sunset integral. If we define
\begin{equation}
\bar\lambda \equiv \frac{4\sigma}{L^2}\lambda, \qquad
l_r^{} \equiv n_r^{} L, \qquad
l_s^{} \equiv n_s^{} L, \qquad
n_n^{} \equiv n_r^{}-n_s^{},
\end{equation}
we find
\begin{align}
\label{sumtheta4}
\sum^{\prime\prime}_{l_r,l_s} e^{-\frac{yl_r^2}{4\sigma\lambda}
-\frac{x l_s^2}{4\sigma\lambda}-\frac{z l_n^2}{4\sigma\lambda}}
& = \sum^{\prime\prime}_{n_r,n_s} e^{-\frac{y}{\bar\lambda}n_r^2
-\frac{x}{\bar\lambda}n_s^2-\frac{z}{\bar\lambda}n_n^2}
\nonumber\\
& = \sum_{n_r^{},n_s^{}} e^{-\frac{y}{\bar\lambda}n_r^2
-\frac{x}{\bar\lambda}n_s^2-\frac{z}{\bar\lambda}n_n^2}
-\sum_{n_r^{}} e^{-\frac{y+z}{\bar\lambda}n_r^2}
-\sum_{n_s^{}} e^{-\frac{x+z}{\bar\lambda}n_s^2}
-\sum_{n_r^{}} e^{-\frac{x+y}{\bar\lambda}n_r^2}
+2
\nonumber\\
& = \theta^{(2)}_0\left(\frac{y}{\bar\lambda},\frac{x}{\bar\lambda}
  ,\frac{z}{\bar\lambda}\right)^3
   -\theta_{30}^{}\left(e^{-\frac{y+z}{\bar\lambda}}\right)^3
   -\theta_{30}^{}\left(e^{-\frac{x+z}{\bar\lambda}}\right)^3
   -\theta_{30}^{}\left(e^{-\frac{x+y}{\bar\lambda}}\right)^3
   +2
\nonumber\\
& \equiv \Theta_0^{}\left(\frac{y}{\bar\lambda},\frac{x}{\bar\lambda}
  ,\frac{z}{\bar\lambda}\right),
\end{align}
which was already used in Eq.~(\ref{resultHrssimple}). Here, the terms
involving $\theta_{30}^{}$ subtract the contributions
with ($n_s=0, n_n=n_r$), ($n_r=0, n_n=-n_s$), and ($n_n=0, n_r=n_s$). 
The constant term corrects for the case when ($n_r=n_s=0$) is
subtracted to often. By taking derivatives {\it w.r.t.} $x,y,z$, we also find
\begin{align}
\label{sumtheta5}
\sum^{\prime\prime}_{l_r,l_s} l_r^2 \, e^{-\frac{yl_r^2}{4\sigma\lambda}
-\frac{x l_s^2}{4\sigma\lambda}-\frac{z l_n^2}{4\sigma\lambda}} & =
3L^2 \,\theta^{(2)}_{02}\left(\frac{y}{\bar\lambda},\frac{x}{\bar\lambda}
  ,\frac{z}{\bar\lambda}\right)
\theta^{(2)}_{0}\left(\frac{y}{\bar\lambda},\frac{x}{\bar\lambda}
  ,\frac{z}{\bar\lambda}\right)^2
\nonumber\\ & \quad
   -3L^2\,\theta_{32}^{}\left(e^{-\frac{y+z}{\bar\lambda}}\right)
     \theta_{30}^{}\left(e^{-\frac{y+z}{\bar\lambda}}\right)^2
   -3L^2\,\theta_{32}^{}\left(e^{-\frac{x+y}{\bar\lambda}}\right)
     \theta_{30}^{}\left(e^{-\frac{x+y}{\bar\lambda}}\right)^2
\nonumber\\
& \equiv 3 L^2 \,\Theta_{02}^{}\left(\frac{y}{\bar\lambda},\frac{x}{\bar\lambda}
  ,\frac{z}{\bar\lambda}\right).
\end{align}
If we introduce the abbreviation
\newcommand{\intHrstheta}{\int\hskip-2.3ex S}
\begin{align}
\intHrstheta  \: & \equiv \: 
\frac{1}{\Gamma(n_1)\Gamma(n_2)\Gamma(n_3)(16\pi^2)^2}
\int_0^1 dx \int_0^{1-x} dy\int_0^\infty d\lambda
\nonumber \\ & \quad
 \times \frac{x^{n_1-1} y^{n_2-1} z^{n_3-1}\lambda^{n_1+n_2+n_3-5}}{\sigma^2} \,
e^{-\lambda Z_{rs}},
\end{align}
we can express the scalar components as
\begin{align}
H^{rs; n_1^{} n_2^{} n_3^{}} = & \intHrstheta \,\Theta_0^{}\left(\frac{y}{\bar\lambda},
     \frac{x}{\bar\lambda},\frac{z}{\bar\lambda}\right),
\nonumber\\
H^{rs; n_1^{} n_2^{} n_3^{}}_1 = & \intHrstheta \,\frac{yz}{\sigma}
     \,\Theta_0^{}\left(\frac{y}{\bar\lambda},
     \frac{x}{\bar\lambda},\frac{z}{\bar\lambda}\right),
\nonumber\\
H^{rs; n_1^{} n_2^{} n_3^{}}_2 = & \intHrstheta \,\frac{xz}{\sigma}
     \,\Theta_0^{}\left(\frac{y}{\bar\lambda},
     \frac{x}{\bar\lambda},\frac{z}{\bar\lambda}\right),
 \end{align}
\begin{align}
H^{rs; n_1^{} n_2^{} n_3^{}}_{21} & = \intHrstheta \,\frac{y^2z^2}{\sigma^2}
     \,\Theta_0^{}\left(\frac{y}{\bar\lambda},
     \frac{x}{\bar\lambda},\frac{z}{\bar\lambda}\right),
\nonumber\\
H^{rs; n_1^{} n_2^{} n_3^{}}_{22} & = \intHrstheta \,\frac{y+z}{2\lambda\sigma}
     \,\Theta_0^{}\left(\frac{y}{\bar\lambda},
     \frac{x}{\bar\lambda},\frac{z}{\bar\lambda}\right),
\nonumber\\
H^{rs; n_1^{} n_2^{} n_3^{}}_{27} & = \intHrstheta \,\frac{L^2}{4\sigma^2\lambda^2}
\left[
-y(y+z) \,\Theta_{02}^{}\left(\frac{y}{\bar\lambda},
     \frac{x}{\bar\lambda},\frac{z}{\bar\lambda}\right)
+yz \,\Theta_{02}^{}\left(\frac{x}{\bar\lambda},
     \frac{y}{\bar\lambda},\frac{z}{\bar\lambda}\right) \right.
\nonumber\\ & \quad
\left. -\,z(y+z) \,\Theta_{02}^{}\left(\frac{z}{\bar\lambda},
     \frac{x}{\bar\lambda},\frac{y}{\bar\lambda}\right)\right],
\end{align}
\begin{align}
H^{rs; n_1^{} n_2^{} n_3^{}}_{23} & = \intHrstheta \, \frac{xyz^2}{\sigma^2}
     \,\Theta_0^{}\left(\frac{y}{\bar\lambda},
     \frac{x}{\bar\lambda},\frac{z}{\bar\lambda}\right),
\nonumber\\
H^{rs; n_1^{} n_2^{} n_3^{}}_{24} & = \intHrstheta \, \frac{-z}{2\lambda\sigma}
     \,\Theta_0^{}\left(\frac{y}{\bar\lambda},
     \frac{x}{\bar\lambda},\frac{z}{\bar\lambda}\right),
\nonumber\\
H^{rs; n_1^{} n_2^{} n_3^{}}_{28} & = \intHrstheta \, \frac{L^2}{8\sigma^2\lambda^2}
\Big[
(2yz-\sigma) \,\Theta_{02}^{}\left(\frac{y}{\bar\lambda},
     \frac{x}{\bar\lambda},\frac{z}{\bar\lambda}\right)
+(2xz-\sigma) \,\Theta_{02}^{}\left(\frac{x}{\bar\lambda},
     \frac{y}{\bar\lambda},\frac{z}{\bar\lambda}\right)
\nonumber\\ & \quad
+(2z^2+\sigma) \,\Theta_{02}^{}\left(\frac{z}{\bar\lambda},
     \frac{x}{\bar\lambda},\frac{y}{\bar\lambda}\right)\Big],
 \end{align}
and
\begin{align}
H^{rs; n_1^{} n_2^{} n_3^{}}_{25} & = \intHrstheta \, \frac{x^2z^2}{\sigma^2}
     \,\Theta_0^{}\left(\frac{y}{\bar\lambda},
     \frac{x}{\bar\lambda},\frac{z}{\bar\lambda}\right),
\nonumber\\
H^{rs; n_1^{} n_2^{} n_3^{}}_{26} & = \intHrstheta \, \frac{x+z}{2\lambda\sigma}
     \,\Theta_0^{}\left(\frac{y}{\bar\lambda},
     \frac{x}{\bar\lambda},\frac{z}{\bar\lambda}\right),
\nonumber\\
H^{rs; n_1^{} n_2^{} n_3^{}}_{29} & = \intHrstheta \, \frac{L^2}{4\sigma^2\lambda^2}
\Big[
xz \,\Theta_{02}^{}\left(\frac{y}{\bar\lambda},
     \frac{x}{\bar\lambda},\frac{z}{\bar\lambda}\right)
-x(x+z) \,\Theta_{02}^{}\left(\frac{x}{\bar\lambda}, 
     \frac{y}{\bar\lambda},\frac{z}{\bar\lambda}\right)
\nonumber\\ & \quad
-z(x+z) \,\Theta_{02}^{}\left(\frac{z}{\bar\lambda},
     \frac{x}{\bar\lambda},\frac{y}{\bar\lambda}\right)\Big].
 \end{align}
 %
 

\section{Numerical results}
\label{Numerics}

As a numerical check of the results presented here, we have evaluated all integrals in terms of modified Bessel
functions as well as theta functions, and checked these for agreement with each other.
We have also verified the expected integral relations by numerical differentiation
{\it w.r.t.} $m_1^2$, $m_2^2$ and $m_3^2$. Furthermore, we have checked that the expected symmetries
under interchange of masses are satisfied. For the sunset integrals, this can be non-trivial
as the permutation symmetries are not explicitly conserved by the analytical methods employed here.
We have also verified that the one-loop results satisfy the integral relations in 
Eq.~(\ref{PVrelations1}) and~(\ref{PVrelations2}). For reference, we present numerical results with
6~digits of precision. Implementations of the full set of sunset integrals are available from the authors 
in {\tt C++} and {\tt Mathematica}.

\begin{figure}
\includegraphics[width=0.49\textwidth]{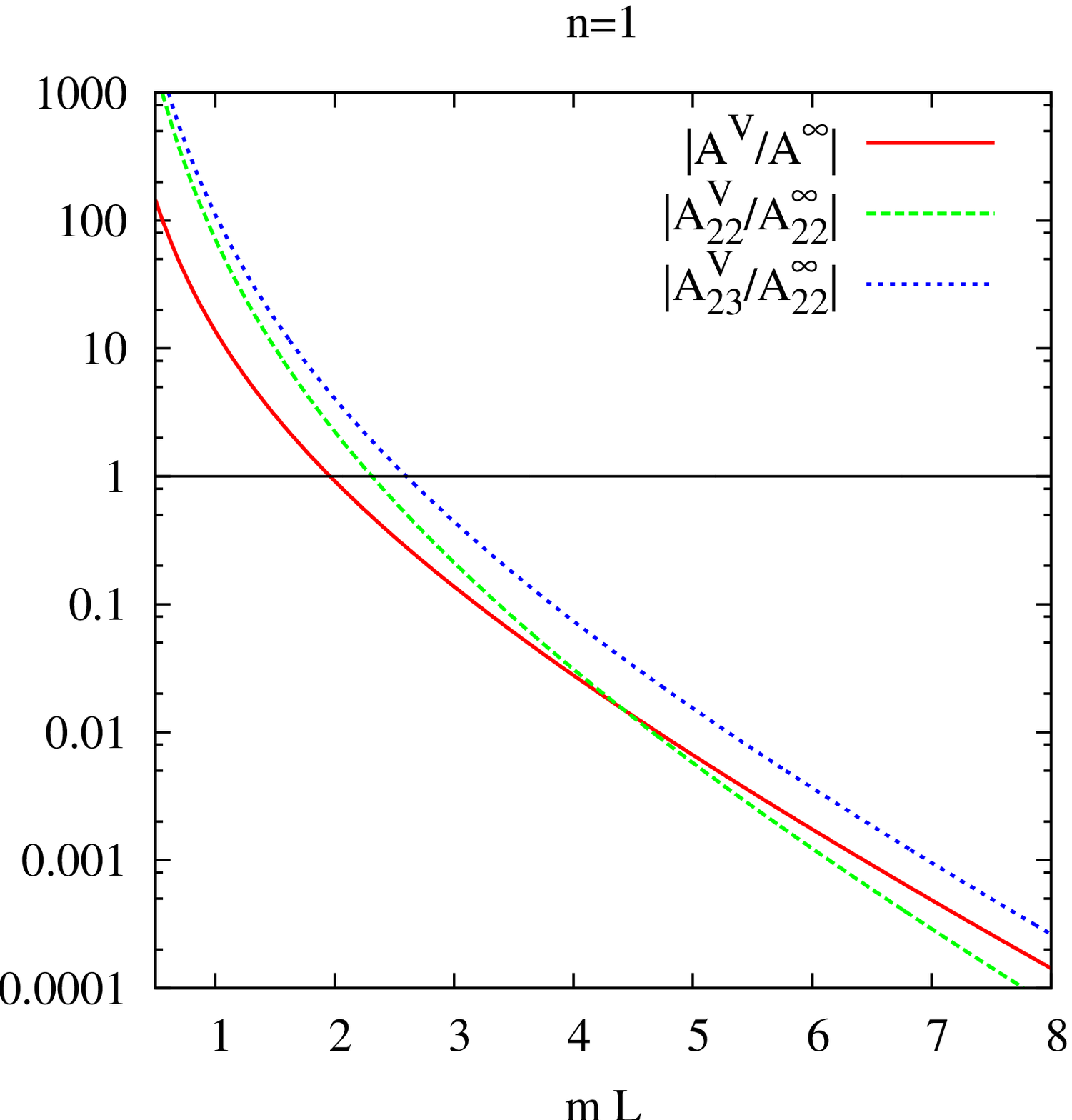}
\includegraphics[width=0.49\textwidth]{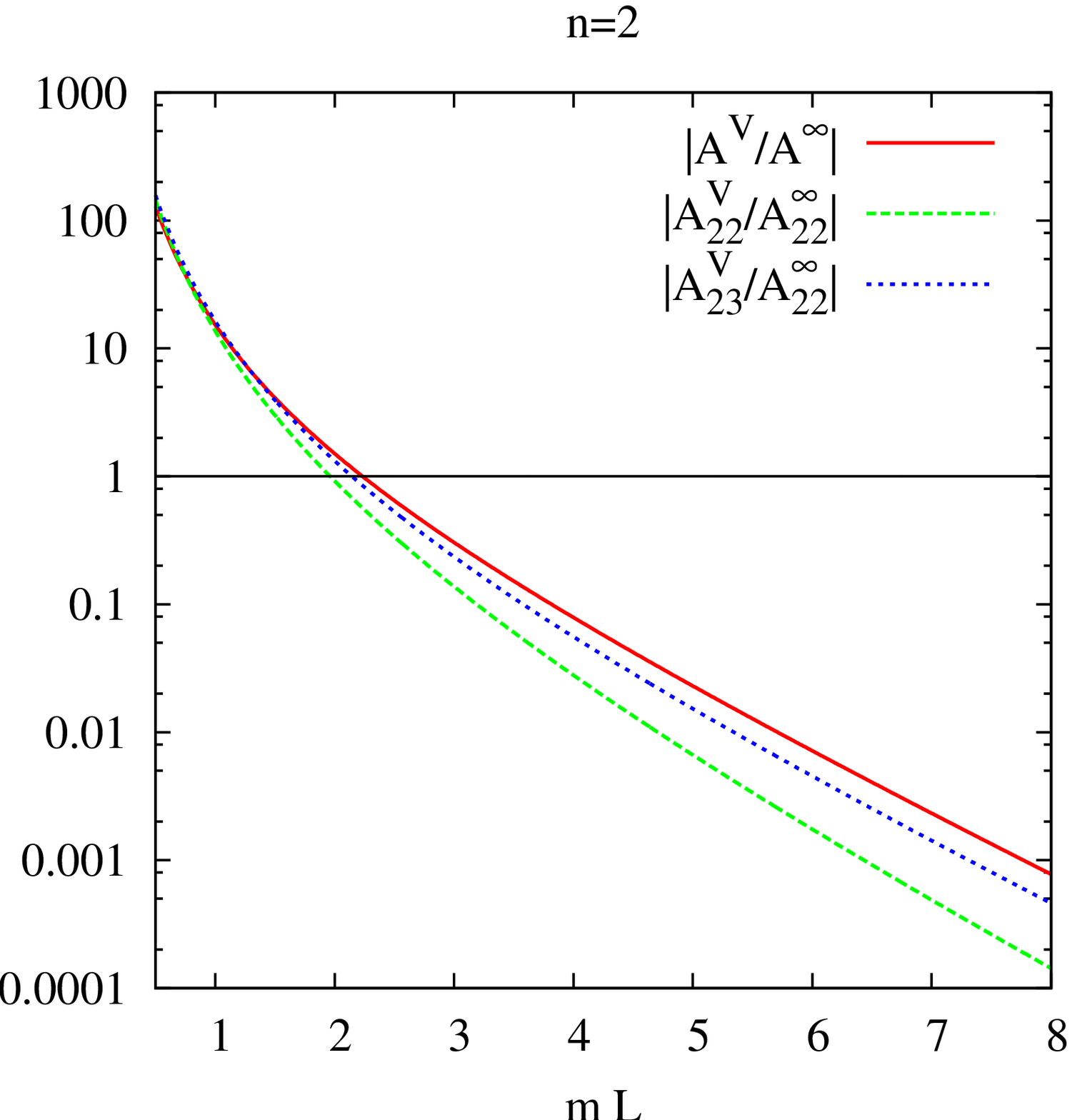}
\caption{\label{fig:tadpole} Ratio of finite-volume corrections to
infinite-volume results for the ``tadpole'' integrals, for $m = 0.1395$~GeV.
The continuum results employ the $\overline{\mathrm{MS}}$ subtraction scheme
with $\mu=0.77$~GeV. 
We compare the ``23'' case to the ``22'' case at infinite volume, as the former has no infinite-volume counterpart.
The left panel shows the results for $n = 1$, the right panel for $n = 2$, see App.~\ref{poles} for the pole configurations of the propagators.
All results are in units of the appropriate powers of GeV.}
\end{figure}

\begin{figure}
\begin{center}
\includegraphics[width=0.49\textwidth]{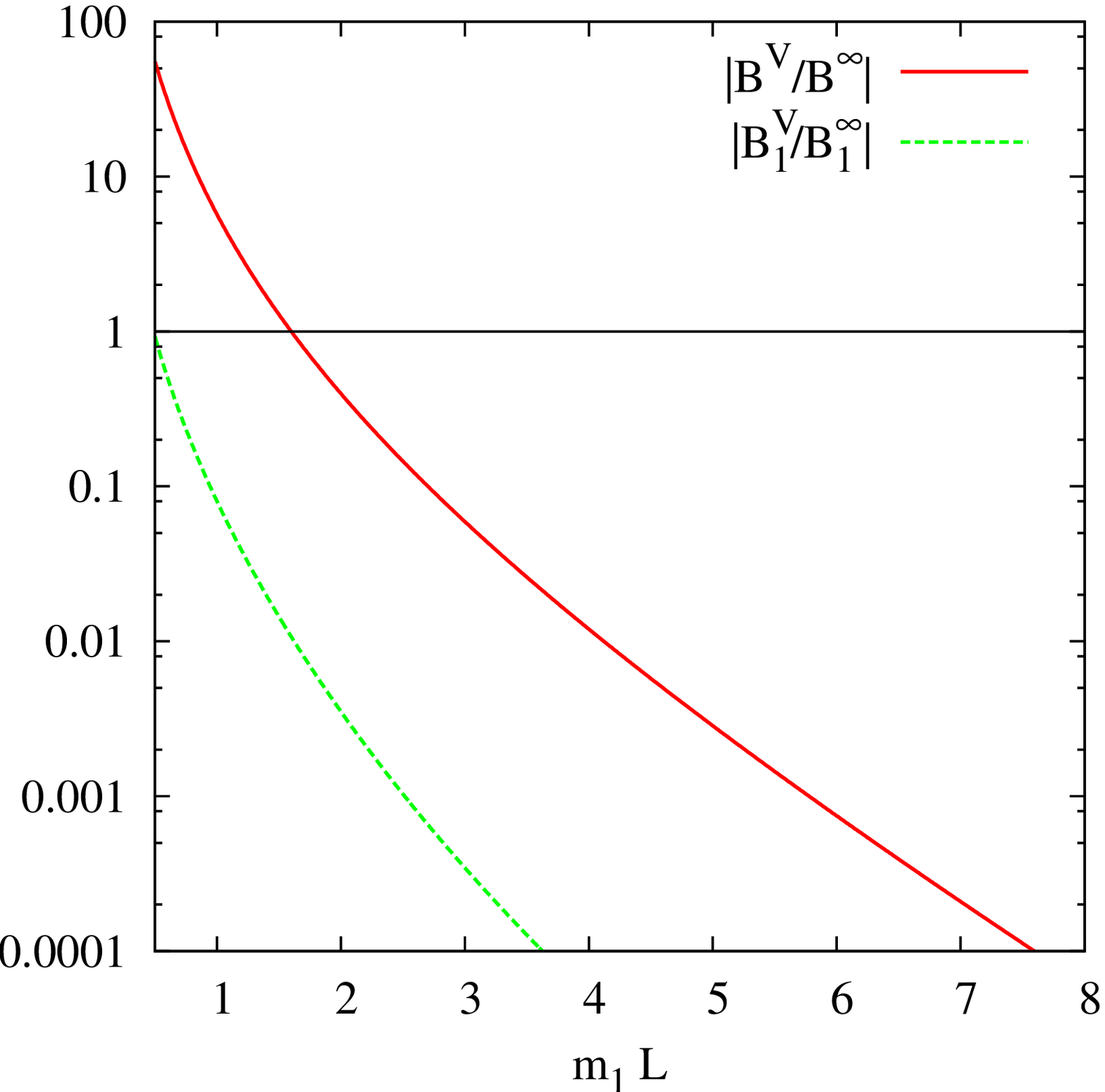} 
\\ \vspace{.5cm}
\includegraphics[width=0.49\textwidth]{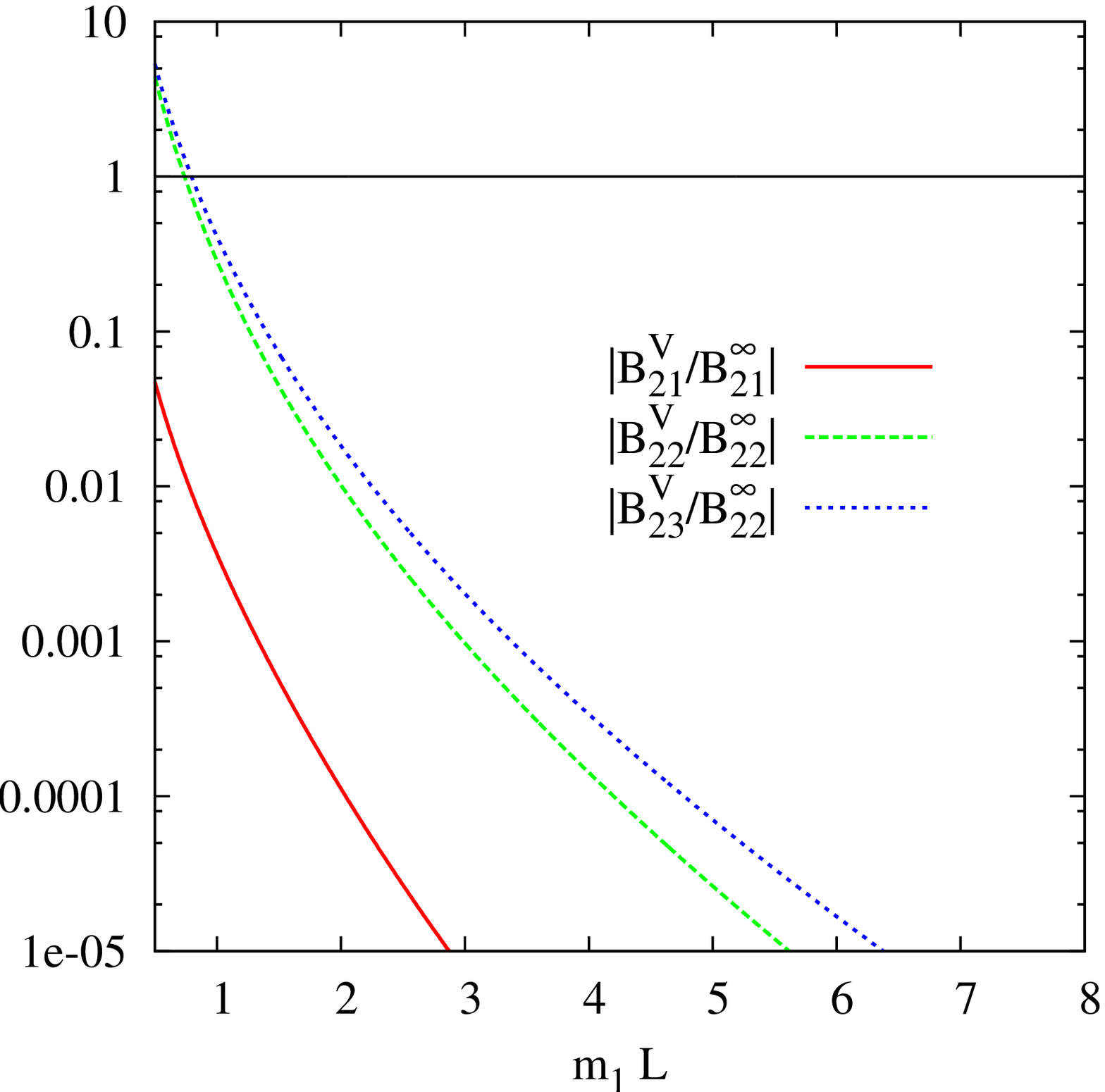}
\includegraphics[width=0.49\textwidth]{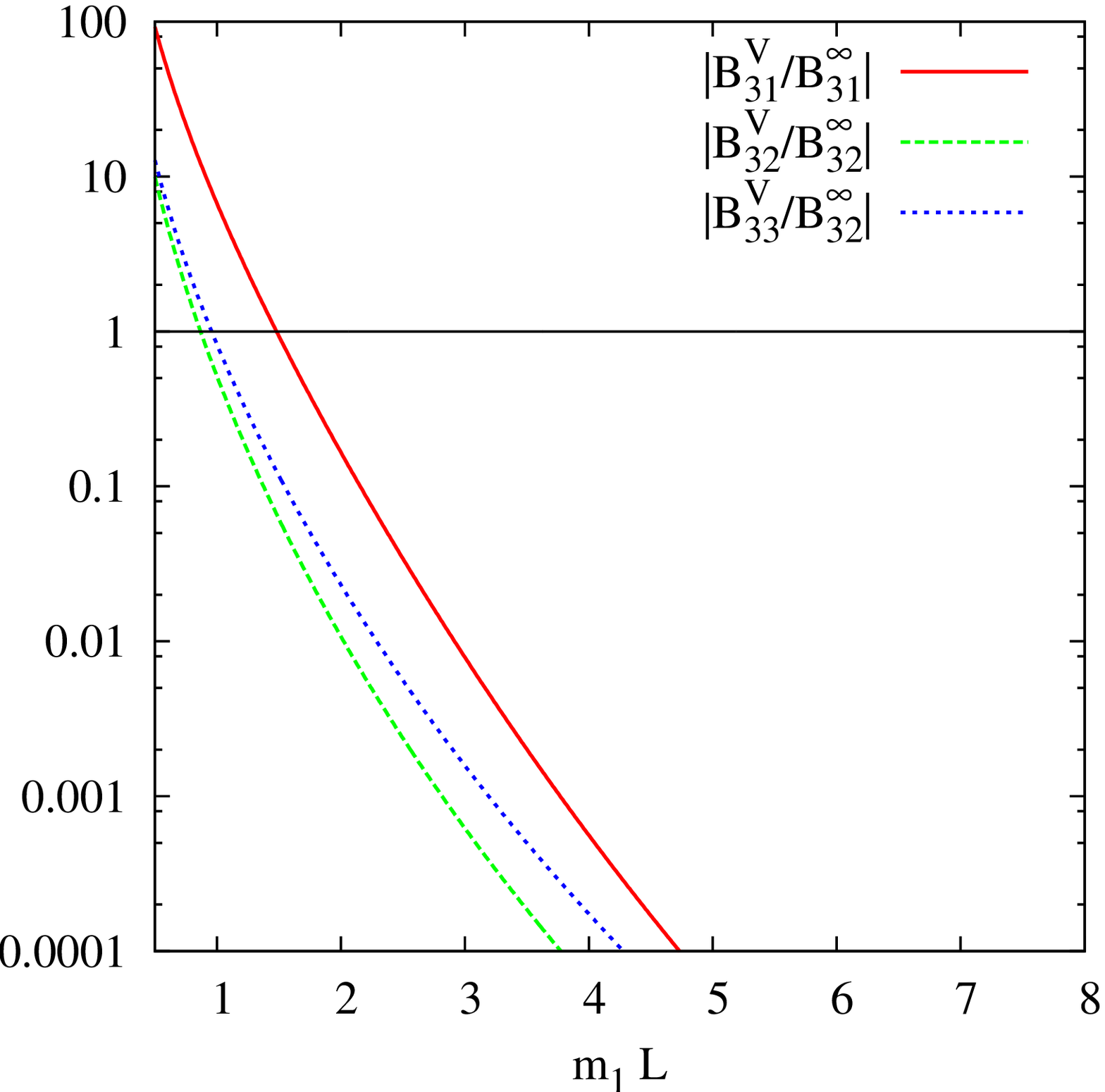}
\end{center}
\caption{\label{fig:bubble1} Ratio of finite-volume corrections to
infinite-volume results for the ``bubble'' integrals, for $m_1 = 0.1395$~GeV, $m_2=0.495$~GeV, and $p^2=m_1^2$.
The continuum results employ the $\overline{\mathrm{MS}}$ subtraction scheme
with $\mu=0.77$~GeV. 
We compare the ``23'' case to the ``22'' case and the ``33'' case to the ``32'' case
at infinite volume, as the former have no infinite-volume counterparts. The top panel shows
$B$ and $B_1$, the bottom left panel shows $B_{21}$, $B_{22}$ and $B_{23}$, and the bottom right panel  
shows $B_{31}, B_{32}$ and $B_{33}$. All results are in units of the appropriate
powers of GeV. Only the case of $n_1^{} = n_2^{} = 1$ is given.}
\end{figure}

Numerical results for the one-propagator or ``tadpole'' integrals, defined in Eq.~(\ref{Adef}),
are given in Tab.~\ref{tab:tadpole}. We note that there is no infinite-volume counterpart of the $A_{23}^V$ 
integral. In Fig.~\ref{fig:tadpole}, we show the ratio of the finite-volume correction to the infinite-volume result 
as a function of~$mL$. For the two-propagator or ``bubble'' integrals, defined in Eq.~(\ref{Bdef}), results for one set of
input parameters are given in Tab.~\ref{tab:bubble}. We only quote the results for $n_1^{} = n_2^{} = 1$.
As evident from Eq.~(\ref{Bfinalresult}), the necessary modifications for the remaining cases are minor.
Fig.~\ref{fig:bubble1} shows the ratio of the finite volume corrections to the corresponding 
infinite-volume integrals as a function of~$m_1^{}L$.

\begin{table}[t]
\begin{center}
\caption{\label{tab:tadpole} 
Numerical results for the one-propagator ``tadpole'' integrals, for $m = 0.1395$~GeV, which corresponds
to $m L\approx 2.12$ ($L = 3$~fm) and $m L\approx 2.83$ ($L = 4$~fm). The corresponding continuum integrals are shown in
the column labeled $L=\infty$. The continuum results employ the $\overline{\mathrm{MS}}$ subtraction scheme
with $\mu=0.77$~GeV. Note that the ``23'' case has no continuum counterpart. All results are given in units of the appropriate
powers of GeV, and the pole configurations $n$ of the propagators are given in App.~\ref{poles}.
}
\vspace{.5cm}
\begin{tabular}{c||c|r|r|r}
& $n$ & \multicolumn{1}{c|}{$L=3$~fm} & \multicolumn{1}{c|}{$L=4$~fm} & \multicolumn{1}{c}{$L=\infty$} \\
\hline\hline
$A^V$ & $1$ &$2.99758\cdot10^{-4}$ & $7.79162\cdot10^{-5}$&$-4.21046\cdot10^{-4}$\\
$A^V$ & $2$ &$1.85663\cdot10^{-2}$ & $5.98396\cdot10^{-3}$&$1.53036\cdot10^{-2}$\\
$A^V_{22}$ & $1$ &$3.81017\cdot10^{-6}$ & $7.16805\cdot10^{-7}$&$2.34818\cdot10^{-6}$\\
$A^V_{22}$ & $2$ &$1.49879\cdot10^{-4}$ & $3.89581\cdot10^{-5}$&$-2.10523\cdot10^{-4}$\\
$A^V_{23}$ & $1$ &$-7.02467\cdot10^{-6}$ & $-1.46116\cdot10^{-6}$& \multicolumn{1}{c}{--}  \\
$A^V_{23}$ & $2$ &$-2.20354\cdot10^{-4}$ & $-6.47885\cdot10^{-5}$& \multicolumn{1}{c}{--} \\
\end{tabular}
\end{center}
\end{table}

\begin{table}[t]
\begin{center}
\caption{\label{tab:bubble} 
Numerical results for the two-propagator ``bubble'' integrals, for $m_1^{} = 0.1395$~GeV, 
$m_2^{} = 0.495$~GeV, and $p^2=m_1^2$, which corresponds
to $m_1^{} L\approx 2.12$ ($L = 3$~fm) and $m_1^{} L\approx 2.83$ ($L = 4$~fm). The corresponding continuum integrals are shown in
the column labeled $L=\infty$. The continuum results employ the $\overline{\mathrm{MS}}$ subtraction scheme
with $\mu=0.77$~GeV. Note that the ``23'' and ``33'' cases have no continuum counterpart. All results are given in units of the appropriate
powers of GeV. Only the case of $n_1^{} = n_2^{} = 1$ is given.
} 
\vspace{.5cm}
\begin{tabular}{c||r|r|r}
 & \multicolumn{1}{c|}{$L=3$~fm} & \multicolumn{1}{c|}{$L=4$~fm} & \multicolumn{1}{c}{$L=\infty$} \\
\hline\hline
$B^V$    &$1.23828\cdot10^{-3}$ & $3.21648\cdot10^{-4}$&$4.02489\cdot10^{-3}$\\
$B^V_1$  &$1.28452\cdot10^{-4}$ & $2.47609\cdot10^{-5}$&$4.97497\cdot10^{-2}$\\
$B^V_{21}$&$3.57770\cdot10^{-5}$ & $5.14256\cdot10^{-6}$& $4.57124\cdot10^{-1}$ \\
$B^V_{22}$&$1.57142\cdot10^{-5}$ & $2.96746\cdot10^{-6}$&$2.11523\cdot10^{-3}$\\
$B^V_{23}$&$-2.87678\cdot10^{-5}$ & $-6.05375\cdot10^{-6}$&  \multicolumn{1}{c}{--} \\
$B^V_{31}$&$1.65184\cdot10^{-5}$ & $1.90690\cdot10^{-6}$& $1.47521\cdot10^{-4}$ \\
$B^V_{32}$&$2.36759\cdot10^{-6}$ & $3.13466\cdot10^{-7}$&$3.23347\cdot10^{-4}$\\
$B^V_{33}$&$-5.22655\cdot10^{-6}$ & $-7.77244\cdot10^{-7}$&  \multicolumn{1}{c}{--} \\
\end{tabular}
\end{center}
\end{table}

\begin{figure}
\includegraphics[width=0.49\textwidth]{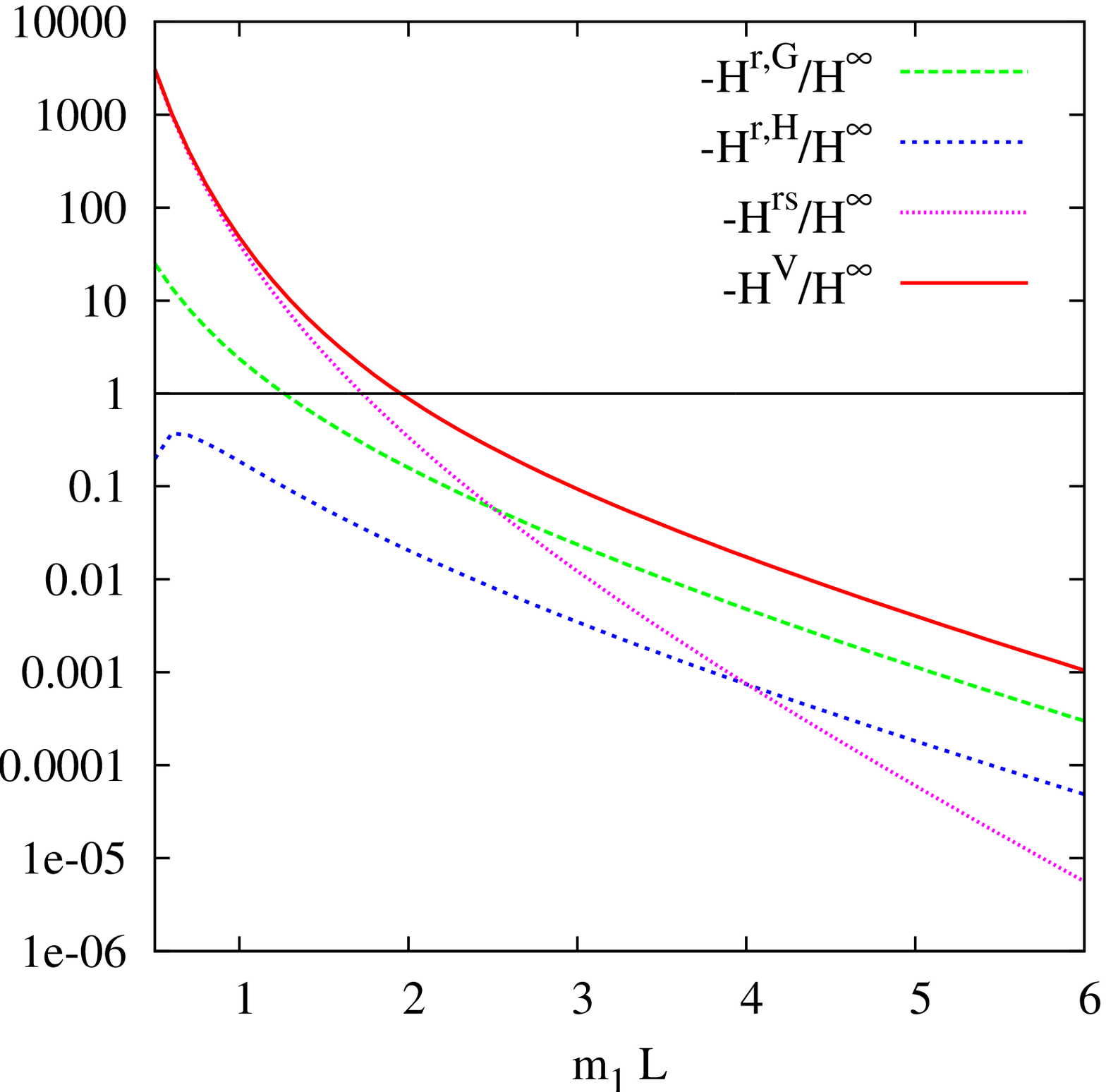}
\includegraphics[width=0.49\textwidth]{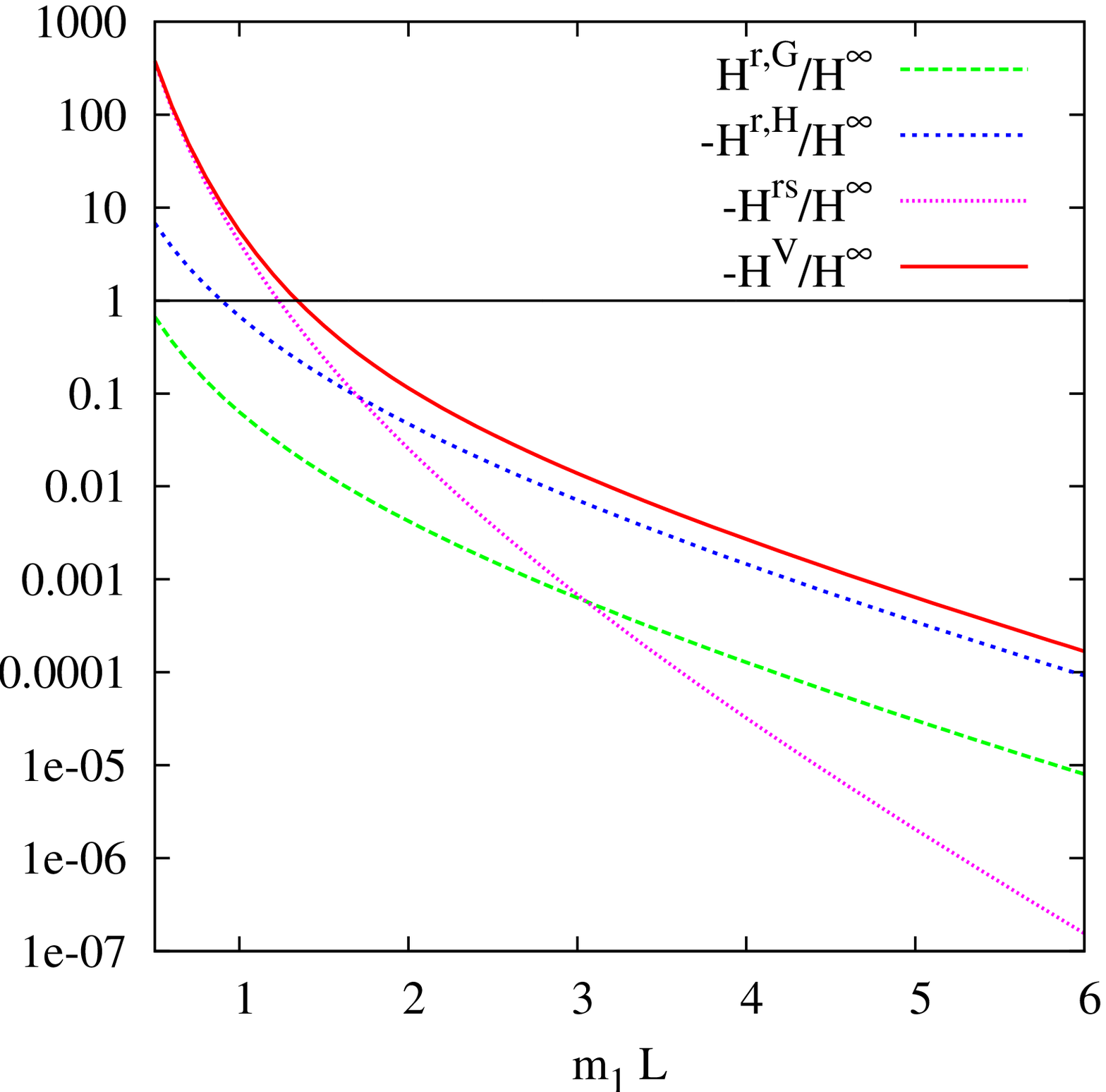}
\caption{\label{fig:sunset1} Ratio of finite-volume corrections to
infinite-volume results for the simplest sunset integrals.
The notation is according to Tab.~\ref{tab:sunset}.
In the left panel $m_1 = m_2 = m_3 = 0.1395$~GeV,
and in the right panel $m_1 = m_2 = 0.1395$~GeV with $m_3=0.495$~GeV. In both cases $p^2 = -m_3^2$.
All results employ the $\overline{\mathrm{MS}}$ scheme with $\mu=0.77$~GeV, and are
given in units of the appropriate powers of GeV. Only the case of $n_1^{} = n_2^{} = n_3^{} = 1$ is shown.}
\end{figure}

We now turn to the main objective of this study, which is an exhaustive evaluation of the sunset integrals at finite volume.
The full expressions for the sunset integrals are defined in Eq.~(\ref{defH}), where
each one is decomposed according to Eq.~(\ref{ssplit}). The components labeled $\sunset{X}_r$ are further
decomposed into a non-locally divergent part and the functions $\sunset{X}_{r,G}$ of Eq.~(\ref{Gintegral}) and
$\sunset{X}_{r,H}$ of Sect.~\ref{Hrbessel} or~\ref{Hrtheta}. The equivalent expressions for $\sunset{X}_s$
and $\sunset{X}_t$ can be obtained from the set of relations given in Eqs.~(\ref{srel}) and~(\ref{trel}).
Finally, the components labeled $\sunset{X}_{rs}$ are given in Sect.~\ref{Hrsbessel} and~\ref{Hrstheta}. 
In order to illustrate the various components of the sunset integrals, we show 
$\sunset{1}_{r,G}$, $\sunset{1}_{r,H}$, $\sunset{s}_{rs}$ and the full result $\sunset{1}$, for two
sets of input parameter values in Fig.~\ref{fig:sunset1}, relative to the infinite-volume results\footnote{
These include the finite parts of the terms containing a non-local divergence.} from Ref.~\cite{ABT2}, which are
\begin{align}
H(m_\pi^2,m_\pi^2,m_\pi^2,-m_\pi^2,\mu^2) & \approx -3.73840 \cdot 10^{-5}~\mathrm{GeV}^{-2},
\nonumber\\
H(m_\pi^2,m_\pi^2,m_K^2,-m_K^2,\mu^2) & \approx -6.74071 \cdot 10^{-5}~\mathrm{GeV}^{-2}.
\end{align}
For reference, we also provide the numerical values of the full sunset integrals as well as the
$G$ and $H$ components in Tab.~\ref{tab:sunset} for a box size of $L = 3$~fm.

\begin{table}[t]
\begin{center}
\caption{\label{tab:sunset} Numerical results for a subset of scalar components of the sunset integrals
with $n_1^{}=n_2^{}=n_3^{}=1$. The contributions $H^{r,G}_i$ are defined in terms of Eq.~(\ref{Gintegral}), using 
the decomposition into scalar components given by Eq.~(\ref{defH}). The expressions for
the $H^{r,H}_i$ are given in Sect.~\ref{Hrbessel} and~\ref{Hrtheta}, and those for $H^{rs}_i$
can be found in Sect.~\ref{Hrsbessel} and~\ref{Hrstheta}. The full results for each scalar component in
the decomposition of Eq.~(\ref{defH}) is given in the column labeled $H^V_i$ (except for cases that
involve a trivial exchange of $m_1^{}$ and $m_2^{}$). As an example, for the simplest sunset integral ($i = 0$)
we have $H^V_{} = H^{r,G}_{} + H^{r,H}_{} + H^{s,G}_{} + H^{s,H}_{} + H^{t,G}_{} + H^{t,H}_{} + H^{rs}_{}$.
All results are for $L=3$~fm, $m_1^{} = 0.1395$~GeV, $m_2^{} = 0.15$~GeV,
$m_3^{} = 0.495$~GeV, $p^2 = -0.16~$GeV$^2$ and $\mu = 0.77$~GeV, given in units of the appropriate
powers of GeV.}
\vspace{.5cm}
\begin{tabular}{c||r|r|r|r}
$i$ & \multicolumn{1}{c|}{$H^{r,G}_i$} & \multicolumn{1}{c|}{$H^{r,H}_i$} & \multicolumn{1}{c|}{$H^{rs}_i$} & 
\multicolumn{1}{c}{$H^V_i$} \\ \hline\hline
0 &$-2.20831\cdot10^{-7}$& $2.02141\cdot10^{-6}$& $5.94236\cdot10^{-7}$ & $4.05528\cdot10^{-6}$ \\
1 & \multicolumn{1}{c|}{--}  & $1.01508\cdot10^{-7}$& $6.66810\cdot10^{-8}$ & $6.04122\cdot10^{-7}$ \\
2 &$-1.10415\cdot10^{-7}$& $5.90020\cdot10^{-7}$& $7.22532\cdot10^{-8}$ & \multicolumn{1}{c}{--}  \\
\hline
21& \multicolumn{1}{c|}{--} & $9.16777\cdot10^{-9}$& $1.58703\cdot10^{-8}$ & $1.97612\cdot10^{-7}$  \\
22&$-2.80694\cdot10^{-9}$& $2.54254\cdot10^{-8}$& $6.99086\cdot10^{-9}$ &$-9.22444\cdot10^{-8}$ \\
27&$ 5.17506\cdot10^{-9}$&$-4.65135\cdot10^{-8}$&$-1.22274\cdot10^{-8}$ &$-6.10707\cdot10^{-8}$ \\
\hline
23&  \multicolumn{1}{c|}{--} & $3.56590\cdot10^{-8}$& $9.04928\cdot10^{-9}$ & $8.30916\cdot10^{-8}$ \\
24&$1.40347\cdot10^{-9}$ &$-7.90209\cdot10^{-9}$&$-9.62049\cdot10^{-10}$&$-1.38446\cdot10^{-8}$ \\
28&$-2.58753\cdot10^{-9}$& $1.44459\cdot10^{-8}$& $1.73731\cdot10^{-9}$ & $2.31182\cdot10^{-8}$ \\
\hline
25& \multicolumn{1}{c|}{--} & $2.63673\cdot10^{-7}$& $1.81386\cdot10^{-8}$ & \multicolumn{1}{c}{--} \\
26&$-8.80371\cdot10^{-9}$&$-6.26120\cdot10^{-8}$& $7.46258\cdot10^{-9}$ & \multicolumn{1}{c}{--} \\
29&$1.72502\cdot10^{-9}$ &$-6.94178\cdot10^{-9}$&$-1.33169\cdot10^{-8}$ & \multicolumn{1}{c}{--} \\
\end{tabular}
\end{center}
\end{table}

\section{Conclusions}
\label{Conclusions}

In conclusion, we have presented a complete treatment of the two-loop sunset integrals
at finite volume. We have also discussed in detail the required one-loop
integrals and shown how to expand these to higher order in $d-4$ when necessary.
As the main result of our work, we have provided complete expressions for the sunset integrals which
are suitable for numerical evaluation. Implementations of the full set of sunset integrals are also available 
from the authors in {\tt C++} and {\tt Mathematica}. The numerical evaluation
has been performed both in terms of modified Bessel functions
and theta functions, which have been shown to be numerically equivalent. Depending on the desired
quantity and precision, one of these methods is typically preferable. For moderate $m_i^{}L$, 
the sunset integrals with two quantized loop momenta are better evaluated in terms of theta functions,
as the number of terms needed in the triple summation over $l_r^2, l_s^2$ and $l_n^2$ in order to
obtain acceptable precision is quite large. For small $m_i^{}L$, the theta-function method is clearly
superior in all cases. For large $m_i^{}L$, the numerical evaluation in terms of modified Bessel functions
is usually faster.

So far, we have not shown any results on the NNLO calculations at finite volume. In the extant NNLO calculations
at infinite volume, many integral relations have been used which are no longer valid at finite volume. Therefore,
these NNLO expressions need to first be recomputed using the more general set of finite-volume sunset integrals presented here.
Work in this direction is in progress~\cite{comingup}.

\acknowledgments

This work is supported, in part, by the European Community SP4-Capacities
``Study of Strongly Interacting Matter'' (HadronPhysics3, Grant Agreement number 283286),
the Swedish Research Council grants 621-2011-5080 and 621-2010-3326 (JB, EB)
and U.S. Dept.\ of Energy grant number DE-FG02-97ER41014, and Helmholtz Association contract VH-VI-417 (TL).


\appendix

\section{Modified Bessel functions}
\label{AppBessel}

Many of the loop integrals encountered at finite volume can be expressed in terms of the modified Bessel 
functions $K_\nu^{}(z)$, and we summarize here the most significant recurring results used in the main text. 
If the integral in question is finite, the propagator factors in the denominator can be conveniently
rewritten using the Gaussian parameterization
\begin{align}
\frac{1}{a^n_{}} = \frac{1}{\Gamma(n)}
\int_0^\infty d\lambda \: \lambda^{n-1}_{} e^{-a\lambda}_{},
\label{prop}
\end{align}
upon which the relevant integrals can be brought into the form
\begin{align}
\mathcal{K}_\nu^{}(Y,Z) = 
\int_0^\infty d\lambda \: \lambda^{\nu-1}_{} e^{-Z \lambda -Y/\lambda}_{}
= 2\left(\frac{Y}{Z}\right)^{\frac{\nu}{2}}_{}
K_\nu^{} \left(2\sqrt{YZ}\right).
\label{lambdaint}
\end{align}
Also, the expansion of the finite-volume integrals to $\mathcal{O}(\varepsilon)$
around $d = 4$ generates the related functions
\begin{align}
\tilde{\mathcal{K}}_{\nu}^{}(Y,Z) & \equiv
\frac{1}{2}\ln\!\left(\frac{Y}{Z}\right) \mathcal{K}_{\nu}^{}(Y,Z)
 + 2\left(\frac{Y}{Z}\right)^{\frac{\nu}{2}}_{}
\tilde K_{\nu}^{} \left(2\sqrt{YZ}\right),
\label{Ktilde}
\end{align}
where $\tilde K_{\nu}^{}(z) \equiv \partial K_{\nu}^{}(z)/\partial\nu$ denotes the
derivative of the modified Bessel functions {\it w.r.t.} the order $\nu$.
Further, differentiation of $\mathcal{K}_{\nu}^{}(Y,Z)$ {\it w.r.t.} $p^2$
involves the functions $\mathcal{K}_{\nu}^\prime(Y,Z)$, given by
\begin{align}
\mathcal{K}_\nu^\prime(Y,Z) & \equiv \frac{\partial \mathcal{K}_\nu^{}(Y,Z)
}{\partial p^2_{}} \:=\: \frac{\partial Z(p^2)}{\partial p^2_{}}
\left(\frac{Y}{Z(p^2)}\right)^{\frac{\nu}{2}}_{} \nonumber \\
& \quad \times
\left[\left(\frac{Y}{Z(p^2)}\right)^{\frac{1}{2}}_{}
\tilde K_{\nu}^\prime \left(2\sqrt{YZ(p^2)}\right)
- \frac{\nu}{2Z(p^2)} \:\tilde K_{\nu}^{}\!\left(2\sqrt{YZ(p^2)}\right)\right],
\label{Kprime}
\end{align}
where $K_\nu^\prime(z) \equiv dK_\nu^{}(z)/dz$. For clarity, the dependence 
on $p^2$ has been made explicit in Eq.~(\ref{Kprime}). The modified Bessel functions satisfy
$K_{-\nu}^{}(z) = K_{\nu}^{}(z)$, as well as the recursion relation
\begin{align}
K_{\nu+1}^{}(z) & = \frac{2\nu}{z}\,K_\nu^{}(z) + K_{\nu-1}^{}(z).
\end{align}
The derivatives are given by
\begin{align}
K_\nu^\prime(z) &\equiv \frac{d}{dz}K_\nu^{}(z) = 
-K_{\nu-1}^{}(z)-\frac{\nu}{z}\,K_\nu^{}(z),
\end{align}
which are also directly provided by standard computer libraries for
the Bessel functions. The $\tilde K_{\nu}(z)\equiv\partial K_\nu(z)/\partial\nu$ 
can be expressed in terms of the $K_\nu^{}$ themselves via
\begin{align}
\tilde K_0^{}(z) & = 0,
\nonumber \\
\tilde K_1^{}(z) & = \frac{1}{z}\,K_0^{}(z),
\nonumber \\
\tilde K_2^{}(z) & = \frac{2}{z}\,K_1^{}(z)+\frac{2}{z^2}\,K_0^{}(z),
\nonumber \\
\tilde K_3^{}(z) & = \frac{3}{z}\,K_2^{}(z)+\frac{6}{z^2}\,K_1^{}(z)
+\frac{8}{z^3}\,K_0^{}(z),
\nonumber\\
\tilde K_n(z) & = \frac{n!}{2}\sum_{k=0}^{n-1}\left(\frac{z}{2}\right)^{k-n}
\frac{K_k(z)}{(n-k)k!},
\end{align}
where higher orders than those given explicitly are not needed
for the present considerations. Finally, for large values of $z$, the modified Bessel
functions behave as
\begin{align}
K_\nu^{}(z) = \sqrt{\frac{\pi}{2z}}\,e^{-z}_{}
+\mathcal{O}\left(\frac{e^{-z}_{}}{z^{3/2}_{}}\right),
\end{align}
which leads to an exponential fall-off for large values of the argument.

\section{Theta functions}
\label{AppTheta}

In the main text, we make use of a variety of theta functions.
For the one-loop integrals, the third Jacobi theta function
\be
\label{thetaJacobi3}
\theta_3(u|\tau) \equiv \sum_n e^{\pi i (\tau n^2 + 2 n u)},
\ee
is needed, for which an alternative definition is
\be
\theta_3(u,q) \equiv \sum_n q^{(n^2)} e^{\pi i  2 n u}
= 1+2 \sum_{n>0} q^{(n^2)} \cos(2\pi n u),
\ee
where $\tau \equiv -\frac{i}{\pi}\log q$. In the literature, the
arguments $q$ and $\tau$ are often suppressed, and the factor
of $\pi$ in the argument of the cosine may also be absent.
The Jacobi theta function is defined for $\mathrm{Im} \, \tau > 0$ or $|q| < 1$, such that the series
converges absolutely.
An important property of $\theta_3$ is the ``modulus symmetry''
\be
\label{modulus}
\theta_3(u+1 | \tau) = \theta_3(u | \tau),
\qquad
\theta_3(u|\tau) = \frac{1}{\sqrt{-i\tau}} \,
e^{-\pi i\frac{u^2}{\tau}} \,
\theta_3\left(\left.\frac{u}{\tau}\right|\frac{-1}{\tau}\right),
\ee
which is also known as Jacobi's imaginary transformation. For small $q$, the summation can be
evaluated directly, and for larger $q$ the second relation in Eq.~(\ref{modulus}) may be used to obtain rapid convergence.

We also need the Riemann theta function in $g$ dimensions, defined by
\be
\label{thetaRiemann}
\theta^{(g)}(z|\tau) \equiv
\sum_{n \in \mathbb{Z}^g_{}}
e^{2\pi i\left(\frac{1}{2}n^T \tau n+n^T z\right)},
\ee
where $n$ denotes a $g$-dimensional column vector with integer components,
$z$ is a complex, $g$-dimensional column vector and $\tau$ is a complex, symmetric matrix with a positive-definite 
imaginary part. The latter requirement ensures that the summation over $n$ converges absolutely.
We note that the most commonly encountered notation is simply $\theta$.
The Riemann theta function also satisfies a modular symmetry, generated by
the transformations
\begin{align}
\label{modular}
\theta^{(g)}(z+y|\tau) & = \theta^{(g)}(z|\tau),
\nonumber\\
\theta^{(g)}(z|\tau) & = \theta^{(g)}(az|a\tau a^T),
\nonumber\\
\theta^{(g)}(z|\tau+b) & = \theta^{(g)}(z+\frac{1}{2}\mathrm{diag}(b)|\tau),
\nonumber\\
\theta^{(g)}(\tau^{-1}z|-\tau^{-1}) & =
\sqrt{\det(-i\tau)} \,
e^{\pi i z^T \tau^{-1} z} \, \theta^{(g)}(z|\tau),
\end{align}
where
$y$ denotes a column vector with integer components, $a$ and $a^{-1}$ are both $g\times g$ matrices with
integer elements, and $b$ is a symmetric $g\times g$ matrix with integer elements as well. The use of these transformations
for the efficient evaluation of the Riemann theta function is explained in Ref.~\cite{RiemannTheta}.
The instances of the Jacobi and Riemann theta functions used in the main text are
\begin{align}
\label{deftheta}
\theta_{30}(q) &\equiv \sum_n q^{(n^2)}
 = \theta_3(u=0, q),
\nonumber\\
\theta_{32}(q) &\equiv \sum_n n^2 q^{(n^2)}e^{-xn^2}
   = q \frac{\partial}{\partial q} \, \theta_3(u=0, q), 
\nonumber\\
\theta_{34}(q) &\equiv \sum_n n^4 q^{(n^2)}
   = \left(q \frac{\partial}{\partial q}\right)^2\theta_3(u=0, q), 
\nonumber\\
\theta^{(2)}_{0}(\alpha,\beta,\gamma) &\equiv \sum_{n_1,n_2}
e^{-\alpha n_1^2 -\beta n_2^2 - \gamma (n_1-n_2)^2},
\nonumber\\
\theta^{(2)}_{02}(\alpha,\beta,\gamma) &\equiv \sum_{n_1,n_2}
 n_1^2 \, e^{-\alpha n_1^2 -\beta n_2^2 - \gamma (n_1-n_2)^2},
\end{align}
where it should be noted that $\theta^{(2)}_0(\alpha,\beta,\gamma)$ is fully symmetric in
the arguments, and that $\theta^{(2)}_{02} = -(\partial/\partial\alpha)\,\theta^{(2)}$.

\section{Integrals in arbitrary dimensions}
\label{AppIntegrals}

When the finite-volume integrals contain a non-local divergence,
the expressions 
\begin{align}
\label{feynmanintegrals}
\int \frac{d^d_{}r}{(2\pi)^d_{}} \,
\frac{1}{(r^2_{} + \Delta)^n_{}} & =
\frac{1}{(4\pi)^\frac{d}{2}_{}}
\frac{\Gamma(n-\frac{d}{2})}{\Gamma(n)} \:
\Delta^{\frac{d}{2}-n}_{}, \\
\int \frac{d^d_{}r}{(2\pi)^d_{}} \,
\frac{r_\mu^{} r_\nu^{}}{(r^2_{} + \Delta)^n_{}} & =
\frac{1}{(4\pi)^\frac{d}{2}_{}}
\frac{\Gamma(n-\frac{d}{2}-1)}{\Gamma(n)} \:
\Delta^{\frac{d}{2}-n+1}_{} \, \frac{\delta_{\mu\nu}}{2},
\end{align}
are used in Euclidean space for arbitrary dimensions $d \equiv 4-2\varepsilon$. 
As detailed in the main text, the expansion of the above results around $\varepsilon = 0$
allows for the non-local divergences to be isolated. We also
recall some further results for arbitrary $d$,
\begin{align}
\int d^d_{}r = r^{d-1}_{} dr\,d\Omega_d^{} & =
\frac{2}{\Gamma(\frac{d}{2})}\,\pi^\frac{d}{2}_{} r^{d-1}_{} dr,
\nonumber\\
\int\frac{d^d_{}\tilde r}{(2\pi)^d_{}} \, e^{-\tilde r^2}_{} & = 
\frac{1}{(4\pi)^{\frac{d}{2}}_{}},
\nonumber\\
\int\frac{d^d_{}\tilde r}{(2\pi)^d_{}} \, \tilde r^2_{} e^{-\tilde r^2}_{} & =
\frac{1}{(4\pi)^{\frac{d}{2}}_{}}\,\frac{d}{2},
\nonumber\\
\int\frac{d^d_{}\tilde r}{(2\pi)^d_{}} \, \tilde r^4_{} e^{-\tilde r^2}_{} & =
\frac{1}{(4\pi)^{\frac{d}{2}}_{}}\,\frac{d}{2}\!\left(\frac{d}{2}+1\right),
\label{simpint}
\end{align}
which are used throughout the main text.

\section{Notation for double poles}
\label{poles}

\begin{table}[b]
\begin{center}
\caption{\label{tab:poles} Table of ``pole configurations'', {\it i.e.} the relationship between the collective index $n$ and
the exponents $n_1^{}$, $n_2^{}$ and $n_3^{}$ of the propagator factors $(p^2 + m_i^2)^{n_i^{}}$ in the sunset integrals.}
\vspace{.5cm}
\begin{tabular}{c||ccc}
$n$ & $n_1^{}$ & $n_2^{}$ & $n_3^{}$ \\ \hline
$1$ & $1$ & $1$ & $1$ \\
$2$ & $2$ & $1$ & $1$ \\
$3$ & $1$ & $2$ & $1$ \\
$4$ & $1$ & $1$ & $2$ \\
$5$ & $2$ & $2$ & $1$ \\
$6$ & $2$ & $1$ & $2$ \\
$7$ & $1$ & $2$ & $2$ \\
$8$ & $2$ & $2$ & $2$
\end{tabular}
\end{center}
\end{table}

In the main text, the notation $A(n,m^2)$ and $B(n_1^{},n_2^{},m_1^2,m_2^2,p^2)$ has been used for the 
one-loop integrals with one and two propagators, respectively. However, we wish to remind the reader that
the established notation in the literature reserves the symbol $A$ for $A(1,m^2)$ and the symbol $B$
for $B(1,1,m_1^2,m_2^2,p^2)$. Along these lines, integrals with three and four propagators are 
usually denoted $C$ and $D$, respectively. 


For the sunset integrals in PQ$\chi$PT, some or all of the 
propagators can appear doubled. This gives eight possible configurations of
single and double poles.
In earlier NNLO work on PQ$\chi$PT, a collective index $n$ was introduced to specify
the pole configuration~\cite{Lahde1,Lahde2,Lahde3}, as a short-hand notation 
for the triplet ($n_1^{},n_2^{},n_3^{}$). The correspondence is shown in Tab.~\ref{tab:poles}.
It should be noted that the cases of $n = 4$ and $n = 6$ are superfluous due to integral relations,
and the case of $n = 8$ appears only in calculations of the flavour-neutral meson properties 
in PQ$\chi$PT. 

\section{Translation to Minkowski conventions}

While we have used the Euclidean formalism throughout, it is also of interest to recall how the expressions
for the one-loop and sunset integrals can be translated to Minkowski conventions. The required substitutions are
\begin{align}
\int \frac{d^qr}{(2\pi)^d} & \longrightarrow \frac{1}{i}\int \frac{d^qr}{(2\pi)^d},
\nonumber\\
\delta_{\mu\nu} & \longrightarrow -g_{\mu\nu}
\nonumber\\
p\cdot q, ~p^2 & \longrightarrow -p\cdot q,~-p^2
\nonumber\\
t_{\mu\nu} & \longrightarrow - t_{\mu\nu}
\nonumber\\
\frac{1}{p^2+m^2} & \longrightarrow -\frac{1}{p^2-m^2},
\end{align}
where $t_{\mu\nu}$ corresponds to the
spatial part of the metric.

\end{document}